\documentclass[12pt]{article}
\usepackage{jheppub}

\makeatletter
\def\@fpheader{\relax}
\makeatother

\preprint{IGC-18/3-1\\[-2.5em]}

\newcommand{\ii}{\mathrm{i}}
\newcommand{\ex}{e}
\newcommand{\pp}{\mathbf{p}}

\newcommand{\qq}{\mathbf{q}}
\newcommand{\xx}{\mathbf{x}}
\newcommand{\yy}{\mathbf{y}}

\usepackage[bb=boondox]{mathalfa}

\usepackage{tocvsec2}
\settocdepth{subsection}

\def\ie{{\it i.e.},\ }
\def\eg{{\it e.g.},\ }

\newcommand{\be}{\begin{equation}}
\newcommand{\ee}{\end{equation}}
\newcommand{\beq}{\begin{equation}}
\newcommand{\eeq}{\end{equation}}
\newcommand{\beqa}{\begin{eqnarray}}
\newcommand{\eeqa}{\end{eqnarray}}
\newcommand{\beqar}{\begin{eqnarray*}}
\newcommand{\eeqar}{\end{eqnarray*}} 

\newcommand{\reef}[1]{(\ref{#1})}
\newcommand{\ssc}{\scriptscriptstyle}
\newcommand{\mt}[1]{\textrm{\tiny #1}}

\newcommand{\vt}{\vartheta}
\newcommand{\vti}{\vartheta_{\!I}}
\newcommand{\vp}{\varphi}

\newcommand{\ket}[1]{|#1\rangle}

\newcommand{\cev}[1]{\reflectbox{\ensuremath{\vec{ \reflectbox{\ensuremath{#1}}}}}}

\newcommand{\veps}{\varepsilon}
\newcommand{\s}{\sigma}
\newcommand{\lp}{\left(}
\newcommand{\rp}{\right)}
\newcommand{\op}{\mathcal{O}}
\newcommand{\cO}{\mathcal{O}}

\newcommand{\pd}{\partial}
\newcommand{\dd}{d} 
\newcommand{\br}{{\bar r}}
\newcommand{\bs}{{\bar s}}
\newcommand{\brp}{{{\bar r}'}}
\newcommand{\bsp}{{{\bar s}'}}
\newcommand{\bp}{|\mathbf{p}|}

\title{Circuit complexity for free fermions}

\author[a]{Lucas Hackl}
\emailAdd{lucas.hackl@psu.edu}
\affiliation[a]{Institute for Gravitation and the Cosmos \& Physics Department,\\ Penn State, University Park, PA 16802, USA}

\author[b]{and Robert C. Myers}
\emailAdd{rmyers@perimeterinstitute.ca}
\affiliation[b]{Perimeter Institute for Theoretical Physics,\\ 31 Caroline St N, Waterloo, ON N2L 2Y5, Canada}

\abstract{
We study circuit complexity for free fermionic field theories and Gaussian states. Our definition of circuit complexity is based on the notion of geodesic distance on the Lie group of special orthogonal transformations equipped with a right-invariant metric. After analyzing the differences and similarities to bosonic circuit complexity, we develop a comprehensive mathematical framework to compute circuit complexity between arbitrary fermionic Gaussian states. We apply this framework to the free Dirac field in four dimensions where we compute the circuit complexity of the Dirac ground state with respect to several classes of spatially unentangled reference states. Moreover, we show that our methods can also be applied to compute the complexity of excited states. Finally, we discuss the relation of our results to alternative approaches based on the Fubini-Study metric, the relevance to holography and possible extensions.
}

\begin{document}
\maketitle


\section{Introduction}

Recently, holographic complexity has been suggested as a new tool with which to gain insight in the role of entanglement in the emergence of spacetime geometry in quantum gravity \cite{Susskind:2014rva,Susskind:2014jwa,Susskind:2014moa, Stanford:2014jda, Brown:2015bva,Brown:2015lvg}. In particular, it has drawn attention to new gravitational observables to probe the bulk spacetime in the holographic theories. The complexity=volume (CV) conjecture suggests that complexity is dual to the volume of an extremal (codimension-one) bulk surface anchored to a certain time slice in the boundary \cite{Stanford:2014jda,Susskind:2014rva}. Alternatively, the  complexity=action (CA) conjecture  identifies the complexity with the gravitational action evaluated on a particular bulk region, known as the Wheeler-DeWitt (WDW) patch, which is again anchored on a boundary time slice \cite{Brown:2015bva,Brown:2015lvg}.\footnote{One can think of the WDW patch as the causal development of the spacelike extremal surface picked out in the CV construction.}

Both of the holographic complexity conjectures point out new classes of interesting gravitational observables and there has been a growing interest in studying these new observables and the corresponding conjectures, \eg \cite{Alishahiha:2015rta,Cai:2016xho,Lehner:2016vdi, Yang:2016awy,Chapman:2016hwi,Carmi:2016wjl, Carmi:2017jqz,Reynolds:2016rvl,Zhao:2017iul,Reynolds:2017lwq, Couch:2017yil,Swingle:2017zcd,Fu:2018kcp}. At present, both conjectures appear to provide viable candidates for holographic complexity, but this research program is still at a preliminary stage. While understanding the properties of the new gravitational observables certainly deserves further study, providing concrete, even qualitative, tests of the two conjectures is hampered because we lack a good understanding of what complexity actually means in the boundary CFT, or in quantum field theory, more generally. Certainly, this lack of understanding stands as an obstacle to constructing a precise translation between the new bulk observables and specific quantities in the boundary theory, \eg in an analogous way that the translation of the replica construction in the boundary yielded a derivation of holographic entanglement entropy \cite{CHM,Lewkowycz:2013nqa,Dong:2016hjy}. Beyond gaining new insights into holographic complexity, developing an understanding of complexity in quantum field theory is an interesting research program in its own right. For example, it may lead to progress in quantum simulations of field theories, \eg \cite{Jordan:2011ne,Jordan:2011ci, Jordan:2014tma,Jordan:2017lea}, or in our understanding of Hamiltonian complexity, \eg \cite{osborne2012hamiltonian, gharibian2015quantum} and  the description of many-body wave functions, \eg \cite{Orus:2013kga,vidal2009entanglement}.

Recently, some preliminary steps were taken to provide a precise definition of circuit complexity in quantum field theories, \eg \cite{Hashimoto:2017fga,Jefferson:2017sdb,Chapman:2017rqy, Yang:2017nfn,Kim:2017qrq,simonR,Khan:2018rzm,prep}. The present paper extends these investigations by examining complexity in a free fermionic quantum field theory. Our current investigation is closely related to the discussions in  refs.~\cite{Jefferson:2017sdb,Chapman:2017rqy}, which studied the ground state complexity of a free scalar field theory.  In particular, ref.~\cite{Jefferson:2017sdb} adapted a geometric approach, which was developed by Nielsen and collaborators \cite{Nielsen:2005mn1,Nielsen:2006mn2,Nielsen:2007mn3}, to evaluate circuit complexity in a scalar field theory, and here we apply Nielsen's approach  to defining the complexity of states in a fermionic field theory. We might note that a possible connection between Nielsen's approach and holographic complexity had been advocated by Susskind \cite{Susskind:2014jwa,Brown:2016wib,Brown:2017jil}, but further, the complexity for the free scalar \cite{Jefferson:2017sdb} was found to show some surprising similarities to holographic complexity, despite the enormous differences between the quantum field theories appearing in these two settings. We should also point out that ref.~\cite{Chapman:2017rqy} developed an alternative approach of defining complexity for the free scalar field theory using the Fubini-Study metric, which matched many results found using Nielsen's approach.\footnote{We also refer the interested reader to ref.~\cite{alvarez1999comment}, which introduces an interesting connection between quantum algorithms and geodesics on the Fubini-Study metric.} Even though, we will focus on Nielsen's approach for the fermionic theory, we will also comment on this alternative approach, as well as point out differences and similarities with the scalar theory. 

The remainder of the paper is organized as follows: In section \ref{preamble}, we provide a brief review of Nielsen's geometric approach to evaluating circuit complexity and we introduce a group theoretic perspective that naturally arises in applying this technique to evaluate the complexity of quantum field theory states. In section \ref{sec:prelude}, we continue to develop this group theoretic approach by first reviewing its application to Gaussian states in free scalar field theories \cite{Jefferson:2017sdb,prep}. This review then sets the stage to extend this technique to examine the complexity of Gaussian states in free fermionic theories, which is discussed in section \ref{twofermi}. As well as discussing the salient features of the application to a general theory of $N$ fermionic degrees of freedom, we present some explicit calculations for the simple case of two fermions. We can then apply the previous analysis to evaluate the complexity of the ground state of a free Dirac field in section \ref{sec:Dirac}. In section \ref{excited1}, we also evaluate the complexity of certain excited states. Section \ref{sec:general} presents a general framework which allows one to evaluate the circuit complexity of arbitrary Gaussian states in any fermionic theory. In section \ref{apply}, we apply this general method to further examine complexity of the free Dirac field. In particular, we investigate how the complexity of the ground state is effected by alternate choices for the reference state, and also the complexity of more general excited states. We close with a brief discussion of our results and of possible future directions in section \ref{discuss}. We leave some additional technical details for appendices. In appendix \ref{app:geodesics}, we discuss a particular class of simple geodesics on general Lie groups, which are relevant in our application of Nielsen's approach to quantum field theories. Appendix \ref{app:minimal} provides a general construction of the minimal geodesics connecting an arbitrary reference state to any desired target state in a fermionic theory.\\

\noindent{\bf Note:} While the present paper was in preparation, ref.~\cite{simonR,Khan:2018rzm} appeared which also address the question of circuit complexity of free fermionic theories. In particular, there is a strong overlap with our study of the ground state complexity in section \ref{sec:Dirac}. However, we would like to note that our approach adopts a more abstract group theoretic formalism, which allows us to prove \eg that our unitary circuits in fact correspond to minimal geodesics, which is lacking in \cite{simonR,Khan:2018rzm}. Further, we evaluate the complexity of the ground state for a variety of different reference states, and we also consider the complexity of various excited states. 

\section{Complexity, Nielsen and group theory} \label{preamble}

The concept of complexity stems from the notion of computational complexity in computer science \cite{CCintroduction,CCintroduction2}.  The question of interest is to ask how much of certain computational resources are required to solve a given task. For a digital computer, we can ask what is minimal number of computational gates required to implement a specific algorithm, \ie a specific map between a certain sets of input bits and output bits. This question readily extends to quantum information science where the question becomes what is the minimal number of gates chosen from some set of elementary unitaries $\{V_I\}$ to implement a unitary transformation $U$, which produces a desired map from some $n$-qubit inputs to the corresponding $n$-qubit outputs \cite{Aaronson:2016vto,watrous}. An implementation of $U$ becomes a string of elementary unitaries, \ie $U=\prod^D_{k=1}V_{I_k}$ where $D$ defines the circuit depth of this particular implementation. The circuit complexity then corresponds to the depth of the optimal construction, \ie the minimal number of gates needed to build $U$. To be even more precise, it is rarely possible to write a given $U$ \emph{exactly} as a finite string of discrete gates $V_I$, but rather only up an error $\epsilon$. Hence the circuit complexity of a unitary transformation $U$ is usually defined with respect to some gate set $\{V_I\}$ and a given tolerance $\epsilon$ as the minimal number of $V_I$ required to implement $U$, up to an error of $\epsilon$.

In the context of holography, or in applying these concepts to quantum field theory, we are interested in quantifying the effort required to prepare a certain target state $\ket{\psi_\mt{T}}$ from a specific reference state $\ket{\psi_\mt{R}}$ by applying a sequence of unitary gates. Here, $\ket{\psi_\mt{R}}$ will be chosen with some notion of simplicity in mind, \eg the degrees of freedom are completely unentangled. Hence the complexity of a family of target states is defined with respect to the reference state $\ket{\psi_\mt{R}}$, as well as the gate set $\{V_I\}$ and the tolerance $\epsilon$.\footnote{Hence the concept of state complexity differs slightly from the computational complexity introduced above. The later requires constructing the optimal $U$ which implements a particular map for many different inputs. With a state complexity, we consider a single fixed input (\ie  the reference state) and construct a new (optimal) circuit for each output (\ie the target states). This differences introduces an ambiguity in the boundary conditions, as explained in the discussion around eq.~\reef{bc44}.} Again, we wish to construct the optimal unitary or shortest circuit which implements
\beq
\ket{\psi_\mt{T}}=U\,\ket{\psi_\mt{R}}~,\label{circuitDef}
\eeq
and the complexity of the state $\ket{\psi_\mt{T}}$ is simply defined as the number of elementary gates comprising this optimal $U$. Of course, generally there will exist infinitely many different sequences of gates which produce the same target state from a given reference state. Hence, our challenge is to identify the optimal circuit from amongst the infinite number of possibilities.

Nielsen and collaborators \cite{Nielsen:2005mn1,Nielsen:2006mn2,Nielsen:2007mn3}, introduced a geometric approach to identify the optimal circuit, which was adapted in \cite{Jefferson:2017sdb} to evaluate the complexity of the ground state of a free scalar field. In contrast to the previous discussion, where $U$ is constructed as a string of discrete gates, this new approach begins with a continuous description of the unitary
\beq
U=\cev{\cal P} \exp\left[-i \int_0^1ds\ H(s) \right]\qquad{\rm where}\ \ 
H(s)=\sum_I Y^I(s)\,\cO_I~,\label{controlY}
\eeq
where the `time-dependent Hamiltonian' $H(s)$ is expanded in terms of a basis of Hermitian operators $\cO_I$, and the $\cev{\cal P}$ indicates a `time' ordering such that the circuit is built from right to left as $s$ increases.\footnote{Note that our notation here differs slightly from that in \cite{Jefferson:2017sdb} where the overall factor of $-i$ was absorbed in the $\cO_I$, which were then anti-Hermitian operators.} Here one might think of the elementary gates taking the form $V_I=\exp[-i\veps \cO_I]$ where $\veps$ is some small parameter, and then the control functions $Y^I(s)$ indicate which gates are being applied at a given time $s$ in the circuit represented by eq.~\reef{controlY}. Further, rather than only considering the complete circuit \reef{controlY}, Nielsen extends this construction to consider trajectories in the space of unitaries,
\beq
U(s)=\cev{\cal P} \exp\left[-i\int_0^s \dd\tilde s\ H(\tilde s)
\right]\ .
\label{row}
\eeq
In this space, the circuits of interest are the trajectories satisfying the boundary conditions $U(s=0)=\mathbb{1}$ and $U(s=1)=U$.\footnote{We define the boundary conditions more precisely below in the discussion around eq.~\reef{bc44}.} In this framework,  $\vec{Y}(s)=(Y^1(s),Y^2(s),\cdots)$ can also be interpreted as the tangent vector of the corresponding trajectory, 
\beq
Y^I(s)\,\op_I=\pd_sU(s)\,U^{-1}(s)~.\label{v12}
\eeq
Let us also note that there is no need to consider a tolerance $\epsilon$ with this continuous description, since the $\vec{Y}(s)$ can always  be adjusted to produce exactly the desired transformation \reef{circuitDef}.

Now Nielsen's approach is to optimize the circuit \reef{controlY} by minimizing a particular \emph{cost} defined by
\beq
{\cal D}(U(t))=\int_0^1\dd s\ F\!\lp U(s),\vec{Y}(s)\rp~,\label{costco}
\eeq
where the cost function $F(U,\vec{Y})$ is a local functional along the trajectory of the position $U(s)$ and the tangent vector $\vec{Y}(s)$. 
Some simple examples would include:
\beqa
F_1(U,\vec Y)&=&\sum_I \left|Y^I\right|\,,\qquad\quad\ 
F_{1p}(U,\vec Y)=\sum_I p_I \left|Y^I\right|\,,
\nonumber\\
F_2(U,\vec Y)&=&\sqrt{\sum_I \lp Y^I\rp^2}\,,\qquad
F_\kappa(U,\vec Y)=\sum_I \left| Y^I\right|^\kappa \,.
\label{eq:Fmetrics}
\eeqa
Given the interpretation of the $Y^I$ as indicating when certain gates appear in the circuit, the $F_1$ measure is the closest to the original definition of simply counting the number of gates in the circuit. In $F_{1p}$, penalty factors $p_I$  are introduced to favour certain directions in the circuit space over others, \ie to give a higher cost to certain classes of gates. Of course, the $F_2$ measure can be recognized as the proper distance in a Riemannian geometry on the space of unitaries. This choice will be the focus of much of our discussion in the following. The $\kappa$ measures $F_\kappa$ were introduced in \cite{Jefferson:2017sdb} because the resulting complexity compared well with results for holographic complexity. Of course, with $\kappa=2$, the $F_\kappa$ measure yields the same optimal trajectories as $F_2$ with a test particle action in the corresponding geometry, while with $\kappa=1$, this reverts back to the $F_1$ measure. We return to discussing the relative merits of these measures in more detail in section~\ref{discuss}. 

In applying the above approach to a free scalar field theory in \cite{Jefferson:2017sdb}, a group theoretic structure was found to naturally appear. To produce a tractable problem, only a limited basis of operators $\op_I$ were used in constructing the unitary circuit \reef{controlY} and these operators naturally formed a closed  algebra, \ie a Lie algebra $\mathfrak{g}$ with $[\op_I,\op_J]=i f_{IJ}{}^K \op_K$. In \cite{Jefferson:2017sdb}, a $\mathrm{GL}(N,\mathbb{R})$ algebra appeared in the construction of the free scalar ground state using a lattice of $N$ bosonic degrees of freedom.\footnote{This was extended to an $\mathrm{Sp}(2N,\mathbb{R})$ algebra in \cite{prep}, as discussed below.} In the following discussion of free fermions, we will be making use of the analogous group structure, which turns out to be $\mathrm{O}(2N)$. One advantage of this group theoretic perspective is that the physical details of the operators $\op_I$ become less important. Rather, we can simply think of the generators in eq.~\reef{controlY} as the elements of the Lie algebra $\mathfrak{g}$ and the circuits are then trajectories in the corresponding group manifold $\mathcal{G}$, without making reference to a specific representation, or rather we can choose whichever representation is most convenient for our calculations.

Let us phrase the preceding description of Nielsen's approach in the corresponding group theoretic language --- see appendix \ref{app:geodesics} for further discussion. In particular, the circuits \reef{controlY} of interest become continuous trajectories $\gamma: [0,1]\to \mathcal{G}$ which connect the identity $\mathbb{1}$ with the desired unitary transformation $U$. In identifying the elementary generators with a basis of the Lie algebra $\mathfrak{g}$, we are presented a natural cost function which is inherited from the geometry of the underlying group structure. That is, we restrict ourselves to a cost function
\begin{align}\label{metA}
\lVert A\rVert=\sqrt{\langle A,A\rangle_{\mathbb{1}}}
\end{align}
that is induced by a positive definite metric $\langle\cdot,\cdot\rangle_\mathbb{1}: \mathfrak{g}\times \mathfrak{g}\to\mathbb{R}$ on the Lie algebra.\footnote{Here, we use the standard identification of the Lie algebra $\mathfrak{g}$ with the tangent space $T_\mathbb{1}\mathcal{G}$ at the identity.} If we extend a circuit $U\to \ex^{-\veps A}\,U$ by applying the gate $\exp[-\veps A]$ from the right, then $\delta U \approx -i\veps A\, U$ and we expect that the length of the circuit should increase by a step $\veps \lVert A\rVert$, irrespective of the precise form of $U$, or equivalently that the tangent vector $A\,U\in T_UG$ has the same length as $A\in T_\mathbb{1}U$. We can therefore extend the metric $\langle\cdot,\cdot\rangle_{\mathbb{1}}$ to arbitrary tangent spaces via right-translation, leading to the right-invariant metric
\begin{align}\label{metB}
\langle X,Y\rangle_U=\langle XU^{-1},YU^{-1}\rangle_{\mathbb{1}}\,.
\end{align}
Using the $F_2$ cost function, the circuit complexity of a given $U\in \mathcal{G}$ is then defined as the minimal path length
\begin{align}\label{metC}
\mathcal{C}_2(U)=\min_{\gamma}\int_0^1\!\!\!dt\,\lVert\dot{\gamma}(t)\rVert\,,
\end{align}
which is nothing else than the \emph{geodesic distance} between $\mathbb{1}$ and $U$ on $\mathcal{G}$, which was turned into a Riemannian manifold by the metric $\langle\cdot,\cdot\rangle_U$. If instead, we wished to consider the $F_{\kappa=2}$ measure, the circuit complexity becomes
\begin{align}\label{metCa}
\mathcal{C}_{\kappa=2}(U)=\min_{\gamma}\int_0^1\!\!\!dt\,\lVert\dot{\gamma}(t)\rVert^2\,,
\end{align}

The group theoretic perspective proves to be quite powerful in evaluating the circuit complexity of simple states in quantum field theory, as was already implicitly seen with the analysis in \cite{Jefferson:2017sdb}. In the following, we will apply the tools of Lie theory and the study of symmetric spaces to examine fermionic Gaussian states. In this case, we can restrict our attention to the group $\mathcal{G}=\mathrm{O}(2N)$ for $N$ fermionic degrees of freedom. Taking $N\to\infty$ then leads to the continuum limit of a fermionic field theory. We will be able solve for the minimal geodesic analytically using the metric $\langle\cdot,\cdot\rangle_\mathbb{1}$, which is compatible with the group structure.\\

\begin{figure}[t]
	\begin{center}
		\includegraphics{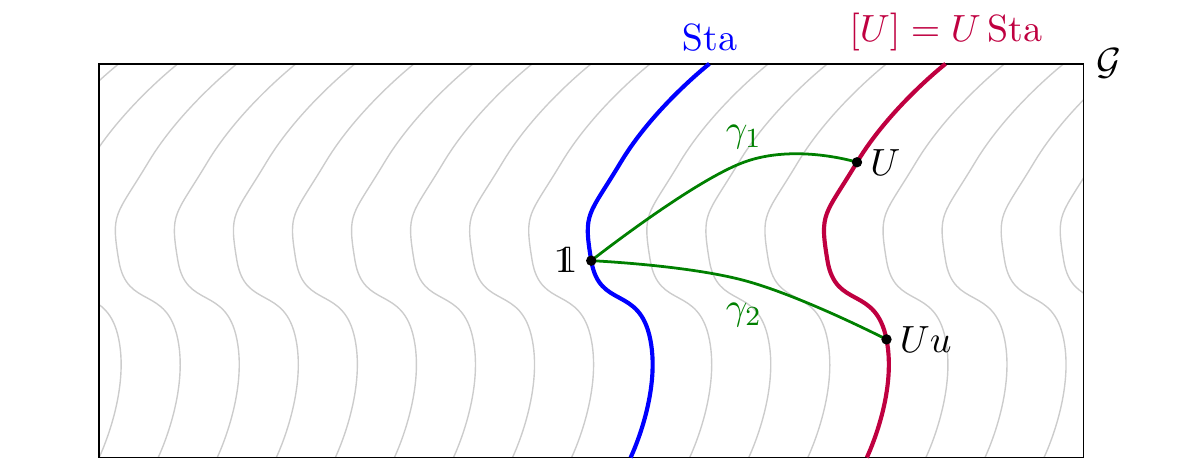}
	\end{center}
	\caption{This figure illustrates the geometry of the Lie group $\mathcal{G}$ with stabilizer subgroup $\mathrm{Sta}$, whose elements $u$ satisfy $u|\psi_{\mathrm{R}}\rangle=|\psi_{\mathrm{R}}\rangle$. This subgroup induces a fibration of $\mathcal{G}$ into equivalence classes given by displaced stabilizers $[U]=U\,\mathrm{Sta}$. The complexity of a target state $|\psi_\mathrm{T}\rangle=U|\psi_\mathrm{R}\rangle$ is then given by the minimal path $\gamma$ to a point on $[U]$, of which we illustrate two examples. $\gamma_1$ goes from $\mathbb{1}$ to $U$ and $\gamma_2$ from $\mathbb{1}$ to $Uu$ where $u\in\mathrm{Sta}$.}
	\label{fig0}
\end{figure}

In closing this section, let us add the following aside: Recall that evaluating the complexity of a given target state amounts to finding the optimal circuit $U$ which produces the desired transformation in eq.~\reef{circuitDef}. However, this prescription typically does not actually fix the boundary condition $U(s=1)$. That is, one will find that there are simple transformations $u$ which leave the reference state invariant, \ie $|\psi_{\mathrm{R}}\rangle=u|\psi_{\mathrm{R}}\rangle$ and then given any unitary $U_0$ satisfying eq.~\reef{circuitDef}, $U=U_0\,u$ will produce the desired transformation as well. This ambiguity is elegantly characterized in our group theoretic approach if we define the stabilizer subgroup
\begin{align}\label{bc44}
\mathrm{Sta}=\{u\in\mathcal{G}\ \mathrm{s.t.}\ u|\psi_{\mathrm{R}}\rangle=|\psi_{\mathrm{R}}\rangle\}\,,
\end{align}
that preserves $|\psi_{\mathrm{R}}\rangle$. We can then define the equivalence relation $U\sim V$ iff $U=V\,u$ with $u\in\mathrm{Sta}$, \ie iff $U|\psi_{\mathrm{R}}\rangle=V|\psi_{\mathrm{R}}\rangle$. Hence the problem of finding the minimal circuit now involves a double extremization. First, we must find the family of geodesics running from $\mathbb{1}$ to all unitaries in the equivalence class $[U]\in \mathcal{G}/\mathrm{Sta}$. Secondly, we must find the shortest geodesic amongst this family. Note that the equivalence class $[U]$ is just given by $U\,\mathrm{Sta}$, where we displace the stabilizer by multiplying with an arbitrary representative $U$ from the left. We illustrate the involved geometry in figure~\ref{fig0}.

In the setting of bosonic and fermionic Gaussian states, we have $\mathrm{Sta}=\mathrm{U}(N)$ and the quotient manifolds $\mathcal{G}/\!\!\sim$ turn out to be given by symmetric spaces \cite{mimura1991topology}, namely type DI corresponding to $\mathrm{Sp}(2N,\mathbb{R})/\mathrm{U}(N)$ for bosons and type CIII corresponding to $\mathrm{SO}(2N)/\mathrm{U}(N)$ for fermionic systems.

\section{Prelude: fermions versus bosons}\label{sec:prelude}

In developing our complexity model for free fermions, we are interested in describing fermionic Gaussian states and the unitary transformations that map Gaussian states onto Gaussian states. 
As we discuss below, this problem naturally involves the group $\mathrm{O}(2N)$ for $N$ fermionic degrees of freedom.  
Nielsen's approach to defining circuit complexity was recently applied for free scalars \cite{Jefferson:2017sdb}, which required understanding the analogous unitary transformations mapping bosonic Gaussian states amongst themselves. A $\mathrm{GL}(N,\mathbb{R})$ structure arose in this analysis but this is only a subgroup of the full $\mathrm{Sp}(2N,\mathbb{R})$ family of transformations, as we review below \cite{prep,BH-BFGaussians}.
Hence it is useful to begin here by comparing and contrasting the bosonic and fermionic Gaussian states.

As emphasized above, the precise representation of the unitary circuits becomes unimportant with our group theoretic perspective.
We use this freedom here to focus on the simple description of the group of transformations mapping  Gaussian states amongst themselves given in terms of their action on the covariance matrix, \eg \cite{BH-BFGaussians}. In particular, we will parametrize Gaussian states in terms of their covariance matrix,
\begin{align}\label{covmat}
	\langle\psi|\,\xi^a\,\xi^b\,|\psi\rangle=\frac{1}{2}(G^{ab}+\ii\, \Omega^{ab})
\end{align}
where $\xi^a\equiv(q_1,\cdots,q_N,p_1,\cdots,p_N)$ describes $N$ degrees of freedom, which may be either bosonic or fermionic. On the right-hand side, $G^{ab}=G^{(ab)}$ is the symmetric part of the correlation matrix on the left, while $\Omega^{ab}=\Omega^{[ab]}$ denotes the antisymmetric part.

\noindent\textbf{Bosons:} For a system of bosonic degrees of freedom, $\Omega^{ab}$ is trivial in that it simply encodes the canonical commutation relations of the $q_i$'s and $p_i$'s. On the other hand, the symmetric two-point function $G^{ab}$ completely characterizes the corresponding Gaussian state  $|\psi\rangle$ --- we are assuming that $\langle\psi|\xi^a|\psi\rangle=0$ here and in the rest of this paper.  Hence, as described in \cite{BH-BFGaussians} and below, a simple description of the group of transformations mapping bosonic Gaussian states amongst themselves is then given in terms of their action on the symmetric covariance matrix.

\noindent\textbf{Fermions:} When we consider eq.~\reef{covmat} for a fermionic system instead, the symmetric part $G^{ab}$ is fixed by the anti-commutation relations amongst the fermionic degrees of freedom while the antisymmetric part $\Omega^{ab}$ completely characterizes the fermionic Gaussian state $|\psi\rangle$. Hence the covariance matrix \reef{covmat} again provides a simple framework to discuss the corresponding group of unitary transformations for fermionic Gaussian states, as we discuss in the following.

\subsection{Single boson} \label{boso}
It is well known that the group of transformations preserving Gaussian states for $N$ bosonic degrees of freedom is  $\mathrm{Sp}(2N,\mathbb{R})$, as explained in \cite{Bianchi:2015fra,prep,BH-BFGaussians}. As above, we assemble the conjugate position and momentum operators as $\xi^a\equiv(q_1,\cdots,q_N,p_1,\cdots,p_N)$. Then the antisymmetric component of the covariance matrix \reef{covmat} becomes
\begin{align}\label{asymb}
	\Omega^{ab}= -\ii\,\langle\psi|\,[\xi^a,\xi^b]\,|\psi\rangle\equiv\left(\begin{array}{cc}
\ \mathbb{0}&\mathbb{1}\\
	-\mathbb{1}&\mathbb{0}
	\end{array}\right)\,,
\end{align}
where $\mathbb{1}$ and $\mathbb{0}$ are $N\!\times\!N$ identity and zero matrices, respectively. This result holds for any Gaussian state (or in fact, any state), since the canonical commutation relations can be written as $[\xi^a,\xi^b]=\ii\, \Omega^{ab}$. The nontrivial component of eq.~\reef{covmat} is then the symmetric two-point function 
\beq
G^{ab}= \langle\psi|\,\{\xi^a,\xi^b\}\,|\psi\rangle\,,
\label{symb}\eeq 
which gives a complete characterization of the Gaussian state  $|\psi\rangle$.

As a simple example, we consider one bosonic degree of freedom, so the group of interest is simply $\mathrm{Sp}(2,\mathbb{R})$ and we have $\xi^a\equiv(q,p)$. Now, another
way to characterize the Gaussian states is in terms of annihilation and creation operators,\footnote{To correctly account for the dimensions of $q$ and $p$, these expressions should include a specific scale, \eg $a=\frac{1}{\sqrt{2}}(\omega_1\, q+\ii\,p/\omega_1)$ yields a properly dimensionless annihilation operator. One effect of the Bogoliubov transformations \reef{bogo} is then to scale this scale, \eg $\omega_1\to \ex^r\omega_1$ with $\vp=\vt=0$ in eq.~\reef{psycho}.  See \cite{prep} for further discussion.} \eg
\beq
a=\frac{1}{\sqrt{2}}(q+\ii\,p)\,, \qquad
a^\dagger=\frac{1}{\sqrt{2}}(q-\ii\,p)\,.
\label{aadag}
\eeq
That is, given these operators, there is a corresponding Gaussian state satisfying $a\,|\psi\rangle=0$. However, there is some freedom in the precise definition the annihilation operator, namely the Bogoliubov transformations,\footnote{Note that we can change $a$ to $\tilde{a}=\ex^{\ii \varphi}a$ without changing the vacuum, which corresponds to a $\mathrm{U}(1)$ subgroup of Bogoliubov transformations that do not change the vacuum. For $N$ bosonic degrees of freedom, there is the freedom of unitarily mixing all $N$ annihilation operators (and creation operators respectively) among themselves, leading to a $\mathrm{U}(N)$ subgroup of different choices of $a_i$ that all define the same vacuum.} 
\begin{align}
	\tilde{a}&=\alpha\, a+\beta\, a^\dagger\,, \label{bogo}\\
	\tilde{a}^\dagger&=\alpha^*\, a^\dagger+\beta^*\, a\,.\nonumber
\end{align}
In order to preserve the commutation relations $[\tilde{a},\tilde{a}^\dagger]=[a,a^\dagger]=1$, the coefficients $\alpha$ and $\beta$ need to satisfy
\begin{align}
	|\alpha|^2-|\beta|^2=1\,.
\end{align}
From this, we can conclude that the most general Bogoliubov transformation (for a single degree of freedom) is given by
\begin{align}\label{psycho}
	\alpha&=\ex^{\ii \varphi}\cosh{r}\,,\\
	\beta&=\ex^{\ii \vartheta}\sinh{r}\,.\nonumber
\end{align}

Now given two pairs of creation and annihilation operators, namely $(a,a^\dagger)$ and $(\tilde{a},\tilde{a}^\dagger)$, they define  two distinct Gaussian states satisfying $a\,|\psi\rangle=0$
and $\tilde a\,|\tilde\psi\rangle=0$. Hence the Bogoliubov transformations \reef{bogo} describe the desired group of transformations mapping the Gaussian states amongst themselves. We can invert eq.~\reef{aadag} to $\tilde\xi^a\equiv(\tilde{q},\tilde{p})$ for the pair $(\tilde{a},\tilde{a}^\dagger)$. Then, the Bogoliubov transformation \reef{bogo} from $(a,a^\dagger)$ to $(\tilde{a},\tilde{a}^\dagger)$ induces a linear transformation $M^a{}_b$ on the space $V^*$ spanned by $\xi^a$ and $\tilde{\xi}^a$, \ie $\xi^a=M^a{}_b\,\tilde{\xi}^b$. Note that we define $M$ to be the inverse transformation that maps $\tilde{\xi}$ into $\xi$. The condition of preserving the commutation relations then translates into
\begin{align}\label{Osym}
	(M\Omega M^\intercal)^{ab}=M^a{}_c\,\Omega^{cd}\,(M^\intercal)_d{}^b=\Omega^{ab}\,,
\end{align}
where $\Omega$ is a symplectic on $V^*$. This expression \reef{Osym} extends trivially to the case of $N$ degrees of freedom
(by simply extending the range of the indices) and then reveals the $\mathrm{Sp}(2N,\mathbb{R})$ group structure noted at the beginning of this section. Of course, we are also interested in the transformation of the symmetric two-point correlator
\beq
\tilde{G}^{ab}=(MGM^\intercal)^{ab}=M^a{}_c\,G^{cd}\,(M^\intercal)_d{}^b\,,
\label{Gsym}
\eeq
which encodes the transformation of the state, namely $\tilde G^{ab}= \langle\tilde\psi|\,\{\xi^a,\xi^b\}\,|\tilde\psi\rangle$, \ie the expectation value of the original operators $\xi^a$ in the transformed state. In particular, in a discussion of the circuit complexity of these states, we can represent the gates and unitary circuits with the appropriate symplectic transformations, and describe their action on the state in terms of the above transformation, \eg see \cite{BH-BFGaussians}.

In our example with $N=1$, the Bogoliubov transformation \reef{bogo} gives the symplectic matrix
\begin{align}
	M\equiv\left(
	\begin{array}{cc}
	\cos (\varphi ) \cosh (r)+\cos (\vartheta ) \sinh (r) &\ \sin (\vartheta ) \sinh (r)- \sin (\varphi ) \cosh (r) \\
	 \sin (\varphi )\cosh (r)+\sin (\vartheta ) \sinh (r) &\ \cos (\varphi ) \cosh (r)-\cos (\vartheta ) \sinh (r)
	\end{array}
	\right)\,.
\end{align}
If we start with an initial state $|\psi\rangle$, whose covariance matrix is $G\equiv\mathbb{1}$, then using eq.~\reef{Gsym}, the transformed state $|\tilde\psi\rangle$ is described by\footnote{This method was already used for circuit complexity in bosonic systems \cite{prep}. Most of the formalism for bosonic (and fermionic) Gaussian states in this paper is based on \cite{Bianchi:2015fra,BH-BFGaussians}.}
\begin{align}
	\tilde{G}^{ab}\equiv\left(
	\begin{array}{cc}
	\cosh (2r)+\cos (\vartheta +\varphi ) \sinh (2 r) & \sin (\vartheta +\varphi ) \sinh (2 r) \\
	\sin (\vartheta +\varphi ) \sinh (2 r) & \cosh (2r)-\cos (\vartheta +\varphi ) \sinh (2 r)
	\end{array}
	\right)\,.
\end{align}
We notice that the final state $|\tilde{\psi}\rangle$ is independent of $(\vartheta-\varphi)$, which corresponds to the $\mathrm{U}(1)$ subgroup where we just multiply creation and annihilation operators with opposite complex phases. As a manifold, we have $\mathrm{Sp}(2,\mathbb{R})=\mathbb{R}^2\times\mathrm{U}(1)$ where $(r,\vartheta+\varphi)$ provide polar coordinates of the plane and $(\vartheta-\varphi)$, the remaining coordinate on the circle $\mathrm{U}(1)$. Since this overall phase is trivial, the space of states $\mathcal{M}_{b,1}$ is properly described by the quotient $\mathcal{M}_{b,1}=\mathbb{R}^2=\mathrm{Sp}(2,\mathbb{R})/\mathrm{U}(1)$. In the general case of $N$ degrees of freedom, this expression for the space of states would become
$\mathcal{M}_{b,\ssc N}=\mathrm{Sp}(2N,\mathbb{R})/\mathrm{U}(N)$, where the $U(N)$ group mixes the various annihilation operators amongst themselves leaving the corresponding Gaussian state unchanged. $\mathcal{M}_{b,\ssc N}$ is also known as the symmetric space of type CI \cite{mimura1991topology}.

For a detailed discussion of the resulting geometry and geodesics, we refer the interested reader to \cite{prep}. However, we add the following comments to conclude our review here: 
For every Gaussian state $|G\rangle$, we can choose a canonical basis $\xi^a\equiv(q_i,p_i)$, such that $G\equiv\mathbb{1}$. This means the bilinear form $G$ does not contain information that is invariant under changing the canonical basis or put simply: ``All Gaussian states look the same if we can choose the right basis for each individual state.'' This changes of course, if we have two Gaussian states $|G\rangle$ and $|\tilde{G}\rangle$ in the same system\footnote{Of course, this is the situation where we are examining circuit complexity of states since we have both the target state and the reference state.} and force ourselves to represent the two-point functions $G$ and $\tilde{G}$ with respect to the same canonical basis. Again, we can choose a basis, such that $G\equiv\mathbb{1}$, but we will not be able to accomplish the same for $\tilde{G}$. The remaining freedom of choosing a canonical basis is described by the group $\mathrm{U}(N)=\mathrm{Sp}(2N,\mathbb{R})\cap\mathrm{SO}(2N)$ consisting of canonical transformation (\ie $M\Omega M^\intercal=\Omega$) that simultaneously orthogonal with respect to $G$ (\ie $MGM^\intercal=G$). The invariant information about the relation between the original state $|\psi\rangle$ and the transformed state $|\tilde{\psi}\rangle$ is completely captured by the eigenvalues of the relative covariance matrix\footnote{Note that one could have just as easily defined $\widehat\Delta=G\,\tilde{g} $ with $\tilde g = \tilde{G}^{-1}$. However, one then has $\widehat\Delta= \Delta^{-1}$ and due to the fact that $\Delta$ is symplectic, the two have the same spectrum. This is discussed in more detail in section \ref{sec:general}.}
\begin{align} \label{Dab}
	\Delta^a{}_b=\tilde{G}^{ac}\,g_{cb}\quad\text{with}\quad g=G^{-1}\,,
\end{align}
\ie $G^{ac}g_{cb}=\delta^a{}_b$. In particular, any quantities that depend on the two states in a $\mathrm{Sp}(2N,\mathbb{R})$-invariant way, \eg their inner product,\footnote{For bosonic states, we find the simple formula $|\langle G|\tilde{G}\rangle|^2=\det\frac{\sqrt{2}\Delta^{1/4}}{\sqrt{\mathbb{1}+\Delta}}$ derived in \cite{BH-BFGaussians}.\label{fn:boson-inner-product}} can be computed purely from $\Delta$. This will apply to the complexity \emph{provided} that we choose a geometry that is $\mathrm{Sp}(2N,\mathbb{R})$-invariant, \eg we do not introduce penalty factors which conflict with the group structure. For our Bogoliubov transformation \reef{bogo}, we have $\mathrm{spec}(\Delta)=(\ex^{2r},\ex^{-2r})$. We say that $|\tilde{\psi}\rangle$ arises from a one-mode squeezing of $|\psi\rangle$ with squeezing parameter $r$. For bosonic Gaussian states, understanding one-mode squeezing is the key to relate any two states. That is, for any two bosonic Gaussian states $|\psi\rangle$ and $|\tilde{\psi}\rangle$ with $N$ degrees of freedom, there exists a normal mode basis $(q_1,\cdots,q_N,p_1,\cdots,p_N)$, such that $|\tilde{\psi}\rangle$ is the result of $N$ independent one-mode squeezing operations in the $N$ different normal modes \cite{BH-BFGaussians,prep}. This is related to the Iwasawa (or KAN) decomposition of 
$\mathrm{Sp}(2N,\mathbb{R})$, \eg see \cite{bump,dutta1995real}.

\subsection{Two fermions} \label{twofermi}

We now turn to the case of fermionic Gaussian states. In this case, the space of Gaussian states for $N$ fermionic degrees of freedom is given by the quotient $\mathcal{M}_{f,\ssc N}=\mathrm{O}(2N)/\mathrm{U}(N)$, which has dimension $N(N-1)$, \eg \cite{BH-BFGaussians}. Of course, this space is a small submanifold within the full $2^N$-dimensional Hilbert space $\mathcal{H}$ of the fermionic system. Further, it is not preserved by general unitary transformations $\mathrm{U}(2^N)$ acting on $\mathcal{H}$, but only the subgroup $\mathrm{O}(2N)$ corresponding to Bogoliubov transformations. That is, the most straightforward way to think of characterizing the fermionic Gaussian states is in terms of the annihilation and creation operators. With $N$ fermionic pairs ($a_i,a^\dagger_i$) satisfying $\{a_i,a_j^\dagger\}=\delta_{ij}$, the corresponding Gaussian state is again defined by $a_i|\psi\rangle=0$ and the Bogoliubov transformations mixing these fermionic operators map Gaussian states to Gaussian states.

In analogy to eq.~\reef{aadag} for the bosons, we begin by defining a set of Hermitian fermionic operators given by
\begin{align}
	q_i=\frac{1}{\sqrt{2}}(a^\dagger_i+a_i)\qquad\text{and}\qquad p_i=\frac{\ii}{\sqrt{2}}(a^\dagger_i-a_i)\,,\label{heavy}
\end{align}
which are commonly referred to as Majorana modes. In contrast to the analogous bosonic operators, they do not consist of conjugate pairs $(q_i,p_i)$, but rather they are governed by the anti-commutation relations: $\{q_i,q_j\}=\delta_{ij}=\{p_i,p_j\}$ and $\{q_i,p_j\}=0$. Turning to the covariance matrix \reef{covmat}, if we choose the Majorana basis $\xi^a\equiv (q_1,\cdots,q_N,p_1,\cdots,p_N)$, the symmetric component becomes simply
\begin{align}\label{symf}
	G^{ab}= \langle\psi|\,\{\xi^a,\xi^b\}\,|\psi\rangle=
\delta^{ab}\,.
\end{align}
This result holds for any Gaussian state since $G$ simply encodes the canonical anti-commutation relations $G^{ab}=\{\xi^a,\xi^b\}\equiv\delta^{ab}$ (which are preserved by the Bogoliubov transformations). Further, as we will see below, this matrix $G^{ab}$ provides a useful positive definite metric. Hence, in the fermionic case, the nontrivial component of eq.~\reef{covmat} is the antisymmetric two-point correlator 
\beq
\Omega^{ab}= -\ii\,\langle\psi|\,[\xi^a,\xi^b]\,|\psi\rangle\,,
\label{asymf}\eeq 
which characterizes the corresponding Gaussian state  $|\psi\rangle$. Given eq.~\reef{heavy} above, we may evaluate this matrix for the state $|\psi\rangle$ annihilated by $a_i$ as
\begin{align}
	\Omega\equiv\left(\begin{array}{cccc}
	\ \mathbb{0}&\mathbb{1}\\
	-\mathbb{1}&\mathbb{0}
	\end{array}\right)\,,
\end{align}
where $\mathbb{1}$ and $\mathbb{0}$ are $N\!\times\!N$ identity and zero matrices, respectively. We note that this $\Omega$  coincides with the form of the symplectic form \reef{asymb} for bosons. 

Now in analogy with our discussion of bosons, a pair $(a_i,a_i^\dagger)$ and $(\tilde{a}_i,\tilde{a}_i^\dagger)$ defines two distinct Gaussian states satisfying $a_i\,|\psi\rangle=0$
and $\tilde a_i\,|\tilde\psi\rangle=0$. Hence understanding the group of transformations mapping fermionic Gaussian states amongst themselves is again understanding the Bogoliubov transformations acting on the fermionic annihilation and creation operators.  It is simplest to work with the Majorana basis, \ie $\tilde\xi^a\equiv(\tilde{q}_i,\tilde{p}_i)$ and $\xi^a\equiv(q_i,p_i)$, where the Bogoliubov transformations act as a linear transformation. Again, we define the inverse transformation $M$, such that $\xi^a=M^a{}_b\,\tilde\xi^b$. The condition of preserving the anti-commutation relations translates into
\begin{align}\label{Gsym2}
	(M\,G\, M^\intercal)^{ab}=M^a{}_c\,G^{cd}\,(M^\intercal)_d{}^b=G^{ab}\,.
\end{align}
Recalling that $G^{ab}\equiv\delta^{ab}$ in the Majorana basis, eq.~\reef{Gsym2} makes evident the $\mathrm{O}(2N)$ group structure, which we referred to above. Of course, the transformation of the states is now encoded in the transformation of the antisymmetric two-point correlator
\beq
\tilde{\Omega}^{ab}=(M\Omega M^\intercal)^{ab}=M^a{}_c\,\Omega^{cd}\,(M^\intercal)_d{}^b\,.
\label{Osym3}
\eeq
Hence in a discussion of the circuit complexity of fermionic Gaussian states, we can represent the unitary circuits and gates with the appropriate orthogonal transformations and their generators, and describe their action on the states in terms of the above transformation.

To make this discussion more concrete, let us consider a simple example. However (as we now show), the simplest case of a single pair, \ie $N=1$, turns out to be trivial. In this case, the most general Bogoliubov transformation is
\begin{align}
\tilde{a}&=\alpha\,a+\beta\,a^\dagger\,,\label{bogo2}\\
\tilde{a}^\dagger&=\alpha^*\,a^\dagger+\beta^*\,a\,.
\nonumber
\end{align}
Demanding that the anti-commutation relation is preserved, \ie
$\{\tilde a,\tilde a^\dagger\}=1$, yields
\begin{align}
|\alpha|^2+|\beta|^2=1\,.
\end{align}
However, fermionic creation and annihilation operators also need to satisfy $\tilde{a}^2=(\tilde{a}^\dagger)^2=0$. Computing this explicitly for above transformation leads to a second requirement
\begin{align}
\tilde{a}^2=\alpha\beta\{a,a^\dagger\}=2\,\alpha\,\beta=0\,.
\end{align}
This means up to an overall phase, the only possible transformations are $\alpha =1, \ \beta=0$ or $\alpha=0,\ \beta=1$. That is, $\tilde{a}=a$ or we swap the role of creation and annihilation operators with $\tilde{a}=a^\dagger$. With $N=1$, the space of Gaussian states is $\mathcal{M}=\mathrm{O}(2)/\mathrm{U}(1)$, where the $\mathrm{U}(1)$ corresponds to the overall complex phase, but this space simply consists of two points.\footnote{It will be a general feature (for any $N$) that the full set of fermionic Gaussian states always consists of two disconnected components corresponding to the $\mathbb{Z}_2$ grading of states with even and odd fermion number. Note that neither of the two components is preferred and which corresponds to an even and odd fermion number depends on one's choice of the vacuum, or alternatively on one's notion of particle.\label{footyABC}}

This means that --- in contrast to a single bosonic degree of freedom --- the squeezing of a single fermionic degree of freedom is trivial. The first non-trivial system consists of two fermionic degrees of freedom, often interpreted as two qubits. With $N=2$, the state manifold will be 
\beq
\mathcal{M}_{f,2}=\mathrm{O}(4)/\mathrm{U}(2)=S^2\cup S^2\,,
\label{homef2}
\eeq
which is two-dimensional. In this case, we consider two pairs fermionic creation and annihilation operators, $(a_1,a_1^\dagger)$ and $(a_2,a_2^\dagger)$. For this example, let us consider the fermionic Bogoliubov transformation
\begin{align}\label{bogo8}
	\tilde{a}_1&=\alpha\, a_1-\beta\, a_2^\dagger\,,\\
	\tilde{a}_2^\dagger&=\beta^*\, a_1+\alpha^*\, a_2^\dagger\,.\nonumber
\end{align}
This is not the most general transformation, but the natural choice if we want to mix $a_1$ with $a_2^\dagger$. In fact, one can show that one can bring any Bogoliubov transformation into this form by mixing $a_1$ with $a_2$, and $\tilde{a}_1$ with $\tilde{a}_2$ via $\mathrm{U}(2)$, which does not change the corresponding Gaussian states, $|\psi\rangle$ and $|\tilde\psi\rangle$.

Further, for eq.~\reef{bogo8}, we may choose $\alpha$  to be real so that the following parametrization works well:
\beq
	\alpha=\cos{\vt}\,,\qquad
	\beta=\ex^{\ii\vp}\,\sin{\vt}\,.
\label{param}\eeq
The induced transformation $M$ that maps $\tilde\xi^a$ into $\xi^a$ can then be written as
\begin{align}\label{two-mode}
\small
\begin{split}
	M&\equiv\left(
	\begin{array}{cccc}
	1 & 0 & 0 & 0 \\
	0 & \cos (\vp) & 0 &-\sin(\vp) \\
	0 & 0 & 1 & 0 \\
	0 & \sin (\vp) & 0&\cos (\vp)
	\end{array}
	\right)\,
	\left(
	\begin{array}{cccc}
	\cos (\vt) & \sin (\vt) & 0 & 0 \\
	-\sin (\vt) & \cos (\vt) & 0 & 0 \\
	0 & 0 & \cos (\vt) & -\sin (\vt) \\
	0 & 0 & \sin (\vt) & \cos (\vt)
	\end{array}
	\right)\,
	\left(
	\begin{array}{cccc}
	1 & 0 & 0 & 0 \\
	0 & \cos (\vp) & 0 &\sin(\vp) \\
	0 & 0 & 1 & 0 \\
	0 & -\sin (\vp) & 0&\cos (\vp)
	\end{array}
	\right)\\
	&=\left(
	\begin{array}{cccc}
	\cos (\vartheta ) &\sin (\vartheta )  \cos (\varphi ) & 0 & \sin (\vartheta ) \sin (\varphi ) \\
	-\sin (\vartheta ) \cos (\varphi ) & \cos (\vartheta ) & -\sin (\vartheta ) \sin (\varphi ) & 0 \\
	0 & \sin (\vartheta ) \sin (\varphi ) & \cos (\vartheta ) & -\sin (\vartheta ) \cos (\varphi ) \\
	-\sin (\vartheta ) \sin (\varphi ) & 0 & \sin (\vartheta ) \cos (\varphi ) & \cos (\vartheta ) \\
	\end{array}
	\right)
\end{split}
\end{align}
Here, we have decomposed $M$ as a series of rotations and so it is clear that $M\in\mathrm{O}(4)$ or rather $M\in\mathrm{SO}(4)$, because we can continuously reach $\mathbb{1}$, and satisfies $MGM^\intercal=G$. The antisymmetric covariance matrix $\tilde{\Omega}=M\Omega M^\intercal$ of the transformed state $|\tilde{\psi}\rangle$ can then be evaluated to be
\begin{align}\label{bounse2}
	\tilde{\Omega}\equiv\left(
	\begin{array}{cccc}
	0 & -\sin (2\vt ) \sin (\vp ) & \cos (2 \vt ) & \sin (2 \vt ) \cos (\vp ) \\
	\sin (2 \vt ) \sin (\vp ) & 0 & -\sin (2\vt ) \cos (\vp ) & \cos (2 \vt ) \\
	-\cos (2 \vt ) & \sin (2 \vt ) \cos (\vp ) & 0 & \sin (2 \vt ) \sin (\vp ) \\
	-\sin (2\vt )\cos (\vp ) & -\cos (2 \vt ) & -\sin (2\vt ) \sin (\vp ) & 0 \\
	\end{array}
	\right)\,.
\end{align}
Note that we get the same state for $\vt=0$ and $\vt=\pi$, which is perhaps half the expected range. This is due to the fact that the transformation with $\vartheta=\pi$ leads to $\tilde{a}_1=-a_1$ and $\tilde{a}_2=-a_2$, which leaves the vacuum invariant. Therefore the state which is most distant\footnote{Of course, we mean `most distant' on the $S^2$ component connected to the identity. We cannot reach the states on the other component along a continuous trajectory without leaving the space of Gaussian states.} from the original Gaussian state $|\psi\rangle$ corresponds $\vt=\pi/2$, which we see trades the annihilation and creation operators, \ie eq.~\reef{bogo8} reduces to $(\tilde{a}_1,\tilde{a}_2)=(-\tilde{a}^\dagger_2,\tilde{a}^\dagger_1)$ with $\vt=\pi/2$ (and $\vp=0$).

As mentioned in eq.~\reef{homef2}, the space of states is given by the quotient $\mathcal{M}_{f,2}=\mathrm{O}(4)/\mathrm{U}(2)=S^2\cup S^2$ because we need to divide by the subgroup ${U}(2)$ associated to mixing creation and annihilations operators among themselves, respectively. In particular, we see that the manifold of fermionic Gaussian states again consists of two disconnected components --- see footnote \ref{footyABC}. We can only continuously deform one state to the other, if they lie in the same component --- unless we are willing to leave the space of Gaussian states. Our choice of Bogoliubov transformations  parametrized by $\vt$ and $\vp$ corresponds to the $S^2$ connected to the identity.

Similar to the bosonic example, we can ask how to encode the invariant relative information between two fermionic Gaussian states $|\Omega\rangle$ and $|\tilde{\Omega}\rangle$. As a preliminary step towards answering this question, let us note that with an appropriate choice of an orthonormal basis $\xi^a\equiv(q_1,q_2,p_1,p_2)$ of Majorana modes, $G\equiv\mathbb{1}$ and the covariance matrix $\Omega$ takes the standard form
\begin{align}\label{canon0}
	\Omega\equiv\left(\begin{array}{cc}
	\mathbb{0}&\mathbb{1}\\
	-\mathbb{1}&\mathbb{0}
	\end{array}\right)\,.
\end{align}
While preserving these forms, we would also like to bring $\tilde{\Omega}$ into a standard form. The allowed transformations are given by the subgroup $\mathrm{U}(2)=\mathrm{O}(4)\cap\mathrm{Sp}(4,\mathbb{R})$, just like for the bosonic case. One can show that the covariance matrix $\tilde{\Omega}$ can be brought into the standard form\footnote{Examining the transformation in eq.~\reef{two-mode}, one finds the final rotation can be eliminated with the phase rotation $(\tilde{a}_1,\tilde{a}_2)\to (\tilde{a}_1,\ex^{-\ii\vp}\tilde{a}_2)$, which of course leaves the $|\tilde \psi\rangle$ unchanged. Further, applying the latter transformation takes $\tilde\Omega$ from eq.~\reef{bounse2} to the canonical form \reef{canon}.}
\begin{align}\label{canon}
	\tilde{\Omega}=\left(
	\begin{array}{cccc}
	0 & 0 & \cos (2\vt) & -\sin (2\vt) \\
	0 & 0 & \sin (2\vt) & \cos (2\vt) \\
	-\cos (2\vt) & -\sin (2\vt) & 0 & 0 \\
	\sin (2\vt) & -\cos (2\vt) & 0 & 0 \\
	\end{array}
	\right)\,,
\end{align}
provided that $|\Omega\rangle$ and $|\tilde\Omega\rangle$ belong to the same connected component. This indicates that the invariant relative information is encoded in $\vt$ alone, \ie the second angle $\vp$ in eq.~\reef{param} is irrelevant.

Following the discussion of the bosonic theories (\eg compare to eq.~\reef{Dab}), we can describe this invariant information about the relation between the two states in terms of the relative (fermionic) covariance matrix\footnote{For fermionic states, we find the formula $|\langle\Omega|\tilde{\Omega}\rangle|^2=\det\frac{\sqrt{\mathbb{1}+\Delta}}{\sqrt{2}\Delta^{1/4}}$ derived in \cite{BH-BFGaussians} which is strikingly similar to the one for bosons from footnote~\ref{fn:boson-inner-product}.}
\begin{align}\label{jangle}	\Delta^a{}_b=\tilde{\Omega}^{ac}\,\omega_{cb}\qquad\text{with}\quad \omega=\Omega^{-1}\,,
\end{align}
\ie $\Omega^{ac}\,\omega_{cb}=\delta^{a}{}_b$.\footnote{We will see in section \ref{sec:general} that the bosonic and fermionic relative convariance matrices $\Delta$ arise in the same way when one labels states by their linear complex structure $J$.} 
The invariant information is then captured in the eigenvalues of this matrix. For our choice of Bogoliubov transformation in eqs.~\reef{bogo8} and \reef{param}, we have $\mathrm{spec}(\Delta)=(\ex^{2\ii \vt},\ex^{2\ii \vt},\ex^{-2\ii \vt},\ex^{-2\ii \vt})$ and as expected, $\vp$ does not appear here. We will later show that for a natural choice of invariant metric on the group, our Bogoliubov transformation that changes $\vt$ continuously from zero to its final value along a path of fixed $\vp$ is the minimal geodesic connecting a reference state $|\psi\rangle$ to a target state $|\tilde{\psi}\rangle$. In particular, the geodesic length will be given by $|2\vt|\in [0,\pi]$. These paths are just the great circles passing through the pole (at $\vt=0$) on the corresponding two-sphere. This means in each linearly independent direction, the maximal path length is $\pi/2$.\footnote{One may be surprised to find $\pi/2$ rather than $\pi$ here. The reason is that at $\pi$, we would reach the group element $M=-\mathbb{1}$, as shown by eq.~\reef{two-mode}, which is as far away from $\mathbb{1}$ as possible. However, the transformed two-point function becomes $\tilde{\Omega}=M\Omega M^\intercal =\Omega$, \ie eq.~\reef{bounse2} reduces to the initial covariance matrix in eq.~\reef{canon0} with $\vt=\pi$, and so the final state is identical to the initial one at $\vt=\pi$. This means the group elements, which take our state as far away as possible from the initial one, are those sitting on the circle at $\vt=\pi/2$, \ie the equator of the connected $S^2$ component. Recall that at $\vt=2\pi$, $M$ returns to the identity, but when we measure the length of the circle covered by $\vt$ running from $0$ to $2\pi$ (with fixed $\vp$) using our metric $\langle\cdot,\cdot\rangle_{\mathbb{1}}$ (see eq.~\reef{gonzo} below), its length is actually $4\pi$. Therefore the resulting distance to the maximally distant states is $\pi$.} However, if with a large number of degrees of freedom, geodesic will be moving along several such paths in orthogonal directions at the same time. In particular, the overall path can become arbitrarily large in the field theory limit where we consider an infinite number of degrees of freedom.

For bosons, we reviewed that any two Gaussian states define a set of normal modes, such that there is a natural transformation built from linearly independent one-mode squeezing operations in these modes. In the case of fermions, we observed that: (a) there are two disconnected components on the manifold of states  (separating states with even and odd fermion number); and (b) one-mode-squeezing is trivial and we need to perform two-mode squeezing operations. Therefore, we can only find normal modes if two Gaussian states lie in the same connected component and these normal modes always come in pairs, so that the two states are related by a collection of independent two-mode squeezing operations. In particular, if we have an odd number of fermionic degrees of freedom, there will always be a single normal mode left that is not squeezed when moving from one state to the other.

\subsection{Gates, circuits and complexity}
\label{gate5}
So far, our discussion of fermionic Gaussian states has been at a fairly abstract level. We have used the covariance matrix $\Omega$ as a convenient parametrization of the manifold of fermionic Gaussian states and the action of Bogoliubov transformations on this space. In particular, much of the discussion focused on the case of two fermionic degrees of freedom. Here, we would like to bring the discussion more closely in line with the continuous description of unitary circuits in eqs.~\reef{controlY} and \reef{row}. In particular, these unitaries will be constructed using some basis of Hermitian operators $\op_I$, which act on the states in the Hilbert space of our fermionic system. Since we are focusing our attention on circuits which map Gaussian states to Gaussian states, \ie which implement Bogoliubov transformations, we will only consider generators that are quadratic operators, in analogy with the study of bosonic Gaussian states in \cite{Jefferson:2017sdb,Chapman:2017rqy,prep}. One may describe these quadratic generators in terms of the annihilation and creation operators, but we find it more convenient to work with the Majorana modes \reef{heavy}, \ie $\xi^a =(q_i,p_i)$. That is, we choose our basis of generators to be the antisymmetric combinations $\op_I= \ii\, \xi^{[a}\xi^{b]}$.\footnote{Of course, the symmetric combinations are trivial, since $\{ \xi^a,\xi^b\}=G^{ab}\equiv\delta^{ab}$.} The antisymmetric form of the indices for these basis generators hints at an $\mathrm{SO}(2N)$ group structure, which is readily confirmed by examining the commutation algebra of the generators.

Let $\hat{K}$ be a general real linear combination of these Hermitian quadratic operators. Such an operator is completely characterized by an antisymmetric matrix $k_{ab}=k_{[ab]}$, 
\begin{align}
\hat{K}=\frac{\ii}{2}\,k_{ab}\ {\xi}^a\,\xi^b\,.
\end{align}
As a Hermitian operator, $\hat{K}$ gives rise to the unitary operator $U(\hat{K})=\ex^{-\ii\hat{K}}$ which acts on our Gaussian states, \ie
$|\psi\rangle\to |\tilde \psi\rangle=U(\hat{K})|\psi\rangle$. However, we wish to understand this transformation through the action of $U(\hat K)$ on the covariance matrix. Hence we consider the corresponding action on the operators $\xi^a$ themselves, \ie
\begin{align}\label{grenade}
\tilde{\xi}^a=U(\hat{K})\, {\xi}^a\, U^\dagger(\hat{K})=\sum^{\infty}_{n=0}\frac{(-)^n}{n!}\,[\ii \hat{K},{\xi}^a]_{(n)}
\end{align}
where we have defined $[\ii\hat{K},\xi^a]_{(n+1)}=[\ii\hat{K},[\ii\hat{K},{\xi}^a]_{(n)}]$ and $[\ii \hat{K}, {\xi}^a]_{(0)}=\xi^a$, and we have used Baker-Campbell-Hausdorff to simplify this expression. With some algebra, we find that the first commutator yields
\beq\label{firstcom}
[\ii\hat{K},{\xi}^a]_{(1)}=[\ii\hat{K},{\xi}^a]=-\frac{1}{2}\,k_{bc}[{\xi}^b {\xi}^c, {\xi}^a]=G^{ac}k_{cb}\,{\xi}^b\,,
\eeq
where we used the anti-commutation relations $\{ \xi^a,\xi^b\}=G^{ab}$. By defining 
\beq\label{magicK}
K^a{}_b=G^{ac}k_{cb}\,,
\eeq
we can write successive commutators as $[\ii\hat{K},{\xi}^a]_{n}=(K^n)^a{}_b\,{\xi}^b$. The action of $U(\hat{K})$ in eq.~\reef{grenade} can therefore be simply expressed as
\begin{align}
\tilde\xi^a=U(\hat{K})\,{\xi}^a\,U^\dagger(\hat{K})=(\ex^{-K})^a{}_b\,{\xi}^b\,,
\end{align}
or alternatively, following the notation introduced in the preceding discussion we have
\beq
\xi^a=M(K)^a{}_b\,\tilde{\xi}^b\qquad{\rm with}\ \ 
M(K)^a{}_b=(\ex^{K})^a{}_b\,.
\eeq
Again, from the antisymmetry of $k_{ab}$, it is obvious that the generator $K$ will be given by an antisymmetric matrix with respect to a basis where $G^{ab}\equiv\delta^{ab}$, which was implicitly chosen in using the Majorana modes for the above. Hence we recognize $M(K)$ as a group element in $\mathrm{SO}(2N)$ with the generator $K^a{}_b=G^{ac}k_{[cb]}\in\mathrm{so}(2N)$.

Recall that in discussing the complexity, we must choose a metric $\langle\cdot,\cdot\rangle_{\mathbb{1}}$ in eq.~\reef{metA} on the Lie algebra, \ie $\mathrm{so}(2N)$ for $N$ fermionic degrees of freedom. This Lie algebra is $(2N-1)N$-dimensional, so the possible metrics correspond to the space of positive definite linear forms described by symmetric $(2N-1)N\,\times\,(2N-1)N$ matrices, which has some very large dimension.\footnote{$\,N(2N-1)(2N^2-N+1)/2$}  However, there is one particularly natural choice that is induced by the anticommutation relations, namely
\begin{align}\label{lion}
	\langle A,B\rangle_{\mathbb{1}}=\mathrm{Tr}(A G B^\intercal g)=A^a{}_bG^{bc}(B^\intercal)_c{}^dg_{da}\,.
\end{align}
This inner product is clearly positive definite because in a basis with $G^{ab}\equiv\delta^{ab}$, we have $\langle A,A\rangle_{\mathbb{1}}=\mathrm{Tr}(AA^\intercal)\geq 0$. This inner product can be recognized to be a canonical Lie algebra structure by realizing that for $A\in\mathrm{so}(2N)$, we have $GA^\intercal g=-A$, so that we can rewrite
\begin{align}\label{gonzo}
	\langle A,B\rangle_{\mathbb{1}}=\mathrm{Tr}(A G B^\intercal g)=-\mathrm{Tr}(AB)\,.
\end{align}
The last expression is well known to be proportional to the negative Killing form, which is a positive definite inner product for semi-simple compact Lie groups \cite{kirillov2008introduction}. Recall that this metric is then extended to the entire group by right translation as in eq.~\reef{metB}. For this choice of metric, the computation of geodesics becomes relatively simple. In fact, in appendix \ref{app:geodesics}, we prove that every geodesic beginning at the identity is given by $\ex^{s A}$ for some fixed $A\in\mathrm{so}(2N)$. For the rest of this paper, we will always refer to this metric, if not indicated otherwise.

Given this key result from appendix \ref{app:geodesics}, we can easily compute the complexity associated to the state produced by such a geodesic
\begin{align}
	\gamma: [0,1]\to\mathrm{SO}(2N): s\mapsto \ex^{s A}\,,
\end{align}
which connects some reference state $|\psi_\mt{R}\rangle$ at $s=0$ to the target state  $|\psi_\mt{T}\rangle=U(\hat{A})\,|\psi_\mt{R}\rangle$ at $s=1$. The key simplification is that the magnitude of the tangent vector along these geodesics is fixed. We can compute explicitly $\dot{\gamma}(s)=A\ex^{sA}$ leading to
\begin{align}
	\lVert\dot{\gamma}(s)\rVert^2=\langle A\ex^{sA},A\ex^{sA}\rangle_{\ex^{sA}}=\langle A\ex^{sA}\ex^{-s A},A\ex^{sA}\ex^{-s A}\rangle_{\mathbb{1}}=\langle A,A\rangle_{\mathbb{1}}=\lVert A\rVert^2\,,
\end{align}
using eq.~\reef{metB}. The result is not very surprising because the trajectory is moving continuously in the direction generated by a single Lie algebra element $A$.

The geodesic trajectory arises naturally in evaluating the complexity using the $F_2$ measure as in eq.~\reef{metC}, in which case it is given by the Riemannian length of the geodesic, 
\begin{align}
	\mathcal{C}_2(\ex^A)=\int_0^1\!ds\, \lVert\dot{\gamma}(s)\rVert=\int_0^1\!ds\, \lVert A\rVert = \lVert A\rVert\,.\label{integralA}
\end{align}
However, the same geodesic appears using the $\kappa$ measure with $\kappa=2$ as in eq.~\reef{metCa}, and then the complexity is given by
\begin{align}
	\mathcal{C}_{\kappa=2}(\ex^A)=\int_0^1\!ds\, \lVert\dot{\gamma}(s)\rVert^2=\int_0^1\!ds\, \lVert A\rVert^2 = \lVert A\rVert^2\,.\label{integralB}
\end{align}
Note that these two results are very simply related, \ie $\mathcal{C}_{\kappa=2}(\ex^A)=\mathcal{C}_2(\ex^A)^2$.

We can make this discussion more explicit by turning the $N=2$ case considered in the previous subsection. In particular, given the two-mode squeezing transformation $M(\vt,\vp)$ in eq.~(\ref{two-mode}), we find the generator to be 
\begin{align}\label{genie}
A(\vt,\vp)=\vt\left(
\begin{array}{cccc}
0 & \cos (\vp) & 0 & \sin (\vp ) \\
-\cos (\vp ) & 0 & -\sin (\vp ) & 0 \\
0 & \sin (\vp ) & 0 & -\cos (\vp ) \\
-\sin (\vp ) & 0 & \cos (\vp ) & 0 \\
\end{array}
\right)\,.
\end{align}
That is, we can write  $M(\vt,\vp)=\ex^{A(\vt, \vp)}$. To gain some intuition for these transformations, we might imagine that $\vp$ is fixed but $\vt$ allowed to vary. Recall that these angular coordinates cover the $S^2$ connected to the identity in eq.~\reef{homef2}. The identity corresponds to say, the north pole (\ie $\vt=0$). Fixing the angle $\vp$ corresponds selecting a direction from amongst the lines of longitude, which describe the different state-changing directions at the identity. Finally varying $\vt$ from zero to say, $\pi/2$ describes a trajectory along this line of longitude from the north pole to the equator. Of course, as described above if we continue along the same great circle, we arrive at the south pole at $\vt=\pi$ and return to the north pole at $\vt=2\pi$.

We can use the above expressions to build a geodesic path from the identity to the group element $\ex^{A(\vt,\vp)}$ given by
\begin{align}\label{hou}
	\gamma(\vt,\vp,s): [0,1]\to\mathrm{SO}(4): s\mapsto \ex^{s\,A(\vt,\vp)}\,.
\end{align}
Further, for the generator $A(\vt, \vp)$ in eq.~\reef{genie}, we can compute the magnitude of the tangent vector using eq.~\reef{gonzo},
\begin{align}\label{hou2}
\lVert A(\vt, \vp)\rVert^2=	\langle A(\vt, \vp),A(\vt, \vp)\rangle_\mathbb{1}=\mathrm{Tr}(A\, G\, A^\intercal \, g)=-\mathrm{Tr}(A^2)=4\,\vt^2\,,
\end{align}
which can then be substituted into either eq.~\reef{integralA} or \reef{integralB} to evaluate the complexity. In particular, the geodesic length of eq.~\reef{hou} is simply given by $\lVert\gamma(\vt,\vp)\rVert=2\vt$. At this point, we have not proven that eq.~\reef{hou} is the minimal geodesic (\ie recall the discussion around eq.~\reef{bc44}), but based on the results of appendix \ref{app:geodesics}, we have shown that the geodesic distance between $\mathbb{1}$ and $\ex^{A(\vt, \vp)}$ is given by $2\vt$.

\section{Complexity for the Dirac field} \label{sec:Dirac}
Before developing systematic methods to compute the circuit complexity of arbitrary fermionic Gaussian states, we can already apply the previous results from section \ref{sec:prelude} discussing two fermionic degrees of freedom to find the complexity of the ground state of a free Dirac fermion. We are applying Nielsen's approach to build the optimal unitary circuit $U$, which accomplishes the transformation $|\psi_\mt{T}\rangle=U\,|\psi_\mt{R}\rangle$. The target state will be the ground state of the Dirac field, $|\psi_\mt{T}\rangle=|0\rangle$. As reference state, we will choose a state where the local fermionic degrees of freedom (at each spatial point on a given time slice) are unentangled, $|\psi_\mt{R}\rangle=|\bar{0}\rangle$. 

We consider a free Dirac field in four-dimensional Minkowski space.\footnote{Here, we closely follow the conventions of \cite{peskin1995introduction}. However, note that we have changed the normalization of the basis spinors by a factor of $\sqrt{m}$, \eg
$\big[u^s(\pp)\big]_\mt{\cite{peskin1995introduction}}=\sqrt{m}\,\big[u^s(\pp)\big]_\mt{here}$.} We introduce the following basis of four-component spinors 
\begin{align}\label{rest}
	u^1(0)=\left(\begin{array}{c}
	1\\0\\1\\0
	\end{array}\right)\,,\quad u^2(0)=\left(\begin{array}{c}
	0\\1\\0\\1
	\end{array}\right)\,,\quad v^1(0)=\left(\begin{array}{c}
	1\\0\\-1\\0
	\end{array}\right)\,,\quad v^2(0)=\left(\begin{array}{c}
	0\\1\\0\\-1
	\end{array}\right)\,.
\end{align}
Boosted spinors can then be found by acting with the boost matrix, \eg
\begin{align}\label{boost}
	u^s(\pp)=\frac1{\sqrt{m}}\left(\begin{array}{cc}
	\sqrt{p\cdot \sigma} & \mathbb{0}\\
	\mathbb{0} & \sqrt{p\cdot \overline{\sigma}}
	\end{array}\right) u^s(0)\,,
\end{align}
where $p\cdot\sigma= E_\pp\,\mathbb{1}-\pp\cdot\vec{\sigma}$ and $p\cdot\overline{\sigma}= E_\pp\,\mathbb{1}+\pp\cdot\vec{\sigma}$, with $E_\pp=\sqrt{m^2+\pp^2}$. Of course, the analogous formula applies for $v^s(\pp)$. We can now write the  the Dirac spinor field (on a fixed time slice, \eg $t=0$) as
\begin{align}
	\psi(\xx)=\int\frac{d^3\pp}{(2\pi)^3}\frac{\sqrt{m}}{\sqrt{2E_\pp}}\sum_{s}\left(a_\pp^s\,u^s(\pp)\,\ex^{\ii \pp\cdot \xx}+b_{\pp}^{s\dagger}\,v^s(\pp)\,\ex^{-\ii\pp\cdot \xx}\right)\,.\label{Dirac}
\end{align}
Clearly, we have four fermionic degrees of freedom per (spatial) momentum $\pp$. Recall that the annihilation and creation operators satisfy
\beq
\{ a_\pp^s,a_\qq^{r\dagger}\}=(2\pi)^3\,\delta^{rs}\,\delta(\pp-\qq)=\{ b_\pp^s,b_\qq^{r\dagger}\}\,.
\eeq
The ground state is the fermionic Gaussian state $|0\rangle$, defined by $a_\pp^s|0\rangle=0=b_\pp^s|0\rangle$, and this will be the target state for which we are evaluating the circuit complexity.

As we indicated above, our desired reference state, $|\psi_\mt{R}\rangle=|\bar{0}\rangle$, will be a Gaussian state where the local fermionic degrees of freedom at each spatial point on a given time slice are unentangled. Therefore, let us now introduce local creation and annihilation operators $(\bar{a}_\xx^s,\bar{a}^{s\dagger}_\xx)$  and $(\bar{b}_\xx^s,\bar{b}^{s\dagger}_\xx)$ satisfying
\beq
\{ \bar{a}_\xx^s,\bar{a}_\yy^{r\dagger}\}=\delta(\xx-\yy)\,\delta^{rs}=\{ \bar{b}_\xx^s,\bar{b}_\yy^{r\dagger}\}\,.
\eeq  
These operators are not completely defined until we make a specific choice on how to express the Dirac field \reef{Dirac} in terms of these local operators as
\begin{align}\label{Dirac2}
	\psi(\xx)=\frac{1}{\sqrt{2}}\sum_s\left(\bar{a}_\xx^su^s(0)+\bar{b}_\xx^{s\dagger}v^s(0)\right)\,.
\end{align}
Our unentangled reference state is then defined by $\bar{a}_\xx^s |\bar{0}\rangle=0=\bar{b}_\xx^s|\bar{0}\rangle$. Note that in this expression, we intentionally chose the rest-frame basis spinors \reef{rest} with $\pp=0$ to find a rotationally invariant reference state $|\bar{0}\rangle$, but we will discuss alternative choices of our reference state in section \ref{sec:alternative}. 

As described in the previous section, the unitary transformation from reference state to the target state, \ie $|\bar{0}\rangle\to |0\rangle= U\,|\bar{0}\rangle$, can be understood in terms of the Bogoliubov transformation relating the annihilation and creation operators with which we define these states. Hence to simplify the latter, we first find the Fourier transformed version of local operators introduced above, 
\begin{align}\label{Four}
	\bar{a}_\pp^s=\int d^3x\,\ex^{-\ii \pp\cdot\xx}\, \bar{a}_\xx^s\qquad\text{and}\qquad \bar{b}_\pp=\int d^3x\,\ex^{-\ii \pp\cdot\xx}\,\bar{b}_\pp^s\,.
\end{align}
The Dirac field is then expressed as
\beq \label{Dirac3}
\psi(\xx)=\int\frac{d^3\pp}{(2\pi)^3}\frac{1}{\sqrt{2}}\sum_s\left(\bar{a}_\pp^s\,u^s(0)\,\ex^{\ii \pp\cdot \xx}+\bar{b}_{\pp}^{s\dagger}\,v^s(0)\,\ex^{-\ii\pp\cdot \xx}\right)\,.
\eeq
We note that the Fourier transform performs a `trivial' Bogoliubov transformation, in that it mixes only the annihilation operators $\bar{a}^s_\xx$ amongst themselves and the same for the $\bar{b}^s_\xx$. As a result, the Gaussian state defined by these new operators is still the unentangled reference state $|\bar{0}\rangle$, \ie $\bar{a}_\pp^s |\bar{0} \rangle= 0= \bar{b}_\pp^s |\bar{0}\rangle$.

Now comparing eqs.~\reef{Dirac} and \reef{Dirac3}, we can immediately identify the Bogoliubov transformation which yields
$(\bar{a}_\pp^s,\bar{a}^{s\dagger}_\pp,\bar{b}_\pp^s,\bar{b}^{s\dagger}_\pp) \to ({a}_\pp^s,{a}^{s\dagger}_\pp,{b}_\pp^s,{b}^{s\dagger}_\pp)$.
In particular, computing the product with the conjugate basis spinors $u^{r\dagger}(\pp)$ and $v^{r\dagger}(-\pp)$ from the left,\footnote{Here, we use the orthogonality relations \cite{peskin1995introduction}: $u^{r\dagger}(\pp)\,v^s(-\pp)=0=v^{r\dagger}(-\pp) \, u^s(\pp)$.} we find 
\begin{align}
	a^r_{\pp}&=\frac{\sqrt{m}}{2\sqrt{E_\pp}}\sum_s\left([{u}^{r\dagger}(\pp)\,u^s(0)]\,\bar{a}^s_{\pp}+[{u}^{r\dagger}(\pp)\,v^s(0)]\,\bar{b}^{s\dagger}_{-\pp}\right)\,,\label{eq:comp1}\\
	{b}^{r\dagger}_{-\pp}&=\frac{\sqrt{m}}{2\sqrt{E_\pp}} \sum_s\left([{v}^{r\dagger}(-\pp)u^s(0)]\,\bar{a}^s_{\pp} +[{v}^{r\dagger}(-\pp)v^s(0)] \,\bar{b}^{s\dagger}_{-\pp}\right)\label{eq:comp2}
\end{align}
The spinor products are most easily evaluated by assuming that $\pp$ points in, \eg the third spatial direction, $\pp=(0,0,p_z)$ and then rotating to a general frame with spinor labels $\bar{r}$ and $\bar{s}$.\footnote{We must point out that this rotation acts on both the momentum and spin at the same time. As a result, the spin labels in eqs.~\reef{prod} and throughout the rest of this section are implicitly oriented along the momentum direction, and we have introduced that `barred' spin labels to denote this orientation. To be precise, $a^{\bs\dagger}_{\pp}$ (or $b^{\bs\dagger}_{\pp}$) with $\bs=1$ creates a particle (an antiparticle) with its spin aligned with the momentum $\pp$, while with $\bs=2$, the spin is oriented in the $-\pp$ direction. Further, we note that the reference state is rotationally invariant and so these rotations leave $|\bar{0}\rangle$ unchanged. \label{footyA1}}  The resulting products are
\begin{align}
	{u}^{\br\dagger}(\pp)\,u^\bs(0)&=\ \frac{\delta^{\br\bs}}{\sqrt{m}} \left(\sqrt{E_\pp+|\pp|}+\sqrt{E_\pp-|\pp|}\right)\,,
\nonumber\\
	{u}^{\br\dagger}(\pp)\,v^\bs(0)& =(-)^\br\frac{\delta^{\br\bs}}{\sqrt{m}}\left(\sqrt{E_\pp+|\pp|}-\sqrt{E_\pp-|\pp|}\right) \,,
\label{prod}\\
	{v}^{\br'\dagger}(-\pp)u^\bs(0)&=(-)^{\br'} \frac{\delta^{\br\bs}}{\sqrt{m}}\left(\sqrt{E_\pp+|\pp|}-\sqrt{E_\pp-|\pp|}\right)\,,
	\nonumber\\
{v}^{\br\dagger}(-\pp)v^\bs(0) &=\ \frac{\delta^{\br\bs}}{\sqrt{m}} \left(\sqrt{E_\pp+|\pp|}+\sqrt{E_\pp-|\pp|}\right)\,.
\nonumber
\end{align}
Note that in the third line, we have introduced the notation $\brp\equiv\br+1$ (mod 2).
Substuting these into eqs.~(\ref{eq:comp1}) and (\ref{eq:comp2}), from before, we find a simple Bogoliubov transformation for pairs of  operators given by
\begin{align}\label{bogo9}
a^\bs_\pp\ &=\alpha^\bs_\pp\  \bar{a}^\bs_\pp-\beta^\bs_\pp\  \bar{b}_{-\pp}^{\bs'\dagger}\,,\\
	{b}_{-\pp}^{\bsp\dagger}&=\beta^\bs_\pp\ \bar{a}^\bs_\pp+\alpha^\bs_\pp\  \bar{b}_{-\pp}^{\bs'\dagger}\,.\nonumber
\end{align}
where
\beqa
\alpha^\bs_\pp&=&\quad\frac{\sqrt{E_\pp+|\pp|}+\sqrt{E_\pp-|\pp|}}{2\sqrt{E_\pp}}\,,
\nonumber\\
\beta^\bs_\pp&=&(-)^{\bs+1}\frac{\sqrt{E_\pp+|\pp|}-\sqrt{E_\pp-|\pp|}}{2\sqrt{E_\pp}}\,.
\label{bogo9a}
\eeqa
Note that there is no sum on $\bs$ in eq.~\reef{bogo9}. Further, it is easy to verify $|\alpha^\bs_\pp|^2+|\beta^\bs_\pp|^2=1$, which ensures that we indeed have a proper fermionic Bogoliubov transformation.
Hence for the annihilation and creation operators are paired according to their momentum and spin (\ie $\bs\in\{1,2\}$), but for each of these pairs the Bogoliubov transformation takes the simple form given in eq.~\reef{bogo8}. In particular, comparing to eq.~\reef{param}, we may set $\cos\vt=\alpha^s_\pp$ and $\vp=0$ for $\bs=1$ (or $\vp=\pi$ for $\bs=2$).

\subsection{Dirac ground state} \label{ground1}
\begin{figure}[t]
\begin{center}
	\includegraphics{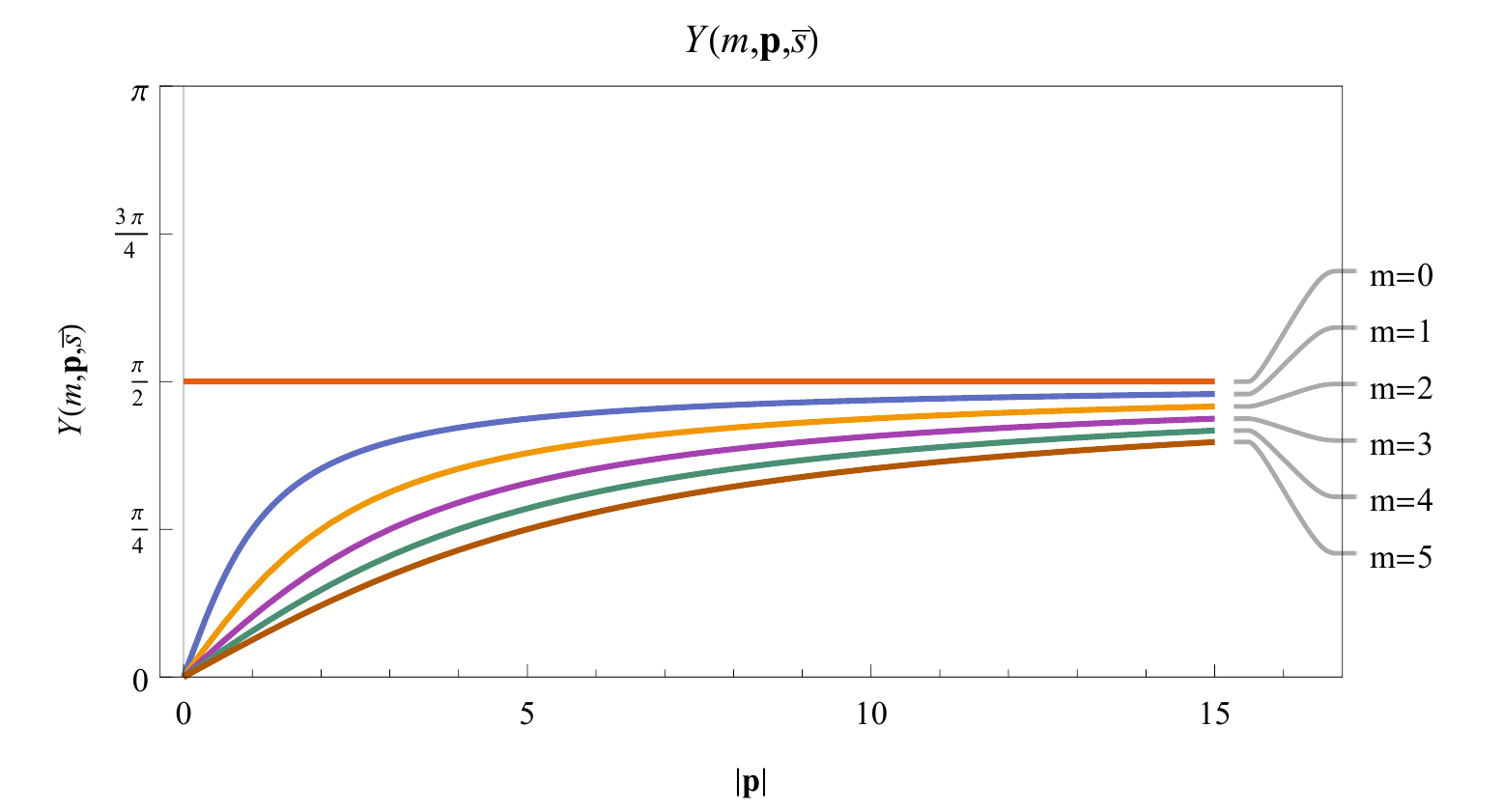}
\end{center}
	\caption{This plot shows the function $Y(m,\pp,\bs)$ in eq.~\reef{march} describing the complexity per mode of a massive Dirac field in its ground state as a function of $|\pp|$. Note that there is a single universal curve if we consider this as a function of $|\pp|/m$.}
	\label{fig1}
\end{figure}

For the complexity to transform the unentangled reference state $|\bar{0}\rangle$ into the fermionic vacuum $|0\rangle$, we recall that the geodesic distance  was given by $2\vt$ in the parameterization of a fermionic two-mode squeezing operations in eq.~\reef{param} --- see discussion around eqs.~\reef{hou} and \reef{hou2}. In particular, there is a generator analogous to that in eq.~\reef{genie} for each pair of modes and the magnitude of this generator is given by
\begin{align}\label{march}
	Y(m,\pp,\bs)=
	2\,\cos^{-1}\!\left[\alpha^\bs_\pp\right]=
	2\,\tan^{-1}\!\left(\frac{|\pp|}{E_\pp+m}\right)=\tan^{-1}\!\left(\frac{|\pp|}{m}\right)\,.
\end{align}
Above, the sign is not fixed by the $\cos^{-1}$ but with choices of $\vp$ above, we ensure that $\sin\vt>0$ and so $
Y(m,\pp,\bs)>0$ in the final expression. Note that for each momentum, the two spins (\ie $\bs=1,2$) give two identical contributions. Figure \ref{fig1} shows this expression as a function of $|\pp|$ for various values of the mass $m$. We note that for large $|\pp|$, the complexity per mode rapidly approaches 
\beq\label{gong9}
Y(m,\pp,\bs)\simeq\frac{\pi}{2}-\frac{m}{|\pp|}+\frac{m^3}{3|\pp|^3}+\mathcal{O}\left(\frac{m^5}{|\pp|^5}\right)\,.
\eeq
A special case is $m=0$ for which the complexity per mode is a fixed constant, \ie $Y(m=0,\pp,\bs)=\pi/2$. That is, $Y$ takes the maximal value for all modes in the theory of the massless free fermion.

To generate the vacuum state $|0\rangle$ from our unentangled reference state $|\bar{0}\rangle$, we are squeezing all of the modes, and we should think of $Y(m,\pp,\bs)$ for various values of $\pp$ and $\bs$ as the components of the tangent vector $\vec Y$ to the geodesic trajectory in the full geometry. The total complexity is then found by integrating over all momenta and summing over the spins with either the $F_2$ or $F_{\kappa=2}$ measures in eq.~\reef{eq:Fmetrics}.\footnote{See also eqs.~\reef{integralA} and \reef{integralB}.} With the $F_2$ measure, we find
\begin{align}\label{tot1}
\mathcal{C}_2\big(|\bar{0}\rangle\to|0\rangle\big)=\sqrt{V\int\!\! \frac{d^3\pp}{(2\pi)^3}\sum_\bs\,{Y}(m,\pp,\bs)^2}\,,
\end{align}
where the spatial volume $V$ appears to normalize the momentum integral. Similarly, for the $\kappa$ measure with $\kappa$=2, we find
\begin{align}\label{tot2}
\mathcal{C}_{\kappa=2}\big(|\bar{0}\rangle\to|0\rangle\big)={V\int\!\! \frac{d^3\pp}{(2\pi)^3}\sum_\bs\,{Y}(m,\pp,\bs)^2}\,.
\end{align}
Note that the integral is over the squares of the individual complexities per mode. The gate generating the minimal circuit corresponds to the sum of individual gates for each mode. As Lie algebra generators, they are orthogonal with respect to our right-invariant metric, such that the total norm of their sum is by an pythogerean sum, or rather integral. Of course, we also have the expected relation $\mathcal{C}_{\kappa=2}= \mathcal{C}_{2}^{\,2}$.

Because $Y(m,\pp,\bs)$ tends to the constant $\pi/2$ at large momenta (as shown in eq.~\reef{gong9}), this total complexity is UV divergent. Choosing a hard cutoff $\Lambda$ for the momentum integral allows us to compute the integral exactly leading to a rather long expression given by
\beqa
	\mathcal{C}_{\kappa=2}\big(|\bar{0}\rangle\to|0\rangle\big)&=&\frac{V}{36 \pi ^2}\left[ 12\,\ii\, m^3 \text{Li}_2\left(1-\frac{2 m}{m-\ii \Lambda }\right)+m^2 \left(12 \Lambda +\ii\,m\, \pi^2\right)\right.
	\nonumber\\
&&\qquad\qquad	+48 \left(\Lambda ^3+\ii\, m^3\right) \tan^{-1}\!\left(\frac{\Lambda }{\sqrt{\Lambda ^2+m^2}+m}\right)^2
	\label{eq:comp-exact}\\
&&	\left.-24 m \left(\Lambda ^2+2\, m^2 \log \left(\frac{2 m}{m-\ii \Lambda }\right)+m^2\right) \tan^{-1}\!\left(\frac{\Lambda }{\sqrt{\Lambda ^2+m^2}+m}\right)\right]\,.
	\nonumber
\eeqa
This expression can be simplified by expanding for large $\Lambda/m$, which yields
\begin{align}\label{eq:comp-asymp}
	\mathcal{C}_{\kappa=2}\big(|\bar{0}\rangle\to|0\rangle\big)\simeq\frac{V}{12}\left[\Lambda^3-\frac{6 m}{\pi}\Lambda^2+\frac{12 m^2}{\pi^2}\Lambda+\frac{4m^3}{\pi} \log\left(\frac{\Lambda}{2m}\right)-\frac{2m^3}{3\pi}+\mathcal{O}(m^4/\Lambda)\right]\,.
\end{align}
This result becomes more and more precise in the massless limit as we can infer from figure \ref{fig1}, where the complexity per mode approaches the constant $\pi/2$. That is, for the massless theory, we have simply $\mathcal{C}_{\kappa=2}\big(|\bar{0}\rangle\to|0\rangle\big)=V\Lambda^3/12$. 

Of course, as noted above, the results for the $F_2$ measure are simply given by taking a square root of the above complexities, \ie $\mathcal{C}_{2} =\sqrt{\mathcal{C}_{\kappa=2}}$. However, the divergence structure produced with the $\kappa=2$ measures matches more closely that found in holographic complexity, \ie we expect that $\mathcal{C}_\mt{holo}\sim V\Lambda^3$ \cite{Carmi:2016wjl}.

At this point, we should add that our results in eqs.~(\ref{eq:comp-exact}) and (\ref{eq:comp-asymp}) agree with those presented in \cite{Khan:2018rzm}, up to an overall normalization constant (\ie, if we multiply our results by $2\pi^2$, the expressions agree). This discrepancy simply arises due to a slight difference in the choice of conventions. Further, let us emphasize that our methods presented in section \ref{sec:general} and appendix \ref{app:minimal} \emph{prove} that our path is the minimal geodesic in the full $\mathrm{SO}(2N)$ group of the fermionic theory, which was left an open question in \cite{Khan:2018rzm}.

We might also consider the $\kappa=1$ measure (or equivalently, the $F_1$ measure) with, 
\begin{align}\label{tot3}
\mathcal{C}_{\kappa=1}\big(|\bar{0}\rangle\to|0\rangle\big)={V\int\!\! \frac{d^3\pp}{(2\pi)^3}\sum_\bs\,|{Y}(m,\pp,\bs)|}\,.
\end{align}
If we again introduce the cutoff $\Lambda$, we can do this integral explicitly and find the relatively simple expression
\begin{align}\label{eq:comp-asymp2}
	\mathcal{C}_{\kappa=1}\big(|\bar{0}\rangle\to|0\rangle\big)=\frac{V}{6\pi^2}\left[2\Lambda^3\tan^{-1}\!\left(\frac{\Lambda}{m}\right)-m\Lambda^2+m^3\log\left(1+\left(\frac{\Lambda}{m}\right)^2\right)\right]\,.
\end{align}
This expression then yields the following large $\Lambda/m$ expansion for the complexity, 
\begin{align}\label{eq:comp-kappa=1-exact}
	\mathcal{C}_{\kappa=1}\big(|\bar{0}\rangle\to|0\rangle\big)\simeq\frac{V}{6\pi}\left[\Lambda^3-\frac{3 m}{\pi}\Lambda^2+\frac{2m^3}{\pi} \log\left(\frac{\Lambda}{m}\right)+\frac{2m^3}{3\pi}+\mathcal{O}(m^5/\Lambda^2)\right]\,.
\end{align}
However, we should note that this measure (as well as the general $\kappa$ measures with $\kappa\ne 2$) is basis dependent \cite{Jefferson:2017sdb} and so implicitly we are choosing the normal mode basis in eq.~\reef{tot3}.

\subsection{Simple excited states} \label{excited1}
We should note that we can also evaluate the complexity of a number of excited states as well. First, we observe that the state $|\tilde{\psi}\rangle=a^{\br\dagger}_\qq|0\rangle$ with a single particle excitation (in a fixed spin state) remains a Gaussian state since it is annihilated by $a^{\br\dagger}_\qq$, \ie $a^{\br\dagger}_\qq |\psi\rangle=(a^{\br\dagger}_\qq)^2|0\rangle=0$, as well as the usual annihilation operators for all of the other spins and momenta. However, this particular state has odd fermion number and is on the disconnected component of the space of Gaussian states --- see footnote \ref{footyABC}. While we can only evaluate the complexity of Gaussian states with even fermion number, we will have to develop our formalism further in the next section to describe the complexity of general states with even fermion number --- see the discussion in section \ref{geee}. However, one simple set of states which we can consider here given the Bogoliubov transformations in eq.~\reef{bogo9} are excited states of the form\footnote{Recall our notation is $\brp=\br+1$ (mod 2). There is no sum over $\br$ here but both creation operators carry the opposite spin labels. Note that this state has vanishing particle number since it involves one particle and one antiparticle. Similarly, it has zero net momentum, but there is a net spin because, \eg with $\br=1$, the particle's spin is oriented in the $+\qq$ direction and the antiparticle with $\brp=2$ also has its spin pointing in the $+\qq$ direction --- see footnote \ref{footyA1}).} 
\beq
\textbf{(A)}\qquad|\tilde{\psi}\rangle = a^{\br\dagger}_\qq\,b^{\br'\dagger}_{-\qq}\,|0\rangle\,.
\label{excite8}
\eeq
The above state is annihilated by $a^{\br\dagger}_\qq$ and $b^{\br'\dagger}_{-\qq}$ (and again by the usual annihilation operators for all of the other modes). Hence eq.~\reef{march} still applies for most of the pairs of modes, but we must reconsider the contribution for the pair labeled by $\pp=\qq$ and $s=r$. However, for this pair of modes, we can simply relabel the annihilation operators $(\tilde{a},\tilde{b})=((-)^{\brp} b^{\brp\dagger}_{-\qq},(-)^\br a^{\br\dagger}_\qq)$. With this choice, eq.~\reef{bogo9} can be rewritten as
\begin{align}\label{bogo11}
\tilde{a}\ &=\tilde{\alpha}\  \bar{a}^\br_\qq-\tilde{\beta}\  \bar{b}_{-\qq}^{\brp\dagger}\,,\\
\tilde{b}^{\dagger}&=\tilde{\beta}\ \bar{a}^\br_\qq+\tilde{\alpha}\  \bar{b}_{-\qq}^{\brp\dagger}\,.\nonumber
\end{align}
with
\beqa
\tilde{\alpha}&=& (-)^{\brp}\beta_{\qq}^\br=
\frac{\sqrt{E_\qq+|\qq|}-\sqrt{E_\qq-|\qq|}}{2\sqrt{E_\qq}}\,,
\label{bogo11a}\\
\tilde{\beta}&=& (-)^\br  \alpha_{\qq}^\br=(-)^\br
\frac{\sqrt{E_\qq+|\qq|}+\sqrt{E_\qq-|\qq|}}{2\sqrt{E_\qq}}
\,.
\eeqa
Hence the Bogoliubov transformation still takes the simple form given in eq.~\reef{bogo8}. In particular, comparing to eq.~\reef{param}, we may set $\cos\tilde{\vt}=\tilde{\alpha}$ and $\vp=\pi$ for $\br=1$ (or $\vp=0$ for $\br=2$).  
\begin{figure}[t]
\begin{center}
	\includegraphics{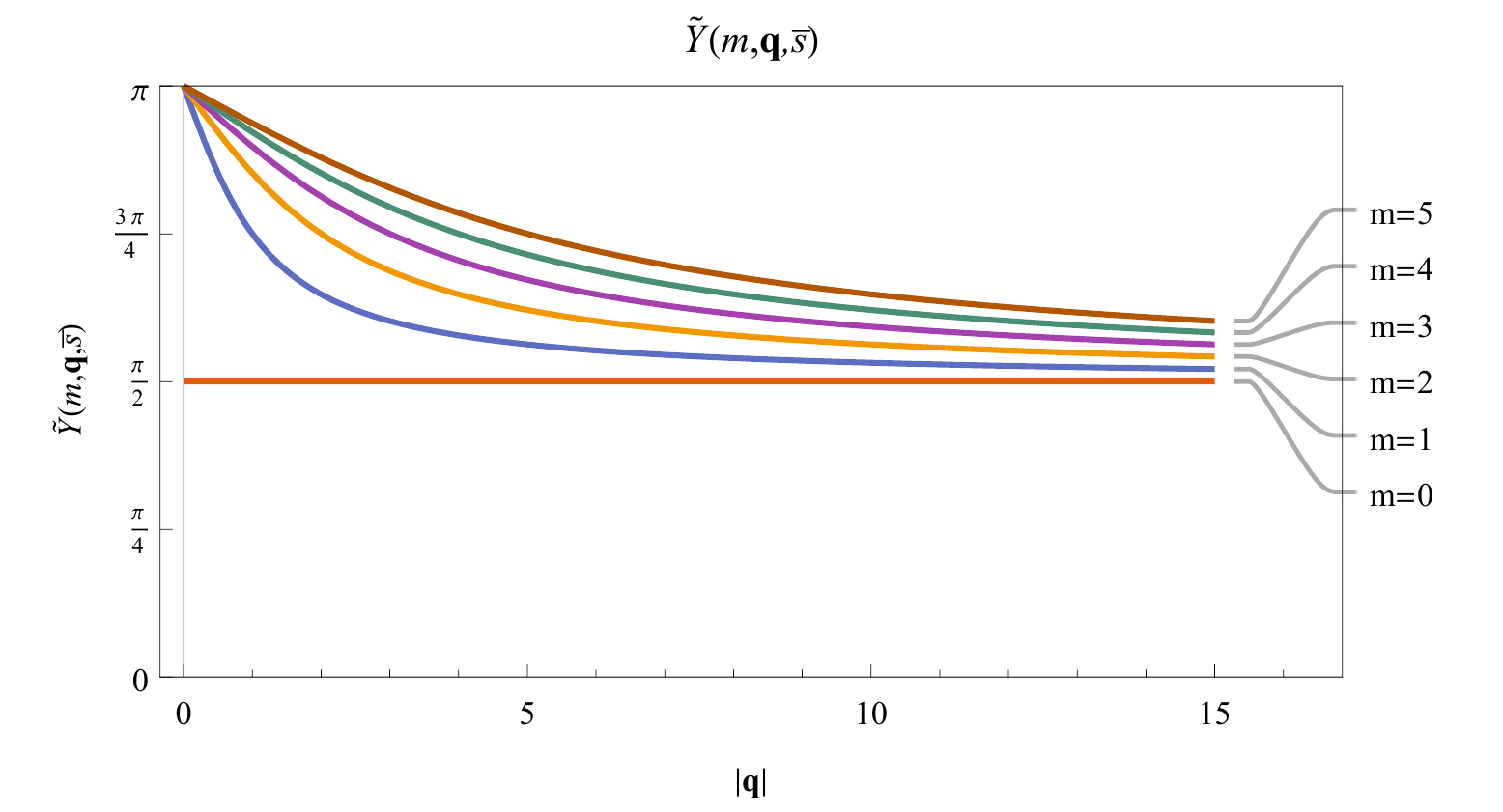}
\end{center}
	\caption{This plot shows the function $\tilde{Y}(m,\qq,\br)$ describing the complexity of the modes excited in the state in eq.~\reef{excite8}, \ie $|\tilde{\psi}\rangle=a^{\br\dagger}_\qq\,b^{\br\dagger}_{-\qq}\,|0\rangle$, as a function of $|\qq|$. \label{fig2}}
\end{figure}

Now the analog of eq.~\reef{march} for the Bogoliubov transformation \reef{bogo11a} for these particular modes is given by
\begin{align}\label{march2}
\tilde Y(m,\qq,\br)=
2\,\cos^{-1}\!\left[\tilde\alpha\right]=
2\,\tan^{-1}\!\left(\frac{{E}_\qq+m}{|\qq|}\right)=\pi-\tan^{-1}\!\left(\frac{|\qq|}{m}\right)\,.
\end{align}
Again, this result is independent of the spin label $\br$ appearing in the state \reef{excite8}. Comparing eqs.~\reef{march} and \reef{march2}, we see that $Y(m,\qq,\br)+\tilde Y(m,\qq,\br)=\pi$.\footnote{Alternatively, comparing eqs.~\reef{bogo11} and \reef{bogo11a}, we have $\cos\tilde{\vt}= \sin{\vt}$ and so $\tilde\vt=\frac{\pi}2-\vt$.} Therefore while $Y\in[0,\pi/2]$, we have $\tilde Y\in[\pi/2,\pi]$. Figure \ref{fig2} shows this expression as a function of $|\qq|$ for various values of the mass $m$. We note that for large $|\qq|$, the complexity per mode rapidly approaches $\pi/2$, which is now the minimal value (and also coincides with the contribution of these modes to the vacuum complexity). Again, $m=0$ is a special case where $\tilde Y(m=0,\qq,\br)=\pi/2$. 

As before, when evaluating the total complexity of our excited state \reef{excite8}, we must integrate the $Y(m,\pp,\bs)$ over all momenta $\pp$, as well as sum over the spin labels $\bs$. However, in this integration only a single contribution, \ie $\pp=\qq$ and $\bs=\br$, differs from that in the vacuum complexity. Hence, for example, with the $\kappa=2$ measure, we have
\beqa\label{tot2a}
\begin{split}
	\mathcal{C}_{\kappa=2}\big(|\bar{0}\rangle\to|\tilde{\psi}\rangle\big)&=\tilde Y(m,\qq,\br)^2- Y(m,\qq,\br)^2+{V\int\!\! \frac{d^3\pp}{(2\pi)^3}\sum_\bs\,{Y}(m,\pp,\bs)^2}
	\\
	&=\tilde Y(m,\qq,\br)^2- Y(m,\qq,\br)^2+\mathcal{C}_{\kappa=2}\big(|\bar{0}\rangle\to|0\rangle\big)\,.
\end{split}
\eeqa
Let us note an important subtlety in arriving at the above expression: At first sight, one may think that since $\tilde{Y}$ only differs for a single momentum mode, this should correspond to a set of measure zero in the integration and hence the complexity should remain unchanged. However, if we are working with a finite volume $V$, the momentum integral would become a discrete sum. Alternatively, each momentum mode occupies a cell of size $(2\pi)^3/V$ in the continuous integration, \ie one can think that exciting a single discrete (physical) mode $\qq$ is properly approximated by exciting all momenta in a cell of size $(2\pi)^3/V$ around $\qq$. These two perspectives are then reconciled by noting that in eq.~\reef{tot2a}, the additional terms do not scale with volume, \ie their contribution is vanishingly small in the limit $V\to\infty$.

Now just as with the vacuum complexity, the complexity of these excited states are UV divergent, as shown in eq.~\reef{eq:comp-asymp}. However, eq.~\reef{tot2a} shows that exciting the particle-antiparticle pair in eq.~\reef{excite8} only makes a finite perturbation of the vacuum complexity. Thus an interesting quantity to consider is the difference between the complexity of our excited state and that of the vacuum state, \ie
\beqa\label{diff2a}
\begin{split}
	\Delta\mathcal{C}_{\kappa=2}\big(|\bar{0}\rangle\to|\tilde{\psi}\rangle\big)&\equiv\mathcal{C}_{\kappa=2}\big(|\bar{0}\rangle\to|\tilde{\psi}\rangle\big)-\mathcal{C}_{\kappa=2}\big(|\bar{0}\rangle\to|0\rangle\big)\\
	&=\tilde Y(m,\qq,\br)^2- Y(m,\qq,\br)^2\\
	&=\pi\,(\pi- 2 \,Y(m,\qq,\br))\,,
\end{split}
\eeqa
which yields a UV finite quantity, \ie the UV divergences in the complexity of the excited state are precisely canceled by those in the vacuum complexity. Note that we used $Y(m,\qq,\br)+\tilde Y(m,\qq,\br)=\pi$ in the final expression. We can construct a similar difference using the $\kappa=1$ measure, which yields
\beqa\label{diff1a}
\begin{split}
	\Delta\mathcal{C}_{\kappa=1}\big(|\bar{0}\rangle\to|\tilde{\psi}\rangle\big)&\equiv\mathcal{C}_{\kappa=1}\big(|\bar{0}\rangle\to|\tilde{\psi}\rangle\big)-\mathcal{C}_{\kappa=1}\big(|\bar{0}\rangle\to|0\rangle\big)\\
	&=\tilde Y(m,\qq,\br)- Y(m,\qq,\br)\\
	&=\pi- 2 \,Y(m,\qq,\br)\,.
\end{split}
\eeqa
Interestingly, both of these differences are equal to one another up to an overall factor of $\pi$. In general, one finds tht $\Delta\mathcal{C}_{\kappa}\big(|\bar{0}\rangle\to|\tilde{\psi}\rangle\big) \propto \pi-2Y(m,\qq,\br)$ but the full expressions are more complex for general $\kappa$. Therefore, since $Y(m,\qq,\br)$ tends to $\pi/2$ for large momenta, $\Delta\mathcal{C}_{\kappa}\big(|\bar{0}\rangle\to|\tilde{\psi}\rangle\big)\to 0$ in the limit of large $\qq$.

One could attempt similar calculations with the $F_2$ measure \reef{tot1}. However, because of the square-root appearing in this expression, one finds that the difference in the complexities is vanishingly small, \ie
\beqa
\Delta\mathcal{C}_{2}\big(|\bar{0}\rangle\to|\tilde{\psi}\rangle\big)&\equiv&\mathcal{C}_{2}\big(|\bar{0}\rangle\to|\tilde{\psi}\rangle\big)-\mathcal{C}_{2}\big(|\bar{0}\rangle\to|0\rangle\big)
\nonumber\\
&\simeq&\frac{1}{2}\frac{\Delta\mathcal{C}_{\kappa=2}}{\sqrt{\mathcal{C}_{\kappa=2}}}
\propto\frac{\pi- 2 \,Y(m,\qq,\br)}{\sqrt{V\Lambda^3}}\,.
\label{diffXX}
\eeqa
Hence this analysis is less interesting for the $F_2$ measure. 

A simple extension of the above discussion would be to excite a finite number of particle-antiparticle pairs in a state of the form
\beq
|\tilde{\psi}\rangle = \prod_i a^{\br_i\dagger}_{\qq_i}\,b^{\brp_i\dagger}_{-\qq_i}\,|0\rangle\,.
\label{excite88}
\eeq
Again, these simple states are characterized by having vanishing particle number and vanishing net momentum. The above calculations extend in a straightforward manner and one would find, \eg 
\beq
\Delta\mathcal{C}_{\kappa=2}\big(|\bar{0}\rangle\to|\tilde{\psi}\rangle\big)=\pi\, \sum_i
(\pi- 2 \,Y(m,\qq_i,\br_i))\,.
\label{diff22a}
\eeq


With the methods developed so far, we can also examine the complexity of some other families of simple excited states. For example, we next consider states where we excite two particles or two antiparticles with the same momentum but opposite spins,
\beq
\textbf{(B)}\qquad
a_\qq^{1\dagger}\,a_\qq^{2\dagger}\,|0\rangle \qquad
{\rm and}
\qquad 
\textbf{(C)}\qquad b_\qq^{1\dagger}\,b_\qq^{2\dagger}\,|0\rangle\,.
\label{excite85}
\eeq
We will focus on the (B) states with two particle excitations in the following, but of course, the  discussion for (C) states would be the same after exchanging $a\leftrightarrow b$ (as well as $\qq\leftrightarrow -\qq$).  

In the new state $a_\qq^{1\dagger}\,a_\qq^{2\dagger}\,|0\rangle$, the sector describing $\qq$ momentum mode has annihilation operators $(\tilde{a}^\br_\qq,\tilde{b}_{-\qq}^\br)=(a^{\br\dagger}_\qq,b^{\br}_{-\qq})$ for $\br=1,2$. Similarly the creation operators are $(\tilde{a}^{\br\dagger}_\qq,\tilde{b}_{-\qq}^{\br\dagger})=(a^{\br}_\qq,b^{\br\dagger}_{-\qq})$, and so for this sector, the Bogoliubov expression \reef{bogo9} becomes
\begin{align}\label{bogo13}
\tilde{a}^{\br\dagger}_\qq\ &=\alpha^\br_\qq\  \bar{a}^\br_\qq-\beta^\br_\qq\  \bar{b}_{-\qq}^{\brp\dagger}\,,\\
	\tilde{b}_{-\qq}^{\brp\dagger}&=\beta^\br_\qq\ \bar{a}^\br_\qq+\alpha^\br_\qq\  \bar{b}_{-\qq}^{\brp\dagger}\,, \nonumber
\end{align}
where $\alpha^\br_\qq$ and $\beta^\br_\qq$ are given by eq.~\reef{bogo9a}. While this transformation does not take our standard form \reef{bogo9}, we see that the creation operators are both given by some (real) linear combination of $\bar{a}^\br_\qq$ and $\bar{b}_{-\qq}^{\brp\dagger}$. Analogously, the annihilation operators are linear combinations of $\bar{a}^{\br\dagger}_\qq$ and $\bar{b}_{-\qq}^{\brp}$ and hence both the target state and the reference state are annihilated by $\bar{b}_{-\qq}^{\brp}$! This contrasts with the standard situation for all of the other momentum modes. Examining eq.~\reef{bogo9}, it is straightforward to show that the target state is not annihilated by $\bar{a}_\pp^{\bs}$, $\bar{b}_{-\pp}^\bsp$ or any linear combination of these operators.
Hence the essential feature of the transformation \reef{bogo13} is that it implicitly swaps $\bar{a}_\qq^\br$ to $\bar{a}_\qq^{\br\dagger}$. Hence since the above transformation does not take our usual form, we instead pair the (annihilation) operators of the reference state as $(\bar{a}_\qq^1,\bar{a}_\qq^2)$ and $(\bar{b}_{-\qq}^{1},\bar{b}_{-\qq}^{2})$. We can then produce the desired target state \reef{excite85} with two transformations of the form in eqs.~\reef{bogo8} and \reef{param}. The first performs the desired swap on the $\bar{a}_\qq^\br$ with $\vartheta_2=\pi/2$ and the second leaves the $\bar{b}_{-\qq}^{\br}$ unchanged with an angle $\vartheta_1=0$, \ie we have
\begin{align}\label{penny14}
	\tilde{Y}(m,\pp,\bar a)=2\vartheta_1=\pi\,,\qquad\quad
	\tilde{Y}(m,\pp,\bar b)=2\vartheta_2=0\,.
\end{align}

Hence, as in eq.~\reef{tot2a} with the $\kappa=2$ measure, we have
\beqa\label{tot33}
\mathcal{C}_{\kappa=2}\big(|\bar{0}\rangle\to|\tilde{\psi}\rangle\big)&=&\tilde Y(m,\qq,\bar a)^2+\tilde Y(m,\qq,\bar b)^2- Y(m,\qq,1)^2\\
&&\qquad  - Y(m,\qq,2)^2+\mathcal{C}_{\kappa=2}\big(|\bar{0}\rangle\to|0\rangle\big)\,.
\nonumber
\eeqa
However, as in eq.~\reef{diff2a}, we may also consider the difference between the complexities of our excited state and the vacuum state, which yields
\beq\label{diff33a}
\Delta\mathcal{C}_{\kappa=2}\big(|\bar{0}\rangle\to|\tilde{\psi}\rangle\big)=\pi^2-2\, Y(m,\qq,1)^2\,,
\eeq
where we used the fact that $Y(m,\qq,\br)$ in eq.~\reef{march} is actually independent of the spin. We must note that the generators implied by the transformation described above in eq.~\reef{penny14} are not the same as the standard two-mode squeezing operators producing eq.~\reef{bogo9}. Hence, we have implicitly made use here of the fact that the $\kappa=2$ measure is independent of the basis of generators. While the same is not true of the $\kappa=1$ measure, we may write 
\beq\label{diff34}
\Delta\mathcal{C}_{\kappa=1}\big(|\bar{0}\rangle\to|\tilde{\psi}\rangle\big)=\pi-2\, Y(m,\qq,1)\,,
\eeq
as long as we align the basis for the excited modes with the generators which produce the above transformation. 
We note that this difference is identical to that found for the previous excited states with the $\kappa=1$ measure, in eq.~\reef{diff1a}.


To close the discussion here, we examine the complexity of a fourth class of simple excited states, 
\begin{align}\label{exciteD}
\textbf{(D)}\qquad |\tilde{\psi}\rangle=a_\qq^{\br\dagger}b^{\br\dagger}_{-\qq}|0\rangle\,.
\end{align}
In these states, we excite one particle and one antiparticle with opposite momenta as in eq.~\reef{excite8}, but here their spins are anti-aligned with each other. For example, setting $\bar{r}=1$, $a^{1\dagger}_\qq$ creates to a particle with its spin aligned to the $+\qq$ direction. However for the antiparticle, we have $b^{1\dagger}_{-\qq}$ which creates an antiparticle whose spin points in the same direction to its momentum $-\qq$. Thus the antiparticle spin oriented in the opposite direction to the spin of the particle, and the state \reef{exciteD} has zero net spin. 

Now using reasoning analogous to that in the previous case, one concludes that both the reference state and the new excited state are annihilated by $\bar{a}^\brp_\qq$ and $\bar{b}^\brp_{-\qq}$. Further the desired transformation must swap $\bar{a}_\qq^\br$ to $\bar{a}_\qq^{\br\dagger}$ and $\bar{b}_{-\qq}^\br$ to $\bar{b}_{-\qq}^{\br\dagger}$.
Hence following the previous reasoning, we  pair the annihilation operators as $(\bar{a}^\br_\qq,\bar{b}^\br_{-\qq})$ and $(\bar{a}^\brp_\qq,\bar{b}^\brp_{-\qq})$. We can then produce the desired target state \reef{exciteD} with two standard Bogoliubov transformations (as in eqs.~\reef{bogo8} and \reef{param}) where the transformation acting on the first pair produces the desired swap
with $\vartheta_1=\pi/2$ and one which leaves the second pair unchanged with $\vartheta_2=0$. Hence, we arrive at essentially the same result as in eq.~\reef{penny14}
\begin{align}\label{penny14x}
	\tilde{Y}(m,\pp,P1)=2\vartheta_1=\pi\,,\qquad\quad
	\tilde{Y}(m,\pp,P2)=2\vartheta_2=0\,.
\end{align}
Further, the results for the complexity are identical to those above for the states in eq.~\reef{excite85}. In particular, if we evaluate the difference between the complexities of this excited state and the vacuum state with the $\kappa=2$ and 1 measures, we find precisely the results in eqs.~\reef{diff33a} and \reef{diff34}, respectively.

At this point, we would like to emphasize that it was essential in our derivation of the complexity of the (A) and (D) families of excited states, in eqs.~\reef{excite8} and \reef{exciteD} that the spin axis of all of the excitations was aligned (or anti-aligned) with the momentum of the given mode. After developing systematic analytical tools in section~\ref{sec:general}, we will be able to generalize these classes in section~\ref{geee} by allowing a spin axis independent of the momentum direction. In contrast, the (B) and (C) families in eq.~\reef{excite85}, the two particles (or antiparticles) combine to form a spin singlet and therefore the result for the complexity should not rely on the alignment of the spin and momentum axes. As a final note, let us add that it is straightforward to extend to the discussion of the complexity of the states in eqs.~\reef{excite85} and \reef{exciteD} to states where we excite a finite number of pairs of particles and antiparticles (in analogy to eq.~\reef{excite88}).

\section{Complexity of general fermionic Gaussian states}\label{sec:general}
We study the circuit complexity of arbitrary fermionic Gaussian states $|\tilde{\Omega}\rangle$ with respect to an arbitrary Gaussian reference state $|\Omega\rangle$. In accord with our previous discussions, the notation here indicates that the state is characterized by $\Omega^{ab}$, the antisymmetric part of the covariance matrix \reef{covmat}. As discussed in section \ref{sec:prelude}, when we apply Nielsen's method by geometrizing the problem of finding the circuit complexity for fermionic systems, we restrict our study to Gaussian states. On the level of Lie groups, this means we are restricting ourselves to a $\mathrm{SO}(2N)$ subgroup of the full $\mathrm{U}(2^N)$ group of unitary transformations, which could act on the states for our system of $N$ fermionic degrees of freedom.

\subsection{Gaussian states from K\"ahler methods}
Recently, it has become apparent that bosonic and fermionic Gaussian states can be characterized in a unified framework \cite{BH-BFGaussians} based on a triangle of structures (K\"ahler methods) consisting of a positive definite metric $G$, a symplectic form $\Omega$ and a linear complex structure $J$. We will review the relevant ingredients of these methods and fix conventions. For the most part, this is a straightforward generalization of our initial warm up exercise with two fermionic degrees of freedom, and much of this analysis was anticipated in the discussion in section \ref{sec:prelude}.

A system with $N$ fermionic degrees of freedom is defined on a Hilbert space $\mathcal{H}=(\mathbb{C}^2)^{\otimes N}$. Linear observables can equivalently described by $N$ pairs of creation and annihilation operators $(a_i,a_i^\dagger)$ or their hermitian counterparts, the Majorana modes $(q_i,p_i)$ in eq.~\reef{heavy}. The latter provides a basis $\xi^a=(q_1,\cdots,q_N,p_1,\cdots,p_N)$ for linear fermionic observables. These form a vector space $\Gamma^*$ which we refer to as the dual phase space,\footnote{This is in direct analogy to the bosonic case, where linear observables are linear phase space functions and thus elements of the dual phase space. The same construction works for fermionic degrees of freedom, as well.} equipped with the positive definite metric $G^{ab}$ that fixes the anticommutation relations as $\lbrace\xi^a,\xi^b\rbrace=G^{ab}$. Recall that the Gaussian states are completely characterized by the antisymmetric covariance matrix $\ii\Omega^{ab}=\langle\psi|\xi^a\xi^b-\xi^b\xi^a|\psi\rangle$ and we will label these states accordingly, \ie   $|\psi\rangle=|\Omega\rangle$, in the following. Hence for our fermionic Gaussian states, eq.~\reef{covmat} becomes
\begin{align}\label{covmat2}
	\langle\Omega|\,\xi^a\,\xi^b\,|\Omega\rangle=\frac{1}{2}(G^{ab}+\ii\Omega^{ab})\,.
\end{align}
Again, this same form also applies for bosonic Gaussian states, however, the roles of $G$ and $\Omega$ are interchanged: For bosons, $G$ labels the state and $\Omega$ fixes the bosonic commutation relations, and as indicated above, $\Omega$ labels the fermionic states while $G$ determines the anticommutation relations for the fermionic degrees of freedom. We also introduce the inverse matrices $g$ and $\omega$ defined by the conditions $G^{ac}g_{cb}=\delta^a{}_b$ and $\Omega^{ac}\omega_{cb}=\delta^a{}_b$.

Mathematically speaking, $G$ represents a \emph{positive definite metric} and $\Omega$ a \emph{symplectic form} on the classical phase space isomorphic to $\mathbb{R}^{2N}$. Together they define a third object
\begin{align}
	J^a{}_b=\Omega^{ac}g_{cb}=-G^{ac}\omega_{cb}\,,
\end{align}
called a \emph{linear complex structure}. Together, they form a triangle of structures that we call \emph{K\"ahler structures} due to its common use in the context of K\"ahler manifolds. The beauty of parameterizing Gaussian states with these structures lies in the fact that this provides a unifying framework for both bosonic and fermionic Gaussian states \cite{BH-BFGaussians}. The linear complex structure $J$ can be used to label both types of states and characterizes them uniquely (up to a complex phase) via the following equation:
\begin{align}
	\frac{1}{2}(\delta^a{}_b-\ii\,J^a{}_b)\,{\xi}^b\,|J\rangle=0\,.
\end{align}
The relative covariance matrix $\Delta$ between a state $|\tilde{J}\rangle$ and $|J\rangle$ can be directly computed as $\Delta=-\tilde{J}\,J$. Again, this is the same formula for bosons and for fermions. However, as we will exclusively focus on fermions for the rest of this paper, we will continue to use the antisymmetric covariance matrix $\Omega$ to label the Gaussian state $|\Omega\rangle$.

\subsection{Geometry of $\mathrm{SO}(2N)$}
We explore the differential geometry of the group $\mathrm{SO}(2N)$ that corresponds to all fermionic squeezing operations that are connected to the identity. The Lie algebra $\mathrm{so}(2N)$ is given by generators $K$ that satisfy
\begin{align}
	K^a{}_c\,G^{cb}=-G^{ac}\,(K^\intercal)_c{}^b\,,
\end{align}
which is equivalent to saying that $K$ is antisymmetric with respect to $G$. As discussed in section \ref{twofermi}, a group element $M=\ex^K$ transforms a Gaussian state as
\beq\label{trans5}
|\Omega\rangle\quad\longrightarrow\quad |\tilde\Omega\rangle=|M\Omega M^\intercal\rangle\,,
\eeq
in accord with eq.~\reef{Osym3}.

Recall that if we choose a target state $|\Omega_\mathrm{T}\rangle$ and a reference state $|\Omega_\mathrm{R}\rangle$, eq.~\reef{circuitDef} still leaves an ambiguity in the desired transformation because there are transformations which leave the reference state unchanged --- see discussion around eq.~\reef{bc44}. Hence we must find the stabilizer subgroup that preserves $|\Omega_\mathrm{R}\rangle$:
\begin{align}\label{stab2}
	\mathrm{Sta}=\{M\in\mathrm{SO}(2N)\big|M\Omega_\mathrm{R} M^\intercal=\Omega_\mathrm{R}\}\,.
\end{align}
Due to the fact that $\Omega_\mathrm{R}$ is a symplectic form, the stabilizer subgroup of the state $|\Omega_\mathrm{R}\rangle$ is given by the intersection of the symplectic and the special orthogonal group which is well known to be $\mathrm{U}(N)$, \ie $\mathrm{Sta}=\mathrm{U}(N)=\mathrm{SO}(2N)\cap\mathrm{Sp}(2N,\mathbb{R})$. Similar to what we saw in section \ref{twofermi},  this $\mathrm{U}(N)$ subgroup corresponds to Bogoliubov transformations which only mix creation and annihilation operators among themselves respectively and which therefore do not change the state being annihilated. The corresponding Lie subalgebra $\mathrm{u}(N)\subset\mathrm{so}(2N)$ is generated by algebra elements $K$ satisfying
\begin{align}\label{switch}
	K\Omega_\mathrm{R}=-\Omega_\mathrm{R}K^\intercal=(K\Omega_{\mathrm{R}})^\intercal{}\,,
\end{align}
which means that $K$ is symmetric with respect to $\Omega_\mathrm{R}$, \ie $(K\Omega_\mathrm{R})^{ab}=(K\Omega_\mathrm{R})^{(ab)}$.

Before we can compute geodesics on $\mathrm{SO}(2N)$, we need to equip it with a geometric structure, namely a right-invariant metric following Nielsen's approach. At this point, we need to make a choice and for a general metric, we would not be able to continue with analytical methods, because even for a right-invariant metric, computing the corresponding geodesics will be very hard. However, for the group $\mathrm{SO}(N)$, there is canonical choice that is compatible with the group structure and built from the metric $G$ that determines the anticommutation relations. As introduced in eq.~\reef{lion}, this choice of metric $\langle\cdot,\cdot\rangle_{\mathbb{1}}$ is given by
\begin{align}
	\langle A,B\rangle_{\mathbb{1}}=\mathrm{Tr}(AGB^\intercal g)=A^a{}_bG^{bc}(B^\intercal)_c{}^dg_{da}\,.
\end{align}
For bosons, such a choice depends on the reference state, but for fermions it is the completely canonical choice that is already induced by the group structure. Not surprisingly, it is proportional to minus the Killing form which is provides a negative definite bilinear form for compact groups. Such a canonical choice does not exist for bosons, because the symplectic group is non-compact and therefore its Killing form is not definite \cite{kirillov2008introduction}. Rather it would give rise to a Lorentzian geometry.

In order to find the minimal geodesic from the identity to some final group element $M$ that prepares a target state $|\Omega_\mathrm{T}\rangle$, we will need to identify the equivalence classes of all group elements preparing the same state. The equivalence relation therefore becomes $M\sim\tilde{M}$ iff $M\Omega_{\mathrm{R}} M^\intercal=\tilde{M}\Omega_{\mathrm{R}} \tilde{M}^\intercal$. The latter can be reformulated as
\begin{align}	\Omega_{\mathrm{R}}=M^{-1}\tilde{M}\Omega_{\mathrm{R}}\tilde{M}^\intercal(M^{-1})^\intercal=(M^{-1}\tilde{M})\Omega_{\mathrm{R}}(M^{-1}\tilde{M})^\intercal\,,
\end{align}
which implies $(M^{-1}\tilde{M})\in\mathrm{Sta}=\mathrm{U}(N)$, the subgroup \reef{stab2} that preserves the reference state $|\Omega_{\mathrm{R}}\rangle$. If we define $u:=M^{-1}\tilde{M}\in\mathrm{U}(N)$, we find $\tilde{M}=M\,u$. This means $M\sim \tilde{M}$ iff there exists a $u\in\mathrm{U}(N)$, such that $\tilde{M}=M\,u$.\par
Similar to the bosonic case, there exists a polar decomposition of any group element $M\in\mathrm{SO}(2N)$, such that $M=Tu$ with
\begin{align}
	T=\sqrt{M\Omega_{\mathrm{R}} M^\intercal\omega_{\mathrm{R}}}\qquad\text{and}\qquad u=T^{-1}M\,.
\end{align}
We need to verify that $T\Omega_{\mathrm{R}}T^\intercal=M\Omega_{\mathrm{R}} M^\intercal$ holds which implies $u\in\mathrm{U}(N)$. We can do this by first confirming that $T^2=M\Omega_{\mathrm{R}}M^\intercal\omega_{\mathrm{R}}$ is symplectic since it satisfies $T^2\Omega_{\mathrm{R}}=\Omega_{\mathrm{R}} (T^2)^\intercal$.  
This implies that its squareroot is also symplectic, \ie it implies that
\beq\label{sympl4}
T\Omega_{\mathrm{R}}=\Omega_{\mathrm{R}} T^\intercal\,,
\eeq
which will be a distinguishing feature to identify $T$. Now to complete the proof, we use the above feature to compute
\begin{align}
	T\Omega_{\mathrm{R}} T^\intercal=T^2\Omega_{\mathrm{R}}=M\Omega_{\mathrm{R}} M^\intercal\omega_{\mathrm{R}}\Omega_{\mathrm{R}}=M\Omega_{\mathrm{R}} M^\intercal\,,
\end{align}
which we wanted to verify. Now using eq.~\reef{trans5}, we write the target state as $|\Omega_{\mathrm{T}}\rangle=|M\Omega_{\mathrm{R}} M^\intercal\rangle$, which implies
\begin{align}\label{top1}	T^2=M\Omega_{\mathrm{R}}M^\intercal\omega_{\mathrm{R}}=\Omega_{\mathrm{T}}\omega_{\mathrm{R}}=\Delta\,,
\end{align}
where we recall from eq.~\reef{jangle} that $\Delta^{a}{}_b=\Omega_{\mathrm{T}}^{ac}(\omega_{\mathrm{R}})_{cb}$ is the relative covariance matrix between the states $|\Omega_{\mathrm{T}}\rangle$ and $|\Omega_{\mathrm{R}}\rangle$. $\Delta$ will have eigenvalues $\ex^{\ii 2\vt}$ with modulus $1$. The square root $\Delta=\sqrt{T}$ then has eigenvalues $\ex^{\ii \vt}$ with $\vt\in[-\pi/2,\pi/2]$.

At this point, we reached a fairly good geometric understanding of the group $\mathrm{SO}(2N)$ as a fiber bundle over its quotient $\mathrm{SO}(2N)/\mathrm{U}(N)$. In particular, we can use the polar decomposition to select a unique point from each fiber, namely the group element $T$ which satisfies $T\Omega_{\mathrm{R}}=\Omega_{\mathrm{R}}T^\intercal$, as in eq.~\reef{sympl4}. In contrast to the symplectic group, there are many Lie algebra elements $A\subset\mathrm{so}(2N)$ that satisfy $\ex^A=T$. However, if we use the standard definition of the logarithmic map that takes $\ex^{\ii 2\vt}$ to a real number $\vt\in(-\pi/2,\pi/2]$, the Lie algebra element $A=\log T=\frac{1}{2}\log\Delta$ becomes unique.

\subsection{Normal modes and two-mode squeezing}

Now we show how for any two fermionic Gaussian states, we can find a set of normal modes, such that one state results from the other by applying independent two-mode squeezing operations onto pairs of normal modes.

Recall the group invariant information about the relation between reference and target state is captured in the relative covariance matrix \reef{jangle}
\begin{align}
	\Delta^a{}_b=\Omega_\mathrm{T}^{ac}\, (\omega_\mathrm{R})_{cb}\,.
\end{align}
In particular, for $\Delta=\mathbb{1}$, reference and target state are the same. Note that $\Delta$ satisfies  $\Delta\Omega_{\mathrm{R}}=\Omega_{\mathrm{R}}\Delta^\intercal$ which is a similar, but different condition than being symplectic. Due to being an element of $\mathrm{SO}(2N)$, its eigenvalues are complex numbers with unit modulus $\ex^{2\ii\vti}$. They can either appear in quadruples $(\ex^{2\ii\vti},\ex^{2\ii\vti},\ex^{-2\ii\vti},\ex^{-2\ii\vti})$ or in pairs $(1,1)$ or $(-1,-1)$. There are in general two classes of spectra which correspond to:
\begin{itemize}
	\item \textbf{Reference and target state cannot be disconnected}\\
	If there is an odd number of pairs $(-1,-1)$ in the spectrum of $\Delta$, the reference and target states are located on disconnected components of the space of fermionic Gaussian states $\mathcal{M}_{f,\ssc N}=\mathrm{O}(2N)/\mathrm{U}(N)$ and they cannot be joined by a geodesic through $\mathrm{SO}(2N)$. 
	\item \textbf{Reference and target state can be connected}\\
	In all other instances, we can find a geodesic that connects the reference state to the target state.
\end{itemize}
We could assign a complexity of infinity to the former class of states, because reference and target state cannot be connected by applying quadratic operators as gates. Instead, one could try to extend the group or analytically continue the formula for the geodesic distance that we will find. However in the following, we will only be considering the complexity for the latter class of states that lie on the same connected component.

After computing the eigenvalues of $\Delta$, we can combine the complex eigenvectors associated to a quadruple of eigenvalues  $(\ex^{2\ii\vti},\ex^{2\ii\vti},\ex^{-2\ii\vti},\ex^{-2\ii\vti})$ to choose a quadruple of real eigenvectors which form an orthonormal basis (in this $I$'th sector with respect to $G$): $(\xi^a)_I = (q,Q,p,P)_I$. In particular, the latter can be chosen such that the associated block of $\Delta$ takes the form
\begin{align}
	\Delta_I\equiv\left(
	\begin{array}{cccc}
	\cos (2 \vti) & -\sin (2 \vti) & 0 & 0 \\
	\sin (2 \vti) & \cos (2 \vti) & 0 & 0 \\
	0 & 0 & \cos (2 \vti) & \sin (2\vti) \\
	0 & 0 & -\sin (2 \vti) & \cos (2 \vti)
	\end{array}
	\right)\,,
\end{align}
which we can recognize from our discussion in section \ref{twofermi}, if we combine the covariance matrices in eqs.~\reef{canon0} and \reef{canon} together in eq.~\reef{jangle}. For eigenvalue pairs $(1,1)$, we have orthonormal eigenvectors $(q,p)_I$, such that the corresponding block in $\Delta$ takes the form
\begin{align}
	\Delta_I\equiv\left(\begin{array}{cc}
	1 & 0\\
	0 & 1
	\end{array}\right)\,,
\end{align}
which is associated to a degree of freedom that was not squeezed. For eigenvalue pairs $(-1,-1)$, we need to count how many such pairs exist. If we have an even number of such pairs, we can actually group them to form quadruples $(\ex^{2\ii\vti},\ex^{2\ii\vti},\ex^{-2\ii\vti},\ex^{-2\ii\vti})$ with $\vti=\pi/2$. However, if we have an odd number of pairs, \ie an even number of eigenvalues $-1$ that cannot be divided by $4$, we are left with one degree of freedom with basis $(q,p)_I$, such that the block in $\Delta_I$ takes the form
\begin{align}
\Delta_I\equiv\left(\begin{array}{cc}
-1 & 0\\
0 & -1
\end{array}\right)\,.
\end{align}
If such a block stands alone and cannot be combined with another one, it implies that reference and target states belong to topologically disconnected components of $\mathcal{M}_{f,\ssc N}$ because there is no one-mode squeezing operation connected to the identity that connects reference to target state. We can always find a basis, in which reference and target state in this sector take the form
\begin{align}
	(\Omega_\mathrm{R})_I\equiv\left(\begin{array}{cc}
	0 & -1\\
	1 & 0
	\end{array}\right)\quad\text{and}\quad (\Omega_\mathrm{T})_I\equiv\left(\begin{array}{cc}
	0 & 1\\
	-1 & 0
	\end{array}\right)\,.
\end{align}
Clearly, this transformation could be implemented by the group element
\begin{align}
	M_I\equiv\left(\begin{array}{cc}
	1 & 0\\
	0 & -1
	\end{array}\right)\in\mathrm{O}(2)
\end{align}
with $\Omega_{\mathrm{T}}=M_I\,\Omega_{\mathrm{R}}\,M_I^\intercal$. However, this group element is not connected to the identity on $\mathrm{O}(2)$ and $\log M_I$ will not give a proper Lie algebra element in $\mathrm{so}(2)$.\footnote{In fact, the logarithm could be computed as $\log M_I=\left(\begin{array}{cc}
	0 & 0\\
	0 & \ii \pi
\end{array}\right)$, which is not a real Lie algebra element, but lies in the complexification of $\mathrm{so}(2)$. In particular, there exists a minimal path from $\mathbb{1}$ to $M_I$ within the complexification of $\mathrm{O}(2)$.}

We prove in appendix \ref{app:minimal} that the shortest geodesic between reference and target states that can be connected is given by
\begin{align}
	\gamma: [0,1]\to\mathrm{SO}(2N): s\mapsto \ex^{sA}\quad\text{with}\quad A=\frac{1}{2}\log\Delta\,.
\end{align}
Right-invariance of the metric implies that the length of this path is just given by the norm of $A$. Using the $F_2$ cost function \reef{eq:Fmetrics}, this implies that we can compute the complexity as
\begin{align}\label{pony1}
	\mathcal{C}_2\big(|\Omega_{\mathrm{R}}\rangle\to|\Omega_{\mathrm{T}}\rangle\big)=\left\lVert \frac{1}{2}\log\Delta\right\rVert=\frac{1}{2}\sqrt{\mathrm{Tr}\left[(\ii\log\Delta)^2\right]}\,.
\end{align}
In terms of our eigenvalue quadruples $(\ex^{2\ii\vti},\ex^{2\ii\vti},\ex^{-2\ii\vti},\ex^{-2\ii\vti})$, we find
\begin{align}
	\mathcal{C}_2\big(|\Omega_{\mathrm{R}}\rangle\to|\Omega_{\mathrm{T}}\rangle\big)=\sqrt{\sum_I (2\vti)^2}\,,\label{complex-contribution}
\end{align}
where $\vti\in[0\,,\pi/2]$. If we choose generators ${\cal O}_I$ in eq.~\reef{controlY} that coincide with the two-mode squeezing operations that generate $\vti$, we can normalize them to satisfy
\begin{align}\label{pony4}
	Y^I=2\,\vti\,.
\end{align}
Such a choice would always be adapted to the pair consisting of reference and target state because $\Delta=\Omega_{\mathrm{T}}\omega_{\mathrm{R}}$ and explicitly given by
\begin{align}
	\mathcal{O}_I=\frac{\log\Delta_I}{\lVert\log\Delta_I\rVert}\,.
\end{align}
The complexity with the $\kappa=2$ measure is then given by
\begin{align}
	\mathcal{C}_{\kappa=2}\big(|\Omega_{\mathrm{R}}\rangle\to|\Omega_{\mathrm{T}}\rangle\big)=\sum_I\,|Y^I|^2=4\sum_{I}\,\vti^{\,2}\,,
\end{align}
while in this basis,\footnote{Recall that for general $\kappa$ measures, \ie $\kappa\ne2$, the complexity is basis dependent \cite{Jefferson:2017sdb} --- see further discussion in section \ref{discuss}.} the $\kappa=1$ measure yields
\begin{align}\label{pony2}
\mathcal{C}_{\kappa=1}\big(|\Omega_{\mathrm{R}}\rangle\to|\Omega_{\mathrm{T}}\rangle\big)=\sum_I\,|Y^I|=2\sum_I\,|\vti|\,.
\end{align}

\section{Applications} \label{apply}

At this point, we have developed general methods to compute the circuit complexity of arbitrary pairs of fermionic Gaussian states as reference and target states. Hence in the present section, we will return to considering the free Dirac field in four dimensions. In section \ref{sec:Dirac}, we already considered the circuit complexity of the ground state and of certain special excited states. Here, we extend these results for the free Dirac field by applying the general method developed in section \ref{sec:general}.  For example, our calculations in section \ref{sec:Dirac} implicitly involved choosing a particularly simple reference state. In section \ref{sec:alternative}, we examine how other choices for the reference state modify the complexity of the ground state. Further in section \ref{geee}, we discuss the complexity of more general excited states. However, we begin below by describing how the general construction of section \ref{sec:general} can be adapted to the continuum quantum field theory of a free Dirac fermion.

In the previous section, we computed the complexity for arbitrary fermionic Gaussian states for systems with a finite number of degrees of freedom, namely $N$, and the key lesson was that the computations simplify for two pure Gaussian states $|\Omega_{\mathrm{R}}\rangle$ and $|\Omega_{\mathrm{T}}\rangle$ of the form
\begin{align}
	\Omega_{\mathrm{R}}=\oplus_{I}\,(\Omega_{\mathrm{R}})_I\,,\qquad\quad
	{\Omega_{\mathrm{T}}}=\oplus_{I}\,(\Omega_{\mathrm{T}})_I\,,
\end{align}
where the $\Omega_{\mathrm{R}}$ and ${\Omega_{\mathrm{T}}}$ are block-diagonal with respect to \emph{the same basis}. This structure immediately implies that the states themselves are tensor products of the form
\begin{align}
	|\Omega_{\mathrm{R}}\rangle=\otimes_I\,|(\Omega_{\mathrm{R}})_I\rangle\,,\qquad |{\Omega_{\mathrm{T}}}\rangle=\otimes_I\,|(\Omega_{\mathrm{T}})_I\rangle\,.
\end{align}
In this case, the relative covariance matrix takes the form
\begin{align}
	\Delta=\oplus_I\,\Delta_I\qquad\text{with}\quad \Delta_I=(\Omega_{\mathrm{T}})_I\ (\omega_{\mathrm{R}})_I\,.
\end{align}
For the Gaussian states on connected component of the state space, each of the $\Delta_I$ has a quadruple of eigenvalues $(\ex^{2\ii\vti},\ex^{2\ii\vti},\ex^{-2\ii\vti},\ex^{-2\ii\vti})$ and 
the overall complexity is then given by, \eg
\begin{align}
	\mathcal{C}_{\kappa}(|\Omega_{\mathrm{R}}\rangle\to|\Omega_{\mathrm{T}}\rangle)=\sum_I\,|2\,\vti|^\kappa\,.
\end{align}

If we are dealing with a continuum quantum field theory, the label $I$ can be continuous by referring for instance to the momentum $\pp$ or the position $\xx$. In this case, the states $|(\Omega_{\mathrm{R}})_I\rangle$ and $|(\Omega_{\mathrm{T}})_I\rangle$ will describe the degrees of freedom for each mode at a given momentum $\pp$ or  position $\xx$, which we expect to be finite in number, \eg the spin $s$ and particle number as in section \ref{sec:Dirac}. 
Then as in eq.~\reef{pony4}, we can recover the complexity per mode with
\begin{align}\label{pony5}
	Y^I=2\,\vti=\frac{1}{2}\sqrt{\mathrm{Tr}\left[(\ii\log\Delta_I)^2\right]}\,,
\end{align}
and we can apply our previous results without any alteration. At this point, we have full control over the individual contributions to the complexity and can study how the overall complexity behaves with various cost functions. We will see that the complexity is in most cases both, IR and UV divergent, but by understanding the individual pieces, we can meaningfully regularize these divergences. For the IR divergence, we put the whole system into a box with volume $V$ and for the UV divergence, we introduce a momentum cutoff $\Lambda$.

For the rest of this section, we will focus on \emph{translationally} invariant states. These states necessarily have a tensor product structure over momentum modes:\footnote{Translational invariance still allows for the possibility of correlating the mode $\pp$ with $-\pp$, which means it would be more precisely to have a tensor product of mode pairs $(\pp,-\pp)$, but we will focus on states that are \emph{actually} tensor product states over all modes $\pp$.}
\begin{align}\label{pony7}
	|\Omega_{\mathrm{R}}\rangle=\otimes_{\pp}\,|\Omega_{\mathrm{R}}(\pp)\rangle_\pp\,,\qquad\quad |{\Omega_{\mathrm{T}}}\rangle=\otimes_{\pp}\,|{\Omega_{\mathrm{T}}}(\pp)\rangle_{\pp}\,.
\end{align}
For our reference state $|\Omega_{\mathrm{R}}\rangle$, we must choose a state that is not just translationally invariant, but also has zero spatial correlation which implies that it should be a tensor product state in position space. This requirement enforces that the covariance matrix that characterizes each momentum mode $|\Omega_{\mathrm{R}}(\pp)\rangle_\pp$ must be the same, \ie
\begin{align}\label{penny14z}
	|\Omega_{\mathrm{R}}\rangle=\otimes_\pp\,|\Omega_0\rangle_\pp=\otimes_\xx\,|\Omega_0\rangle_\xx\,,
\end{align}
Therefore the reference state $|\Omega_{\mathrm{R}}\rangle$ is completely characterized by the finite dimensional covariance matrix $\Omega_0$, which describes the correlations within the degrees of freedom associated with each mode. The fact that these correlations look the same when studied in either momentum or position space is a consequence of $|\Omega_{\mathrm{R}}\rangle$ being translationally invariant without spatial correlations.

Examining translationally invariant states of the Dirac field in more detail, we begin by noting that for each momentum mode $\pp$,
the Dirac field has the four components $\psi(\pp)=(\psi_1(\pp),\psi_2(\pp),\psi_3(\pp),\psi_4(\pp))^\intercal$
with $\psi_i(\pp)=\int d^3x\, \ex^{-\ii \pp\cdot \xx}\,\psi(\xx)$. The anticommutation relations are given by
\begin{align}\label{lobe}
\{\psi_i(\pp),\psi^\dagger_j(\qq)\}=(2\pi)^3\,\delta_{ij}\,\delta^{(3)}(\pp-\qq)\,.
\end{align}
Thus, we can associate the Hilbert space $(\mathbb{C}^{2})^4$ to each momentum mode $\pp$. From the expansion in eq.~\reef{Dirac}, we have
\begin{align}
\psi(\pp)=\frac{\sqrt{m}}{\sqrt{2E_\pp}}\sum_s\left(a_{\pp}^s\,u^s(\pp)+b_{-\pp}^{s\dagger}\,v^s(-\pp)\right)\,.\label{expansion}
\end{align}
Now following eq.~\reef{heavy} for each momentum $\pp$, we can define four pairs of Majorana modes  
\begin{align}\label{eq:xi-momentum-basis}
Q_i(\pp)=\frac{1}{\sqrt{2}}\left(\psi_i^\dagger(\pp)+\psi_i(\pp)\right)\,,\qquad 
P_i(\pp)=\frac{\ii}{\sqrt{2}}\left(\psi_i^\dagger(\pp)-\psi_i(\pp)\right)\,.
\end{align}
In the notation of section \ref{sec:prelude}, we assemble these modes as $\xi^a(\pp)=\big(Q_i(\pp),P_i(\pp)\big)$, which then satisfy
\beq
\{\xi^a(\pp),\xi^b(\qq)\}=(2\pi)^3\,\delta^{ab}\,\delta^{(3)}(\pp-\qq)\,.
\eeq
The ground state $|0\rangle$ of the Dirac field has a covariance matrix \reef{covmat2} given by
\begin{align}
\langle 0|\xi^a(\pp)\xi^b(\qq)|0\rangle=\frac{1}{2}\left(G^{ab}(\pp)+\ii\Omega^{ab}(\pp)\right)\times (2\pi)^3\,\delta^{(3)}(\pp-\qq)\,.\label{definition}
\end{align}
With respect to our basis $\xi^a$ above, the symmetric component is simply $G^{ab}(\pp)\equiv\delta^{ab}$. What remains is to compute the antisymmetric component $\Omega^{ab}(\pp)$, which is a real linear form. For a pure state, $\Omega^{ab}(\pp)$ is a symplectic form compatible with $G^{ab}(\pp)$, which is equivalent to saying that with respect to the above basis, the matrix $\Omega^{ab}(\pp)$  has eigenvalues $\pm \ii$.

Recall that $\Omega^{ab}$ is the component of the covariance matrix that characterizes the fermionic Gaussian states. As in eq.~\reef{pony7}, the Dirac vacuum $|0\rangle$ can be written as a tensor product over sectors for each momentum $\pp$, \ie
\begin{align}\label{boil2}
|0\rangle=\bigotimes_{\pp\in\mathbb{R}^3}|\Omega(m,\pp)\rangle_\pp\,,
\end{align}
where we made the dependence on the mass $m$ explicit. We can evaluate the covariance matrix $\Omega^{ab}(m,\pp)$, which will encode the relevant properties of the complexity in a given mode $\pp$. Computing $\Omega^{ab}(\pp)$ explicitly takes some work: We begin by explicitly evaluating the spinors $u^\bs(\pp)$ and $v^\bs(-\pp)$, \eg see eq.~\reef{boost}.\footnote{Recall from footnote \ref{footyA1}, that we evaluate $u^\bs(\pp)$ and $v^\bs(-\pp)$ by boosting the spinors in eq.~\reef{rest} along the $z$-axis with $\tilde p_z=|\pp|$ and then rotate the spinor to align the momentum (and spin) with the direction of $\pp$.}   We then substitute these expressions into the eq.~(\ref{expansion}) and then write out the left-hand side of eq.~(\ref{definition}) in terms of the creation and annihilation operators of $|0\rangle$. Using their algebra, we can then simplify the right-hand side and extract $\Omega^{ab}(m,\pp)$. What we find is rather simple and can be expressed using $\pp=(p_x,p_y,p_z)$:
\begin{align}\label{capital}
\Omega^{ab}(m,\pp)\equiv\left(
\begin{array}{cccccccc}
0 & \frac{p_y}{E_\pp} & 0 & 0 & -\frac{p_z}{E_\pp} & -\frac{p_x}{E_\pp} & \frac{m}{E_\pp} & 0 \\
-\frac{p_y}{E_\pp} & 0 & 0 & 0 & -\frac{p_x}{E_\pp} & \frac{p_z}{E_\pp} & 0 & \frac{m}{E_\pp} \\
0 & 0 & 0 & -\frac{p_y}{E_\pp} & \frac{m}{E_\pp} & 0 & \frac{p_z}{E_\pp} & \frac{p_x}{E_\pp} \\
0 & 0 & \frac{p_y}{E_\pp} & 0 & 0 & \frac{m}{E_\pp} & \frac{p_x}{E_\pp} & -\frac{p_z}{E_\pp} \\
\frac{p_z}{E_\pp} & \frac{p_x}{E_\pp} & -\frac{m}{E_\pp} & 0 & 0 & \frac{p_y}{E_\pp} & 0 & 0 \\
\frac{p_x}{E_\pp} & -\frac{p_z}{E_\pp} & 0 & -\frac{m}{E_\pp} & -\frac{p_y}{E_\pp} & 0 & 0 & 0 \\
-\frac{m}{E_\pp} & 0 & -\frac{p_z}{E_\pp} & -\frac{p_x}{E_\pp} & 0 & 0 & 0 & -\frac{p_y}{E_\pp} \\
0 & -\frac{m}{E_\pp} & -\frac{p_x}{E_\pp} & \frac{p_z}{E_\pp} & 0 & 0 & \frac{p_y}{E_\pp} & 0 \\
\end{array}
\right)\,,
\end{align}
where as usual, $E_\pp=\sqrt{\pp^2+m^2}$. Note that here $\Omega^{ab}(m,\pp)$ is an eight-by-eight matrix because for each each momentum mode, it describes correlations over both spin and particle number.\footnote{Hence in the following discussion, the corresponding relative covariance matrix (see eq.~\reef{penny11} or \reef{penny12}) will have eight eigenvalues $\ex^{\pm2\ii\vti}$. This contrasts with the previous discussion in section \ref{sec:Dirac}, where implicitly the spin was treated as a separate quantum number and we had a single quadruple of eigenvalues, as in section \ref{sec:general}.}

With eq.~\reef{capital} in hand, it is easy to take various limits. For example, we can consider the rest frame ($\pp=0$) and the massless limit ($m=0$ or equivalently $|\pp|\to\infty$). In the next section, we will use these expressions to define the covariance matrix $\Omega_0$ appearing in the reference state \reef{penny14z}. For example, in this newly developed language, the reference state $|\bar{0}\rangle$ appearing in section~\ref{sec:Dirac} can be described  as
\begin{align}
	|\bar{0}\rangle=\otimes_\pp \,|\Omega(M,0)\rangle_\pp\,.
\end{align}
That is, in this state, we put every single momentum mode $\pp$ into the same state corresponding to zero-momentum mode of Dirac ground state with covariance matrix $\Omega(M,0)$. Note that we have introduced a new mass scale $M$ here, but in fact it turns out that $\Omega(M,0)$ is a special case which independent of $M$ --- see further discussion in section \ref{sec:alternative}.

\subsection{Alternative reference states}\label{sec:alternative}

Above, we discussed that requiring our reference state $|\Omega_{\mathrm{R}}\rangle$ be translationally invariant and also have no spatial correlations enforces that $|\Omega_{\mathrm{R}}\rangle$ take the simple form given in eq.~\reef{penny14z}. Therefore, the reference state is completely determined by a single (finite dimensional) covariance matrix $\Omega_0$, which fixes the correlations for each momentum mode. We can then study the contribution to the complexity for each momentum mode $\pp$ by considering the geodesic length from some reference state 
$|\Omega_0\rangle_\pp$ to $|\Omega(m,\pp)\rangle_\pp\in (\mathbb{C}^2)^{4}$. The minimal geodesic for the full theory then moves in this normal mode submanifold and the full complexity of the vacuum state combines all of these contributions for each momentum sector.  As alluded to above, we specify the reference covariance matrix $\Omega_0$ with eq.~\reef{capital} and a specific choice of a reference momentum $\qq$ and a reference mass $M$, \ie we set $\Omega_0^{ab}=\Omega^{ab}(M,\qq)$.\footnote{Of course, as well as substituting $\pp\to\qq$ and $m\to M$ in eq.~\reef{capital}, we also replace $E_\pp\to\tilde E_\qq=\sqrt{\qq^2+M^2}$.} We emphasize that we are using the same fixed momentum $\qq$ for all of the momentum modes. In the notation of eq.~\reef{boil2}, the reference state can be written as
\begin{align}\label{coil3}
|M,\qq\rangle	=\bigotimes_{\pp\in \mathbb{R}^3}|\Omega(M,\qq)\rangle_\pp\,.
\end{align}
In comparison to section \ref{sec:Dirac}, we are replacing eq.~\reef{Dirac2} with 
\begin{align}\label{Dirac2a}
\psi(\xx)=\frac{1}{\sqrt{2}}\sum_\bs\left(\bar{a}_\xx^\bs\,  \tilde{u}^\bs(\qq)+ \bar{b}_\xx^{\bs\dagger}\, \tilde{v}^\bs(\qq)\right)\,.
\end{align}
where the basis spinors defined as above, except the tilde superscript indicates the replacement $m\to M$. 

Now in principle, for each quadruple of momentum modes, we would compute the geodesic between $|\Omega(M,\qq)\rangle$ to $|\Omega(m,\pp)\rangle$. However, given our analysis in the previous section, we know that we must replace eq.~\reef{march} with $Y(m,\pp,\bs)=2\vt$ where
\begin{align}\label{pony6}
2\vt=\frac{1}{2\sqrt{2}}\sqrt{\mathrm{Tr}\left[(\ii\log{\Delta})^2\right]}\quad\text{with}\quad \Delta^a{}_b=\Omega^{ac}(m,\pp)\,\omega_{cb}(M,\qq)\,,
\end{align}
using eq.~\reef{pony5}. The normalization factor is different here because in this construction $\Delta$ has eight (rather than four) eigenvalues $\pm 2\ii \vt$. That is, just as in eq.~\reef{march}, the complexity per mode $Y(m,\pp,\bs)$ is independent of the spin $\bs$, and the trace in eq.~\reef{pony6} effectively sums over the spins as well.

\subsubsection{Rotational invariant reference state} \label{rotor}

We begin here with the simple choice $\qq=0$, which produces a rotationally invariant reference state.
In this particular case, the mass scale $M$ of the reference state does not enter in any way, \ie the only nonvanishing entries in $\Omega^{ab}(\qq=0, M)$ reduce to $\pm\tilde E_{\qq=0}/M=\pm 1$.\footnote{Therefore, the covariance matrix in the rest frame (for massive fermions) is always the same independent of the mass.} This means, the $Y(m,\pp,\bs)$ can only depend on the mass and momentum of the mode that we are considering. Let us construct $\Delta$:
\begin{align}\label{penny12}
\Delta^a{}_b=\Omega^{ac}(m,\pp)\,\omega_{cb}(M,0)\equiv\left(
\begin{array}{cccccccc}
\frac{m}{E_\pp} & 0 & -\frac{p_z}{E_\pp} & -\frac{p_x}{E_\pp} & 0 & 0 & 0 & -\frac{p_y}{E_\pp} \\
0 & \frac{m}{E_\pp} & -\frac{p_x}{E_\pp} & \frac{p_z}{E_\pp} & 0 & 0 & \frac{p_y}{E_\pp} & 0 \\
\frac{p_z}{E_\pp} & \frac{p_x}{E_\pp} & \frac{m}{E_\pp} & 0 & 0 & \frac{p_y}{E_\pp} & 0 & 0 \\
\frac{p_x}{E_\pp} & -\frac{p_z}{E_\pp} & 0 & \frac{m}{E_\pp} & -\frac{p_y}{E_\pp} & 0 & 0 & 0 \\
0 & 0 & 0 & \frac{p_y}{E_\pp} & \frac{m}{E_\pp} & 0 & -\frac{p_z}{E_\pp} & -\frac{p_x}{E_\pp} \\
0 & 0 & -\frac{p_y}{E_\pp} & 0 & 0 & \frac{m}{E_\pp} & -\frac{p_x}{E_\pp} & \frac{p_z}{E_\pp} \\
0 & -\frac{p_y}{E_\pp} & 0 & 0 & \frac{p_z}{E_\pp} & \frac{p_x}{E_\pp} & \frac{m}{E_\pp} & 0 \\
\frac{p_y}{E_\pp} & 0 & 0 & 0 & \frac{p_x}{E_\pp} & -\frac{p_z}{E_\pp} & 0 & \frac{m}{E_\pp} \\
\end{array}
\right)
\end{align}
The corresponding eigenvalues appear with a multiplicity of four and are explicitly given by
\begin{align}\label{wong1}
\mathrm{spec}(\Delta)=\frac{m\pm\ii |\pp|}{E_\pp}=\ex^{\pm 2 \ii \vt}\,.
\end{align}
Note that this corresponds to two quadruples of $(\ex^{2\ii \vt},\ex^{2\ii \vt},\ex^{-2\ii \vt},\ex^{-2\ii \vt})$ associated to the two spin degrees of freedom $s=1,2$. We can recall from eq.~\reef{complex-contribution} that the contribution to the complexity of each spin is given by
\begin{align}\label{harvey}
Y(m,\pp,\bs)=2\vt=\tan^{-1}\!\left(\frac{|\pp|}{m}\right)\,,
\end{align}
which completely agrees with the result found in eq.~\reef{march}. Of course, this is not surprising, because $\qq=0$ corresponds to choosing the same reference state as the one we considered before, \ie eq.~\reef{Dirac2a} completely agrees with eq.~\reef{Dirac2} since the mass does not play a role at $\qq=0$ and the basis spinors reduce to those given in eq.~\reef{rest}.

\begin{figure}[t]
	\begin{center}
	\includegraphics{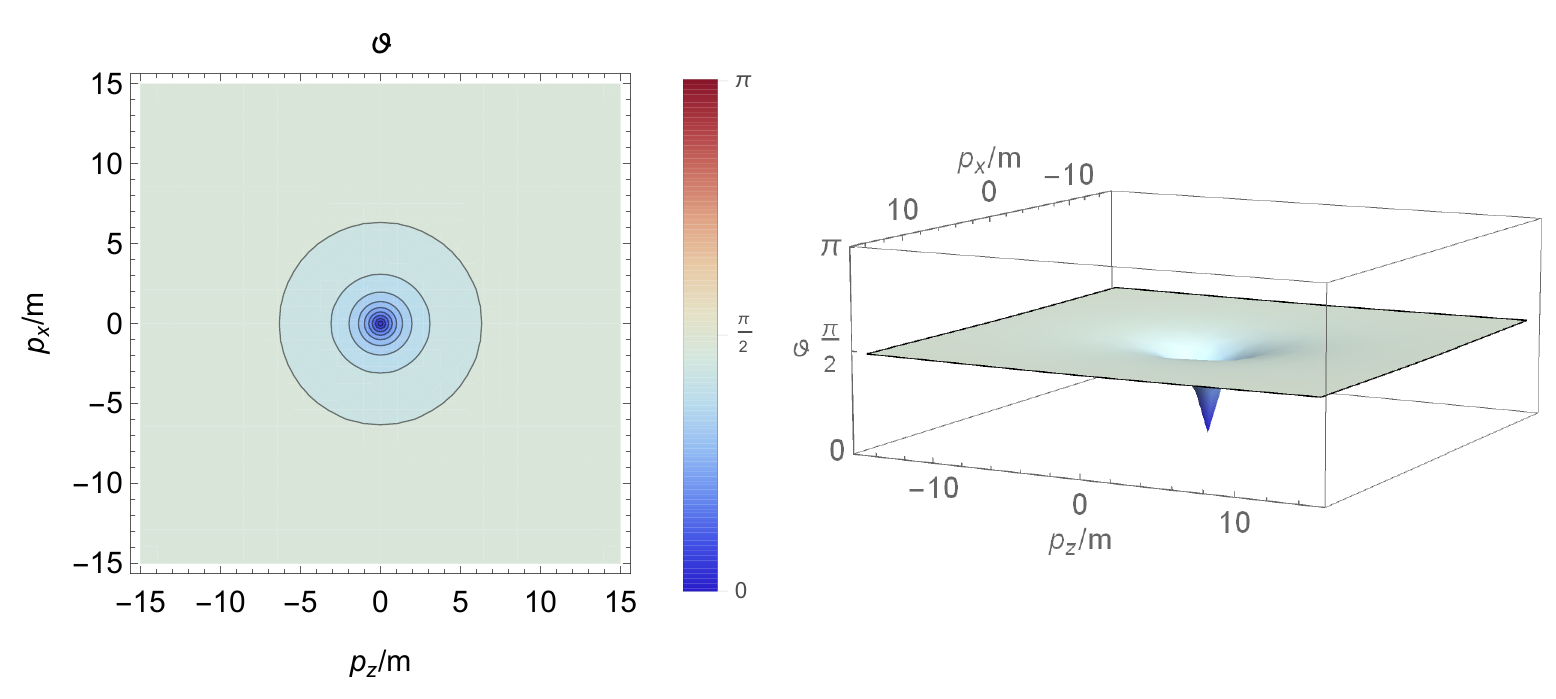}
	\end{center}
	\caption{This plot shows the function $Y(m,\pp,\bs)$ for $\pp=(p_x,0,p_z)$ and a rotationally reference state $|M,\qq=0\rangle$.}
	\label{fig3}
\end{figure}

Figure \ref{fig3} shows a three-dimensional plot of the function $Y$ given by eq.~\reef{harvey} --- see also figure \ref{fig1}. The complexity for the $\kappa=2$ and $\kappa=1$ measures are given in eqs.~\reef{eq:comp-exact} and \reef{eq:comp-kappa=1-exact}, respectively. Recall as shown in eq.~\reef{gong9}, that $Y(m,\pp,\bs)\to\frac\pi2$ in the limit
of large momentum and hence both of these complexities are UV divergent, as shown in eqs.~\reef{eq:comp-asymp} and \reef{eq:comp-asymp2}. More specifically, the leading divergences are 
\beq\label{coil2}
\mathcal{C}_{\kappa=2} \simeq \frac{V\Lambda^3}{12}\left(1-\frac{6m}{\pi\Lambda}+\frac{12m^2}{\pi^2\Lambda^2}\right)\qquad{\rm and}\qquad
\mathcal{C}_{\kappa=1} \simeq \frac{V\Lambda^3}{6\pi}\left(1-\frac{3m}{\pi\Lambda}\right)\,,
\eeq
as given in eqs.~\reef{eq:comp-asymp} and \reef{eq:comp-kappa=1-exact}. In both cases, the next correction is a divergence of order $\log\left(\Lambda/m\right)$.

\subsubsection{Massless reference state}
\begin{figure}[t]
	\begin{center}
		\includegraphics{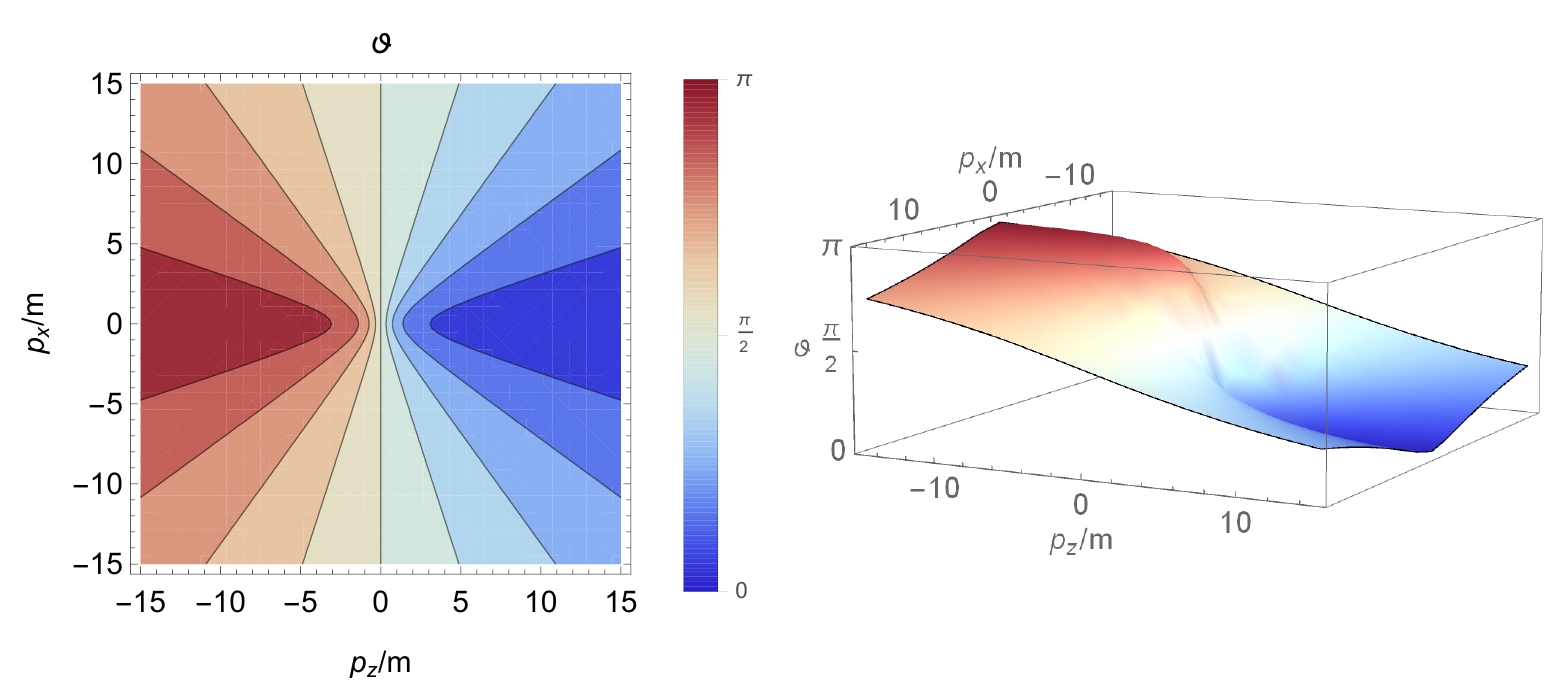}
	\end{center}
	\caption{This plot shows the function $Y(m,\pp,s)$ for $\pp=(p_x,0,p_z)$ and a massless reference state $|M=0,\qq=(0,0,q)\rangle$.}
	\label{fig4}
\end{figure}

We can also choose a reference state that corresponds to spinors associated to a massless state with momentum $\qq$ in a given direction. Without loss of generality, we choose the positive $z$-direction, namely $\qq=(0,0,q)$. In this case, the complexity of the momentum mode $(p_x,p_y,p_z)$ should also depend on the angle that $\pp$ has with the $z$-axis. This reference state is therefore not rotationally invariant, but only invariant under the little group of a massless particle. The explicit form of $\Delta$ is given by
\begin{align}
\Delta^a{}_b=\Omega^{ac}(m,\pp)\,\omega_{cb}(0,\qq)\equiv\left(
\begin{array}{cccccccc}
\frac{p_z}{E_\pp} & -\frac{p_x}{E_\pp} & \frac{m}{E_\pp} & 0 & 0 & -\frac{p_y}{E_\pp} & 0 & 0 \\
\frac{p_x}{E_\pp} & \frac{p_z}{E_\pp} & 0 & -\frac{m}{E_\pp} & -\frac{p_y}{E_\pp} & 0 & 0 & 0 \\
-\frac{m}{E_\pp} & 0 & \frac{p_z}{E_\pp} & -\frac{p_x}{E_\pp} & 0 & 0 & 0 & -\frac{p_y}{E_\pp} \\
0 & \frac{m}{E_\pp} & \frac{p_x}{E_\pp} & \frac{p_z}{E_\pp} & 0 & 0 & -\frac{p_y}{E_\pp} & 0 \\
0 & \frac{p_y}{E_\pp} & 0 & 0 & \frac{p_z}{E_\pp} & -\frac{p_x}{E_\pp} & \frac{m}{E_\pp} & 0 \\
\frac{p_y}{E_\pp} & 0 & 0 & 0 & \frac{p_x}{E_\pp} & \frac{p_z}{E_\pp} & 0 & -\frac{m}{E_\pp} \\
0 & 0 & 0 & \frac{p_y}{E_\pp} & -\frac{m}{E_\pp} & 0 & \frac{p_z}{E_\pp} & -\frac{p_x}{E_\pp} \\
0 & 0 & \frac{p_y}{E_\pp} & 0 & 0 & \frac{m}{E_\pp} & \frac{p_x}{E_\pp} & \frac{p_z}{E_\pp} \\
\end{array}
\right)\,.
\end{align} 
Again, we find two quadruples of equal eigenvalues, which are given by
\begin{align}
\mathrm{spec}(\Delta)=\frac{p_z\pm \ii\sqrt{m^2+p_x^2+p_y^2}}{E_\pp}=\ex^{\pm2\ii \vt}\,.
\end{align}
Hence using eq.~\reef{complex-contribution}, we can extract the contribution to the complexity to be
\begin{align}\label{coil5}
Y(m,\pp,\bs)=2\vt=\frac{\pi}{2}-\tan^{-1}\!\left(\frac{p_z}{\sqrt{m^2+p_x^2+p_y^2}}\right)\,.
\end{align}
We give a three-dimensional plot of this expression in figure \ref{fig4}.

We expect that the complexity will again be UV divergent. In order to compute the leading order contribution, we need to take the limit $|\pp|\to\infty$. This time, there is no universal limit given by a single constant, but rather the limit will depend on the angle $\theta$ between $\pp$ and $\qq$:\footnote{That is, 
$\cos\theta=p_z/|\pp|$ in the present case where $\qq$ is aligned with the (positive) $z$-axis.}
\begin{align}
Y(m,\pp,s)&=\frac{\pi}{2}-\tan^{-1}\!\left(\frac{\bp\,\cos{\theta}}{\sqrt{m^2+\bp^2\sin^2{\theta}}}\right)
\nonumber\\
&\simeq \theta +\frac{m^2 \cot (\theta )}{2\, \bp^2}+\mathcal{O}\left(\frac{m^4}{\bp^4}\right)\,.
\end{align}
We can compare this expression with eq.~\reef{gong9} for the restframe reference state.
The leading order contribution to the complexity can be computed by simply substituting this limit for $Y$ in the desired integral. For example, using eq.~\reef{tot2} for the $\kappa=2$ measure, we find 
\begin{align}
\mathcal{C}_{\kappa=2}\left(|M=0,\qq\rangle\to|0\rangle\right)
&\simeq \frac{2\,V}{(2\pi)^3}\int_0^\Lambda\!\! d\bp \int_0^{2\pi}\!\!d\phi\int_0^\pi\!\! d\theta\ \bp^2\sin\theta\left(\theta^2 +\frac{m^2\,\theta\cot(\theta)}{\bp^2} \right)
\nonumber\\
&=\frac{V\Lambda^3}{6\pi^2}\left[(\pi^2-4)-6\,\frac{m^2}{\Lambda^2} \right] \,.
\label{grey2}
\end{align}
Similarly, using eq.~\reef{tot3} for the $\kappa=1$ measure produces
\begin{align}\label{coil33}
\mathcal{C}_{\kappa=1}\left(|M=0,\qq\rangle\to|0\rangle\right)\simeq \frac{V\Lambda^3}{6\pi}+{\mathcal O}(Vm^4/\Lambda)\,.
\end{align}
Note that the leading divergence above is identical to that found for the reference state with $\qq=0$ while the leading divergence in $\mathcal{C}_{\kappa=2}$ is about 20\% larger,  \eg compare with eq.~\reef{coil2}. The latter shows that for the restframe reference state, the leading corrections were ${\mathcal O}(Vm\Lambda^2)$ but above, we see that the corrections vanish at this order for the massless reference state. This comparison can be understood from the limiting function for a general reference state $|M,\qq\rangle$, which we will discuss next.

\subsubsection{Massive reference state}
We now consider our most general reference state \reef{coil3}, corresponding to a Gaussian state given by a tensor product over identical states $|\Omega(M,\qq)\rangle$, with a fixed reference momentum $\qq$ and mass $M$.\footnote{However, let us note that we will find below that the overall scale is not important. That is, the complexity will be determined by the normalized vector $\hat\qq=\qq/M$.} The calculation of complexity can be simplified by choosing again the momentum $\qq$ to be along the $z$-direction, such that we have $\qq=(0,0,q)$. The explicit form of the relative covariance matrix $\Delta$ is then given by
\begin{align}
\begin{split}\label{coil4}
\Delta^a{}_b&=\Omega^{ac}(m,\pp)\,\omega_{cb}(M,\qq)\\
&\equiv\left(
\begin{array}{cccccccc}
\frac{m M+p_z q}{E_\pp \tilde{E}_\qq} & -\frac{p_x q}{E_\pp \tilde{E}_\qq} & \frac{m q-M p_z}{E_\pp \tilde{E}_\qq} & -\frac{M p_x}{E_\pp \tilde{E}_\qq} & 0 & -\frac{p_y q}{E_\pp \tilde{E}_\qq} & 0 & -\frac{M p_y}{E_\pp \tilde{E}_\qq} \\[6pt]
\frac{p_x q}{E_\pp \tilde{E}_\qq} & \frac{m M+p_z q}{E_\pp \tilde{E}_\qq} & -\frac{M p_x}{E_\pp \tilde{E}_\qq} & \frac{M p_z-m q}{E_\pp \tilde{E}_\qq} & -\frac{p_y q}{E_\pp \tilde{E}_\qq} & 0 & \frac{M p_y}{E_\pp \tilde{E}_\qq} & 0 \\[6pt]
\frac{M p_z-m q}{E_\pp \tilde{E}_\qq} & \frac{M p_x}{E_\pp \tilde{E}_\qq} & \frac{m M+p_z q}{E_\pp \tilde{E}_\qq} & -\frac{p_x q}{E_\pp \tilde{E}_\qq} & 0 & \frac{M p_y}{E_\pp \tilde{E}_\qq} & 0 & -\frac{p_y q}{E_\pp \tilde{E}_\qq} \\[6pt]
\frac{M p_x}{E_\pp \tilde{E}_\qq} & \frac{m q-M p_z}{E_\pp \tilde{E}_\qq} & \frac{p_x q}{E_\pp \tilde{E}_\qq} & \frac{m M+p_z q}{E_\pp \tilde{E}_\qq} & -\frac{M p_y}{E_\pp \tilde{E}_\qq} & 0 & -\frac{p_y q}{E_\pp \tilde{E}_\qq} & 0 \\[6pt]
0 & \frac{p_y q}{E_\pp \tilde{E}_\qq} & 0 & \frac{M p_y}{E_\pp \tilde{E}_\qq} & \frac{m M+p_z q}{E_\pp \tilde{E}_\qq} & -\frac{p_x q}{E_\pp \tilde{E}_\qq} & \frac{m q-M p_z}{E_\pp \tilde{E}_\qq} & -\frac{M p_x}{E_\pp \tilde{E}_\qq} \\[6pt]
\frac{p_y q}{E_\pp \tilde{E}_\qq} & 0 & -\frac{M p_y}{E_\pp \tilde{E}_\qq} & 0 & \frac{p_x q}{E_\pp \tilde{E}_\qq} & \frac{m M+p_z q}{E_\pp \tilde{E}_\qq} & -\frac{M p_x}{E_\pp \tilde{E}_\qq} & \frac{M p_z-m q}{E_\pp \tilde{E}_\qq} \\[6pt]
0 & -\frac{M p_y}{E_\pp \tilde{E}_\qq} & 0 & \frac{p_y q}{E_\pp \tilde{E}_\qq} & \frac{M p_z-m q}{E_\pp \tilde{E}_\qq} & \frac{M p_x}{E_\pp \tilde{E}_\qq} & \frac{m M+p_z q}{E_\pp \tilde{E}_\qq} & -\frac{p_x q}{E_\pp \tilde{E}_\qq} \\[6pt]
\frac{M p_y}{E_\pp \tilde{E}_\qq} & 0 & \frac{p_y q}{E_\pp \tilde{E}_\qq} & 0 & \frac{M p_x}{E_\pp \tilde{E}_\qq} & \frac{m q-M p_z}{E_\pp \tilde{E}_\qq} & \frac{p_x q}{E_\pp \tilde{E}_\qq} & \frac{m M+p_z q}{E_\pp \tilde{E}_\qq}
\end{array}
\right)
\end{split}
\end{align}
After some extended calculations, we find the two quadruples of identical eigenvalues corresponding to
\begin{align}
\mathrm{spec}(\Delta)=\frac{(m M+p_z q)\pm \ii\sqrt{(p_x^2+p_y^2)(M^2+q^2)+(Mp_z-mq)^2}}{E_\pp \tilde{E}_\qq}\,.
\end{align}
Let us note that as expected from the covariance matrix \reef{coil4}, these eigenvalues are only functions of the dimensionless ratio $q/M$ (rather than of $q$ and $M$ independently). Hence to simplify the following expressions, we introduce $\hat q=q/M$.
From here, we can use the same steps as above to find the general function of the complexity, namely
\begin{align}\label{coil6}
Y(m,\pp,\bs)&=\frac{\pi}{2}-\tan^{-1}\!\left(\frac{m+p_z\,\hat{q}}{\sqrt{(\hat{q}^2+1)(p_x^2+p_y^2)+(p_z-m\,\hat{q})^2}}\right)\,.
\end{align}
We can verify that in the limit $q/M\to 0$ (restframe) or $q/M\to \infty$ (massless reference state), we find the expected results in eqs.~\reef{harvey} and \reef{coil5}, respectively.

\begin{figure}[t]
\begin{center}
	\includegraphics{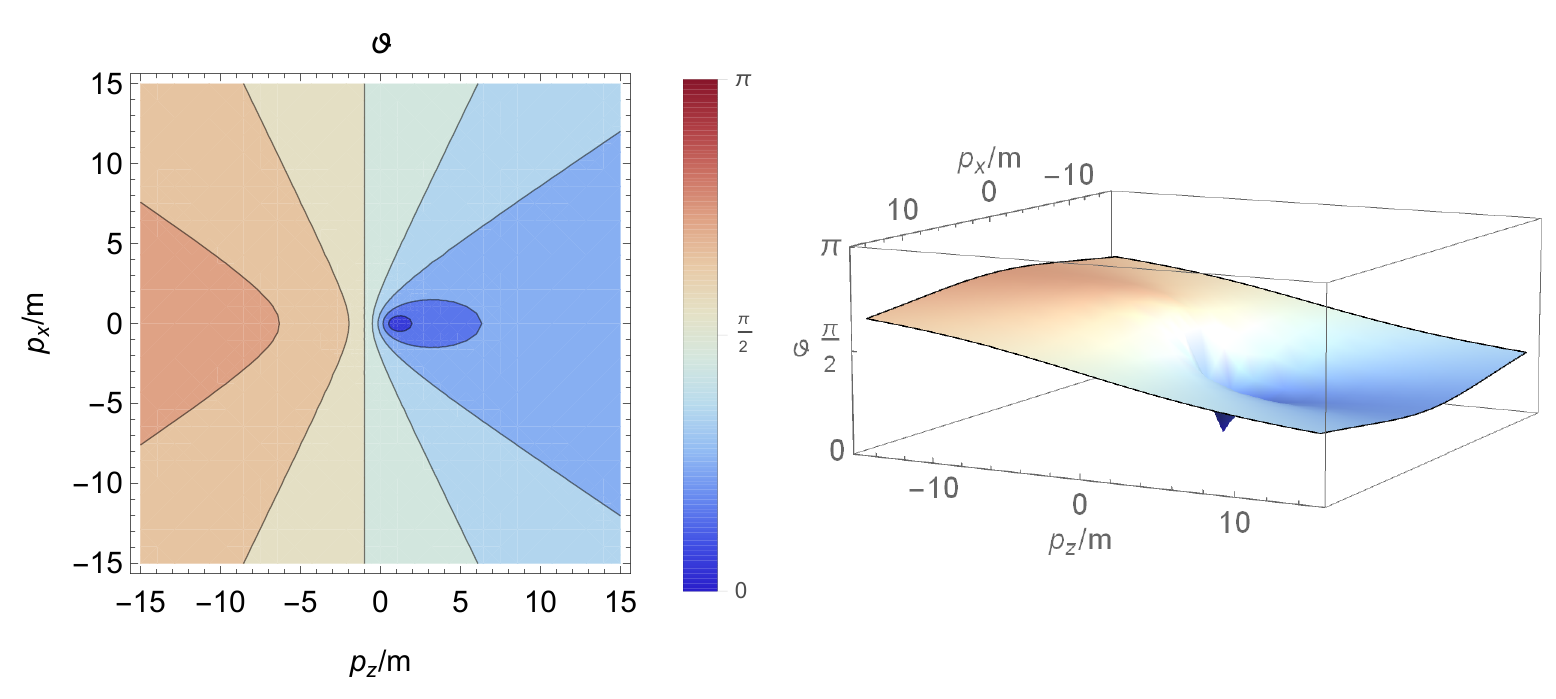}
\end{center}
	\caption{This plot shows the function $Y(m,\pp,s)$ for $\pp=(p_x,0,p_z)$ for a massive reference state $|M,\qq=(0,0,M)\rangle$.}
	\label{fig5}
\end{figure}
We illustrate $Y(m,\pp,\bs)$ with a three-dimensional plot in figure \ref{fig5}, for $\hat{q}=q/M=1$. A convenient choice is to study the complexity as function of the dimensionless vectors $\hat{\pp}=\pp/m$ and $\hat{\qq}=\qq/M$. In particular, given eq.~\reef{coil6}, we can write the complexity per mode as $Y(\hat\pp,\hat\qq)$ --- as in the two previous examples, $Y$ is independent of the spin $s$. Note that we allow these vectors to be infinitely large for $m\to 0$ or $M\to 0$ corresponding to point on the two-sphere at infinity. Clearly, $Y$ takes its minimum value (equal to zero) at the point where $\hat{\pp}=\hat{\qq}$. We can always choose a plane, such that both points $\hat{\qq}$ and $\hat{\pp}$ lie on it. If we now use the (inverse) stereographic projection to map the plane onto a half-sphere (of unit radius) touching the origin, then $Y$ becomes just the geodesic distance between the projected points on the sphere. We illustrate this geometry in figure \ref{fig6}.
\begin{figure}[t]
	\begin{center}
		\includegraphics{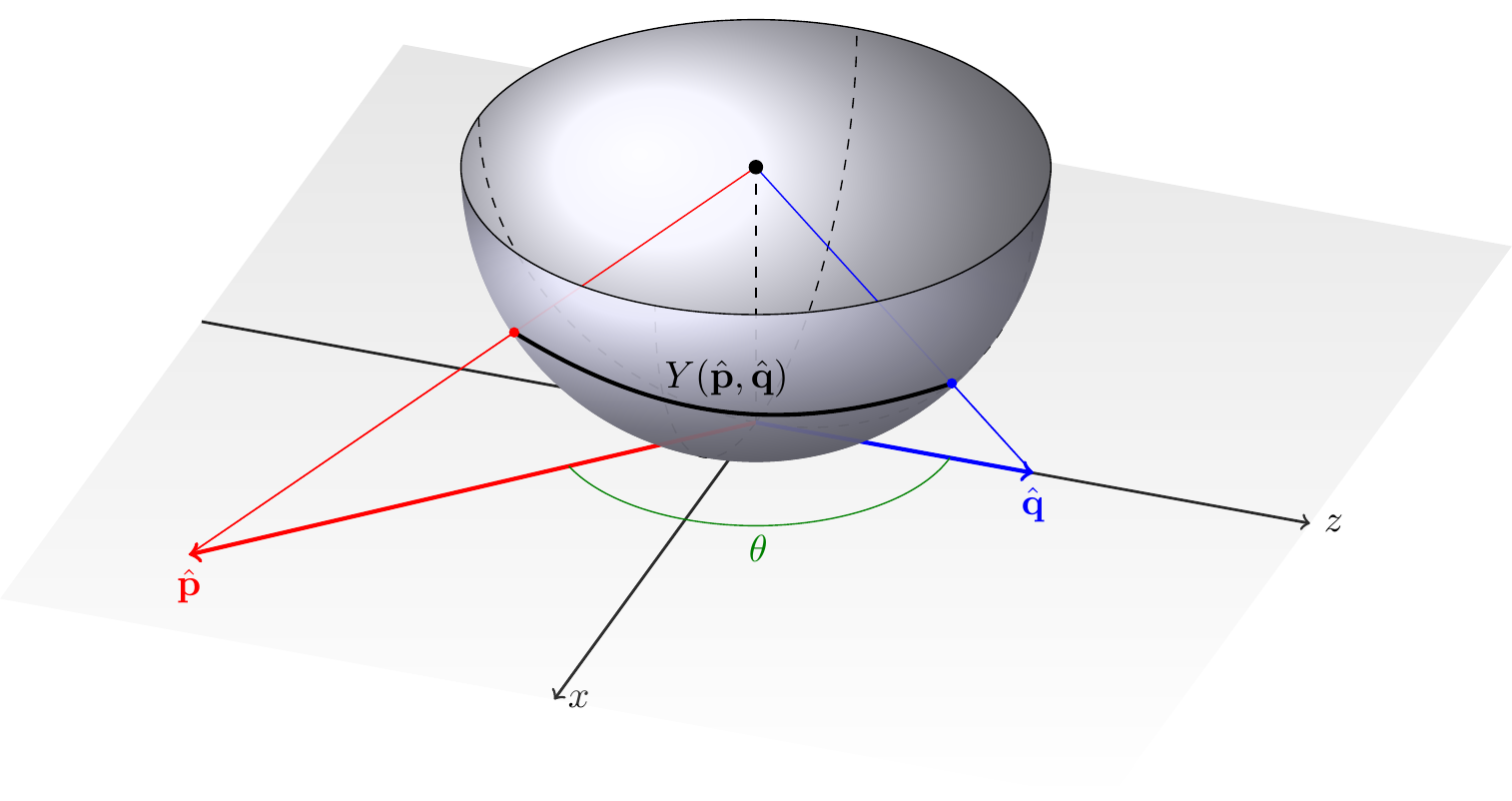}
	\end{center}
	\caption{This figure illustrates the geometry of the  complexity $Y(\hat \pp,\hat \qq)$ between a reference state $|\Omega(M,\qq)\rangle$ and the target state $|\Omega(m,\pp)\rangle$. By using the inverse stereographical projection of the plane onto a half-sphere of unit radius, the complexity can be identified with the geodesic distance on the sphere between the two projected points. We also indicate the angle $\theta$ between $\hat{q}$ and $\hat{p}$.}
	\label{fig6}
\end{figure}

In order to compute the leading order UV contribution to the complexity, we need to take the limit $|\pp|\to\infty$. Again, this limit will depend on the angle $\theta$ between $\pp$ and $\qq$, \ie 
$\cos\theta=p_z/|\pp|$ in the case where $\qq$ is aligned with the (positive) $z$-axis. Again, the other relevant quantity will be $\hat{q}=q/M$. The asymptotics for $|\pp|\to\infty$ are given by
\begin{align}
\begin{split}
Y(m,\pp,\bs)&\simeq \frac{\pi }{2}-\tan ^{-1}\left(\frac{\hat{q} \cos (\theta )}{\sqrt{\left(\hat{q}^2+1\right) \sin ^2(\theta )+\cos ^2(\theta )}}\right)\\
&\qquad\,-\frac{1}{\sqrt{\left(\hat{q}^2+1\right) \sin ^2(\theta )+\cos ^2(\theta )}}\,\frac{m}{\bp}\\
&\qquad\,+\frac{\hat{q}^3\, \sin^2(\theta)\cos(\theta)}{2 \left[\left(\hat{q}^2+1\right) \sin ^2(\theta )+\cos ^2(\theta )\right]^{3/2}}\,\frac{m^2}{\bp^2}+\mathcal{O}\left(\frac{m^3}{\bp^3}\right)\,,
\end{split}
\end{align}
which can be used to identify the leading UV divergences in the complexity. Considering the the $\kappa=2$ measure, we substitute the above expression into  eq.~\reef{tot2} and find 
\begin{align}
\mathcal{C}_{\kappa=2}\left(|M,\qq\rangle\to|0\rangle\right)
&= \frac{2\,V}{(2\pi)^3}\int_0^\Lambda\!\! d\bp\int_0^{2\pi}\!\!d\phi\int_0^\pi\!\! d\theta\ \bp^2\sin\theta\ 
\ Y(m,\pp,\bs)^2
\nonumber\\
\begin{split}\label{frank1}
&\simeq\frac{V\Lambda ^3}{12\pi^2} \bigg[\pi^2-8+\frac{8}{\hat{q}}\,\tan ^{-1}\!\hat{q}+4 \left( \tan ^{-1}\!\hat{q}\right)^2\\
&\qquad\quad -\frac{6\pi\,m}{\Lambda\,\hat{q}}\,\tan ^{-1}\!\hat{q}+\frac{12m^2}{\Lambda^2}\,\left(\frac{2}{\hat{q}}\,\tan ^{-1}\!\hat{q}-1\right)\\
&\qquad\quad 
+\mathcal{O}\left(m^3/\Lambda^3\right)\Big]\,.
\end{split}
\end{align}
This function neatly interpolates between the previous results two results, namely between eq.~\reef{coil2} for the restframe reference state with $\hat{q}=q/M\to 0$, and eq.~\reef{grey2} for the massless reference state with $q/M\to\infty$. Alternatively, using eq.~\reef{tot3} for the $\kappa=1$ measure, we find
\begin{align}\label{frank2}
\mathcal{C}_{\kappa=1}\left(|M,\qq\rangle\to|0\rangle\right)\simeq \frac{V\Lambda^3}{6\pi}\left[1-\frac{3m}{\pi\,\Lambda \hat{q}}\,\tan ^{-1}\!\hat{q}+\mathcal{O}(m^2/\Lambda^2)\right]\,.
\end{align}
It is straightforward to show that in the limit $q/M\to0$, the above reduces to the corresponding expression in eq.~\reef{coil2}. Similarly in the limit $q/M\to\infty$, one finds that the subleading correction vanishes above, which is in agreement with the result in eq.~\reef{coil33}.

\subsection{More excited states} \label{geee}

The formalism developed in section \ref{sec:general} allows us to compute the complexity between any two fermionic Gaussian states that belong to the same connected component of the state space, \ie they must both have even or odd fermion number (see footnote \ref{footyABC}). A key feature which distinguishes fermions from bosons is that states with individual particle excitations, \eg  $\prod_i a^{\br_i\dagger}_{\pp_i} |0\rangle$, are still Gaussian states. That is, as a result of the fermion anticommutation relations, these states are annihilated by the operators $a^{\br_i\dagger}_{\pp_i}$. Hence we can apply our techniques to compute the complexity of fermionic Gaussian states to such excited states (provided that the number of excitations is even so that they are on the same connected component as the reference state). We already considered several simple examples of such excited states in section~\ref{excited1}, but with the new methods at our disposal, we can approach this question systematically here. Thoughout the following analysis, we will use $|\bar{0}\rangle$ as the reference state, with $\Omega_0 =\Omega(M,\qq=0)$.

\subsubsection{Excitations with a single momentum}
We will begin by analyzing excitations in a single mode $\pp$, for which we will consider two and four excitations. In section \ref{excited1}, we have already considered the complexity of simple examples of these states, as given in eqs.~\reef{excite8}, \reef{excite85} and \reef{exciteD}. The elaboration here will have to do with the spin. So far, we only considered spins of the individual (anti)particles are aligned along the axis defined by the corresponding momentum --- see  footnote \ref{footyA1}. Previously, we denoted spins with this orientation with the spin labels $\br,\bs$. However, we now introduce the labels $r,s\in\{1,2\}$ to denote spins aligned along an arbitrary axis that is not related to the momentum axis, but rather is fixed in the rest frame of the fermions. Without loss of generality, we will choose this axis to be the $z$-axis in the following.\footnote{Our conventions in the following will be that both $a^{r\dagger}_\pp$ and $b^{r\dagger}_\pp$ create excitations whose spins are oriented along the positive (negative) $z$-axis with $r=1$ (with $r=2$).}  

As in section~\ref{sec:alternative}, the strategy for evaluating the complexity is: Given our excited state $|\tilde{\psi}\rangle$, we  first compute the corresponding covariance matrix. Since the excitations involve a single momentum $\qq$, we need only focus on that sector and the corresponding $\tilde\Omega(m,\qq)$ is computed with respect to the basis $\xi^a(\qq)$ introduced in eq.~\reef{eq:xi-momentum-basis}. We then evaluate the relative covariance matrix $\Delta(\qq)=\tilde\Omega(m,\qq)\,\omega(M,0)$, where $\omega(M,0)$ is the inverse of $\Omega_0=\Omega(M,0)$ which means we continue to use the rotational invariant reference state as in section~\ref{sec:Dirac}. We then evaluate the spectrum of $\Delta(\qq)$, which reveals the change in the complexity of the excited state. The tedious part of this computation lies in evaluating $\tilde\Omega(m,\qq)$ which is best accomplished by expanding $\xi^a(\qq)$ in terms of creation and annihilation operators and keeping in mind that for the excited state, some of the original creation operators now annihilate $|\tilde{\psi}\rangle$.

\begin{figure}[t]
\begin{center}
	\includegraphics{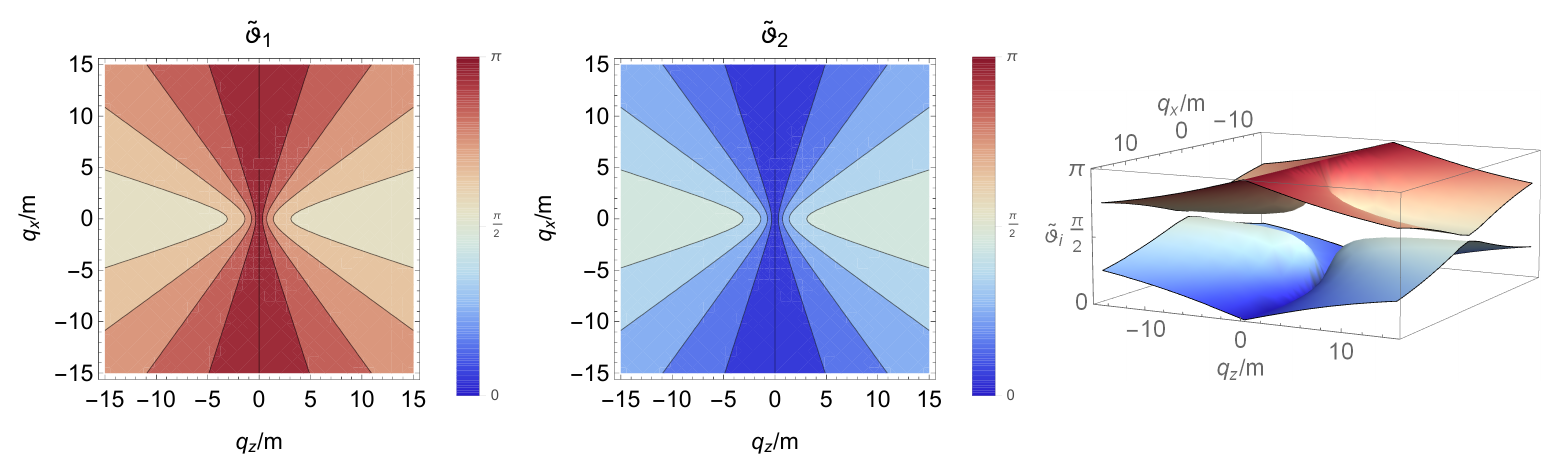}
\end{center}
	\caption{This plot shows the angles $2\tilde{\vartheta}_i$ of the two eigenvalue quadruples $(\ex^{2\ii\tilde{\vartheta}_i},\ex^{2\ii\tilde{\vartheta}_i},\ex^{-2\ii\tilde{\vartheta}_i},\ex^{-2\ii\tilde{\vartheta}_i})$ of $\Delta(\qq)$ in the state $a_\qq^{r\dagger}b_{-\qq}^{r\dagger}|0\rangle$ for $\qq=(q_x,0,q_z)$.}
	\label{fig6}
\end{figure}

The first class of excited states which we consider here take the form
\beq
\textbf{(E)}\qquad|\tilde\psi\rangle=a_{\qq}^{r\dagger}\, b_{-\qq}^{r\dagger}\,|0\rangle\,,
\label{excite8z}
\eeq
with arbitrary $\qq$. These states are similar to those from eqs.~\reef{excite8} and \reef{exciteD}, but as described above 
in the present state, the spins are aligned with the $z$-axis in the rest frame. In fact, it is straightforward to see that the (A) and (E) families coincide for the special case where the momentum $\qq$ is oriented along the $z$-axis. The relative covariance matrix is given by:
\begin{align}\label{penny11}
\begin{split}
\Delta(\qq)=\tilde\Omega(m,\qq)\,\omega(M,0)\qquad\qquad\qquad\qquad\qquad\qquad\qquad\qquad\qquad\qquad\qquad\qquad\qquad\qquad\qquad\qquad\\
\tiny
\equiv\left(
\begin{array}{cccccccc}
\frac{q_z^2 }{(E_\qq+m) E_\qq}-1 & \frac{q_x q_z }{(E_\qq+m) E_\qq} & \frac{q_z}{E_\qq} & 0 & 0 & \frac{q_y q_z }{(E_\qq+m) E_\qq} & 0 & 0 \\[6pt]
\frac{q_x q_z }{(E_\qq+m) E_\qq} & 1-\frac{ q_z^2}{(E_\qq+m) E_\qq} & 0 & \frac{q_z}{E_\qq} & -\frac{q_y q_z}{(E_\qq+m) E_\qq} & 0 & 0 & 0 \\[6pt]
-\frac{q_z}{E_\qq} & 0 & \frac{q_z^2 }{(E_\qq+m) E_\qq}-1 & \frac{q_x q_z }{(E_\qq+m) E_\qq} & 0 & 0 & 0 & \frac{q_y q_z }{(E_\qq+m) E_\qq} \\[6pt]
0 & -\frac{q_z}{E_\qq} & \frac{q_x q_z }{(E_\qq+m) E_\qq} & 1-\frac{q_z^2}{(E_\qq+m) E_\qq} & 0 & 0 & -\frac{q_y q_z}{(E_\qq+m) E_\qq} & 0 \\[6pt]
0 & -\frac{q_y q_z}{(E_\qq+m) E_\qq} & 0 & 0 & \frac{q_z^2 }{(E_\qq+m) E_\qq}-1 & \frac{q_x q_z }{(E_\qq+m) E_\qq} & \frac{q_z}{E_\qq} & 0 \\[6pt]
\frac{q_y q_z }{(E_\qq+m) E_\qq} & 0 & 0 & 0 & \frac{q_x q_z }{(E_\qq+m) E_\qq} & 1-\frac{q_z^2}{(E_\qq+m) E_\qq} & 0 & \frac{q_z}{E_\qq} \\[6pt]
0 & 0 & 0 & -\frac{q_y q_z}{(E_\qq+m) E_\qq} & -\frac{q_z}{E_\qq} & 0 & \frac{q_z^2 }{(E_\qq+m) E_\qq}-1 & \frac{q_x q_z }{(E_\qq+m) E_\qq} \\[6pt]
0 & 0 & \frac{q_y q_z }{(E_\qq+m) E_\qq} & 0 & 0 & -\frac{q_z}{E_\qq} & \frac{q_x q_z }{(E_\qq+m) E_\qq} & 1-\frac{q_z^2}{(E_\qq+m) E_\qq} \\[6pt]
\end{array}
\right)
\end{split}
\end{align}
The eigenvalues of $\Delta(\qq)$ appear in two quadruples $(\ex^{2\ii\tilde{\vartheta}_i},\ex^{2\ii\tilde{\vartheta}_i},\ex^{-2\ii\tilde{\vartheta}_i},\ex^{-2\ii\tilde{\vartheta}_i})$ given by
\begin{align}\label{donkong}
\ex^{\pm 2\ii \tilde{\vartheta}_1}&=-\frac{\sqrt{m^2+q_x^2+q_y^2}\pm \ii \,q_z}{E_\qq}\,,\qquad
\ex^{\pm 2\ii \tilde{\vartheta}_2}=\frac{\sqrt{m^2+q_x^2+q_y^2}\pm \ii \,q_z}{E_\qq}\,.
\end{align}
Hence the angles $\tilde{\vartheta}_i$ are given by\footnote{The angles are chosen consistently with eq.~\reef{donkong} such that $0\le2\tilde{\vt}_i\le\pi$.}
\begin{align}
2\tilde{\vartheta}_1&=\pi-\sin^{-1}\!\left(\frac{|q_z|}{E_\qq}\right)\,,\qquad\quad
2\tilde{\vartheta}_2=\sin^{-1}\!\left(\frac{|q_z|}{E_\qq}\right)\,.
\end{align}
and we plot the values $2\tilde{\vartheta}_i$ in figure~\ref{fig6}. In particular, we notice that with $q_x=q_y=0$, the spin axis and the momentum axis are aligned and our results agree with eq.~\reef{march2} for excited states in the (A) class, \ie in eq.~\reef{excite8}. Following the analysis in section \ref{excited1}, we then evaluate the difference in the complexity of the excited state and that of the vacuum. With the $\kappa=1,2$ measures, we find
\begin{align}\label{bingo}
\Delta\mathcal{C}_{\kappa=2}(|\bar{0}\rangle\to |\psi\rangle)&=(2\tilde{\vt}_1)^2+(2\tilde{\vt}_2)^2-2\,Y(m,\qq,\bs)^2\\
\Delta\mathcal{C}_{\kappa=1}(|\bar{0}\rangle\to |\psi\rangle)&=\pi-2\,|Y(m,\qq,\bs)|\,,
\nonumber
\end{align}
where $Y(m,\qq,\bs)$ is given in eq.~\reef{march}, and we have used $2\tilde{\vartheta}_1+2\tilde{\vartheta}_2=\pi$ in the second expression. It is interesting to observe that our result here for $\Delta\mathcal{C}_{\kappa=1}$ is precisely the same as that found in eq.~\reef{diff1a} for the excited states in eq.~\reef{excite8}. Of course, we must reiterate that the $\kappa=1$ measure is basis dependent and implicitly, for $\Delta\mathcal{C}_{\kappa=1}$, we are aligning the basis in the $\qq$ sector here with the generators which produce the above transformations.

\begin{figure}[t]
\begin{center}
	\includegraphics{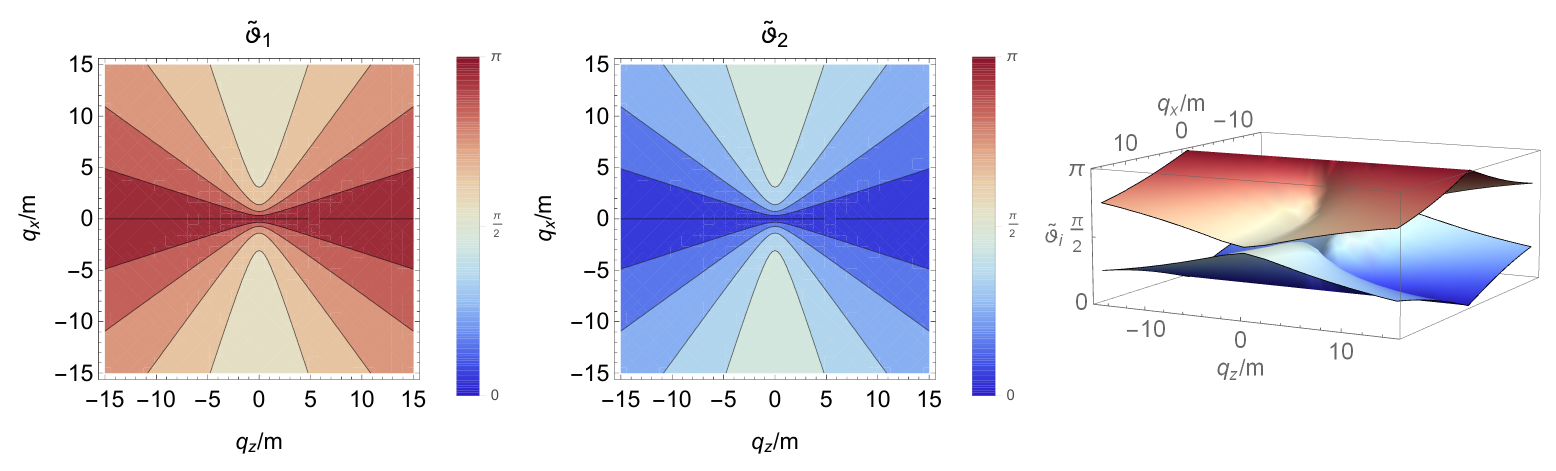}
\end{center}
	\caption{This plot shows the angles $2\tilde{\vartheta}_i$ of the two eigenvalue quadruples $(\ex^{2\ii\tilde{\vartheta}_i},\ex^{2\ii\tilde{\vartheta}_i},\ex^{-2\ii\tilde{\vartheta}_i},\ex^{-2\ii\tilde{\vartheta}_i})$ of $\Delta(\qq)$ in the state $a_\qq^{r\dagger}b_{-\qq}^{r'\dagger}|0\rangle$ for $\qq=(q_x,0,q_z)$. Note that the result is identical with the plots shown in figure~\ref{fig6} where $q_x$ and $q_z$ are swapped.}
	\label{fig8}
\end{figure}

Next as a generalization of the (D) states in eq.~\reef{exciteD}, we consider 
\beq
\textbf{(F)}\qquad|\tilde\psi\rangle=a_{\qq}^{r\dagger}\, b_{-\qq}^{r'\dagger}\,|0\rangle\,,
\label{excite8z}
\eeq
where, as before, we use the convention that $r'$ to refers to the opposite spin label as $r$, \ie $r'\equiv r+1\,(\text{mod}\,2)$. Again, we are considering the spins to be oriented along the $z$-axis, independent of the orientation of the momentum $\qq$. We can assume $q_y=0$ without loss of generality, \ie we orient the spatial axes so that the momentum lies in the $xz$-plane. This choice simplifies the computation of the relative covariance matrix, for which we find
\begin{align}
\begin{split}
\Delta(\qq)=\tilde\Omega(m,\qq)\,\omega(M,0)\qquad\qquad\qquad\qquad\qquad\qquad\qquad\qquad\qquad\qquad\qquad\qquad\qquad\qquad\qquad\qquad\\
\tiny
\equiv\left(
\begin{array}{cccccccc}
0 & -\frac{q_x}{E_\qq} & \frac{q_x^2}{(E_\qq+m) E_\qq}-1 & -\frac{q_x q_z }{(E_\qq+m) E_\qq} & 0 & 0 & 0 & 0 \\
\frac{q_x}{E_\qq} & 0 & -\frac{q_x q_z }{(E_\qq+m) E_\qq} & 1-\frac{q_x^2}{(E_\qq+m) E_\qq} & 0 & 0 & 0 & 0 \\
\frac{q_x^2}{(E_\qq+m)E_\qq}-1 & -\frac{q_x q_z }{(E_\qq+m) E_\qq} & 0 & \frac{q_x}{E_\qq} & 0 & 0 & 0 & 0 \\
-\frac{q_x q_z }{(E_\qq+m) E_\qq} & 1-\frac{q_x^2}{(E_\qq+m) E_\qq} & -\frac{q_x}{E_\qq} & 0 & 0 & 0 & 0 & 0 \\
0 & 0 & 0 & 0 & 0 & -\frac{q_x}{E_\qq} & \frac{q_x^2}{(E_\qq+m) E_\qq}-1 & -\frac{q_x q_z }{(E_\qq+m) E_\qq} \\
0 & 0 & 0 & 0 & \frac{q_x}{E_\qq} & 0 & -\frac{q_x q_z }{(E_\qq+m) E_\qq} & 1-\frac{q_x^2}{(E_\qq+m) E_\qq} \\
0 & 0 & 0 & 0 & \frac{q_x^2}{(E_\qq+m) E_\qq}-1 & -\frac{q_x q_z }{(E_\qq+m) E_\qq} & 0 & \frac{q_x}{E_\qq} \\
0 & 0 & 0 & 0 & -\frac{q_x q_z }{(E_\qq+m) E_\qq} & 1-\frac{q_x^2}{(E_\qq+m) E_\qq} & -\frac{q_x}{E_\qq} & 0 \\
\end{array}
\right)
\end{split}
\end{align}
The eigenvalues of $\Delta(\qq)$ appear in two quadruples $(\ex^{2\ii\tilde{\vartheta}_i},\ex^{2\ii\tilde{\vartheta}_i},\ex^{-2\ii\tilde{\vartheta}_i},\ex^{-2\ii\tilde{\vartheta}_i})$ given by
\begin{align}\label{rush}
\ex^{\pm 2\ii \tilde{\vartheta}_1}&=-\frac{\sqrt{m^2+q_z^2}\pm \ii \,\sqrt{q_x^2+q_y^2}}{E_\qq}\,,\qquad
\ex^{\pm 2\ii \tilde{\vartheta}_2}=\frac{\sqrt{m^2+q_z^2}\pm \ii \,\sqrt{q_x^2+q_y^2}}{E_\qq}\,,
\end{align}
where we reinstated $q_y$ with the substitution $q_x\to\sqrt{q_x^2+q_y^2}$. Thus, the two angles $\tilde{\vartheta}_i$ are given by
\begin{align}\label{bingo2}
2\tilde{\vartheta}_1=\pi-\sin^{-1}\!\left(\frac{\sqrt{q_x^2+q_y^2}}{E_\qq}\right)
\,,\qquad\quad
2\tilde{\vartheta}_2&=\sin^{-1}\!\left(\frac{\sqrt{q_x^2+q_y^2}}{E_\qq}\right)\,.
\end{align}
and we plot $2\tilde{\vartheta}_i$ in figure~\ref{fig8}. Let us observe that with $q_x=0=q_y$, the spin axis is aligned with the momentum axis and the above angles coincide with those in eq.~\reef{penny14} for the excited states in class (D) (since in this case, the two classes coincide). Furthermore, we can compare these results with those for the (E) excited states. For example, the eigenvalues in eq.~\reef{donkong} match those above in eq.~\reef{rush} if we swap $q_z\leftrightarrow \sqrt{q_x^2+q_y^2}$. Again, we can evaluate the difference in the complexities of the excited state and the Dirac vacuum with the $\kappa=1,2$ measures to find the same result as in eq.~\reef{bingo}, where the $\tilde{\vt}_i$ appearing in $\Delta\mathcal{C}_{\kappa=2}$ are given by eq.~\reef{bingo2}. We note that once more that $\Delta\mathcal{C}_{\kappa=1}$ is precisely the same as found in eq.~\reef{diff1a} for the (A) states in eq.~\reef{excite8} --- again, this result relies on the fact that $2\tilde{\vartheta}_1+2\tilde{\vartheta}_2=\pi$ for the new excited states. We reiterate that the $\kappa=1$ measure is basis dependent and implicitly, for $\Delta\mathcal{C}_{\kappa=1}$, we are aligning the basis in the $\qq$ sector here with the generators which produce the above transformations.

Another interesting class of excited states is given by 
\beq
\textbf{(G)}\qquad|\tilde\psi\rangle=a_{\qq}^{r\dagger}\,a_{\qq}^{r'\dagger}\, \,b_{-\qq}^{r\dagger}\,b_{-\qq}^{r'\dagger}\,|0\rangle\,,
\label{excite88a}
\eeq
where we excite every degree of freedom in $\qq$ sector. Physically, this means we have two particles with momentum $\qq$ but opposite spins, and two antiparticles with with momentum $-\qq$, also with both spins. These states are similar to a special case with two pairs excited for the same momentum $\qq$ in eq.~\reef{excite88}, \ie $a_{\qq}^{\br\dagger} a_{\qq}^{\brp\dagger} b_{-\qq}^{\br\dagger} b_{-\qq}^{\brp\dagger}|0\rangle$. Again, as our notation indicates, the difference is in the orientation of the spins, but we return to this point below.
In evaluating the corresponding covariance matrix $\tilde\Omega(m,\qq)$, for this state \reef{excite88a} we are swapping \emph{all} creation operators with annihilation operators (for this momentum mode $\qq$) and this swap will just reverse the overall sign of the covariance matrix from that in eq.~\reef{capital}, \ie $\tilde\Omega(m,\qq)=-\Omega(m,\qq)$. The reason for the sign change can be understood best by recalling that the eigenspaces of the matrix $J=\Omega g$ correspond to the spaces spanned by creation operators (corresponding to eigenvalues $+\ii$) and annihilation operators (eigenvalues $-\ii$). Hence if we swap the role of the operators, we need to go from $J\to\tilde J=-J$, which implies to $\Omega\to\tilde\Omega=-\Omega$. We find that the eigenvalues of $\Delta=\tilde\Omega(m,\qq)\,\omega(M,0)$ have the opposite sign, \ie there are two four-fold degenerate eigenvalues,
\begin{align}
\mathrm{spec}(\Delta)=-\frac{m\pm\ii |\qq|}{E_\qq}=\ex^{\pm 2\ii \tilde{\vartheta}}\,,
\end{align}
which of course, is the same as in eq.~\reef{wong1} up to an overall sign change. Accordingly, we have
\begin{align}\label{goal}
\tilde{Y}(m,\qq,s)=2\tilde{\vartheta}=\pi-\mathrm{tan}^{-1}\left(\frac{|\qq|}{m}\right)\,,
\end{align}
where as usual, we chose the angle to lie in the range $0\le2\tilde{\vt}\le\pi$.

\begin{table}
	\caption{In this table, we summarize the results for the complexity of various excited states in a single mode $\qq$. The first four cases were considered in section \ref{excited1}. The relative covariance matrix $\Delta(\qq)$ will have two eigenvalue quadruples $(\ex^{2\ii \tilde{\vt}_i},\ex^{2\ii \tilde{\vt}_i},\ex^{-2\ii \tilde{\vt}_i},\ex^{-2\ii \tilde{\vt}_i})$ which encode the angles needed to evaluate the complexity or the difference of the complexity from that of the ground state. Recall barred spin labels $\br,\bs$ denote spins aligned along the momentum axis, while unbarred labels $r,s$ denote spins aligned along a fixed axis in the rest frame. Given such a fixed axis, we specify the momentum $\qq$ by the magnitude of the tangential component $q_\parallel$ and of the orthogonal component $q_\bot$. Recall that 
		cases	 (A) and (D) coincide with cases (E) and (F), respectively, when $q_\bot=0$. Further, since the states in (B), (C) and (G) form spin singlets, the spins can actually be oriented along any axis.}
	\begin{center}
		\begin{tabular}{c | c | c | c | c}
			& \textbf{Excited states} & $2\tilde{\vt}_1$ & $2\tilde{\vt}_2$ & \textbf{Section}\\
			\hline
			\hline
			& & & & \\[-12pt]
			\textbf{(A)} & $a_{\qq}^{\br\dagger}b_{-\qq}^{\brp\dagger}|0\rangle$ & $\pi-\tan^{-1}\!\left(\frac{|\qq|}{m}\right)$ & $\tan^{-1}\left(\frac{|\qq|}{m}\right)$ & \ref{excited1}\\[5pt]
			\textbf{(B)} &$a_{\qq}^{1\dagger}a_{\qq}^{2\dagger}|0\rangle$ &  $\pi$  &$0$&  \ref{excited1}\\[5pt]
			\textbf{(C)} & $b_{\qq}^{1\dagger}b_{\qq}^{2\dagger}|0\rangle$ &  $\pi$  &$0$& \ref{excited1}\\[5pt]
			\textbf{(D)} & $a_{\qq}^{\br\dagger}b_{-\qq}^{\br\dagger}|0\rangle$ &  $\pi$  &$0$& \ref{excited1}\\[4pt]
			\hline
			& & & & \\[-12pt]
			\textbf{(E)} &$a_{\qq}^{r\dagger}b_{-\qq}^{r\dagger}|0\rangle$ & $\pi-\sin^{-1}\!\left(\frac{q_\parallel}{E_\qq}\right)$ & $\sin^{-1}\!\left(\frac{q_\parallel}{E_\qq}\right)$ & \ref{geee}\\[5pt]
			\textbf{(F)} & $a_{\qq}^{r\dagger}b_{-\qq}^{r'\dagger}|0\rangle$ & $\pi-\sin^{-1}\left(\frac{q_\bot}{E_\qq}\right)$ & $\sin^{-1}\left(\frac{q_\bot}{E_\qq}\right)$  & \ref{geee}\\[5pt]
			\textbf{(G)} & $a_{\qq}^{1\dagger}a_{\qq}^{2 \dagger} b_{-\qq}^{1\dagger}b_{-\qq}^{2 \dagger}|0\rangle$ & $\pi-\tan^{-1}\!\left(\frac{|\qq|}{m}\right)$ & $\pi-\tan^{-1}\!\left(\frac{|\qq|}{m}\right)$ & \ref{geee}
		\end{tabular}
	\end{center}
	\label{sum-table}
\end{table}

Clearly, in eq.~\reef{goal}, we have $\tilde{Y}(m,\qq,s)=\pi-Y(m,\qq,\bs)$ where $Y(m,\qq,\bs)$ is the complexity of the corresponding modes in the Dirac ground state given eq.~\reef{march}. We already plotted this function in figure~\ref{fig2}, since the same expression appeared in evaluating the complexity of the state $a_{\qq}^{\br\dagger}b_{-\qq}^{\brp\dagger}|0\rangle$ --- see eq.~\reef{excite8}.  Hence, we will find the same complexity here for eq.~\reef{excite88a} as for the excited state $a_{\qq}^{\br\dagger} a_{\qq}^{\brp\dagger} b_{-\qq}^{\br\dagger} b_{-\qq}^{\brp\dagger}|0\rangle$, which is the special case of eq.~\reef{excite88} noted above. As mentioned above, the difference between the two states is the orientation of the spin axes of the individual particles (and antiparticles), however, we are finding that the complexity does not depend on these details. The reason for this is that in both cases, the full state is a spin singlet (\ie the net spin is zero) and so it is invariant under rotations of the spin axis. In fact then, the two states are identical and so it is a confirmation of our methods that we find the same complexity in either case irrespective of the details of the construction of the state. The above discussion also provides us with an alternative perspective on how to arrive at this same result.

Further, the same reasoning can be applied to the (B) and (C) families of states in eq.~\reef{excite85}. In either case, the pair of excited particles (or antiparticles) form a spin singlet. Hence the states would actually be identical with the spins oriented along any axis, \ie they need not be along the momentum axis as in the discussion in section \ref{excited1}. Of course then, with an alternate choice of spin axis, the complexity would remain the same as in eqs.~\reef{diff33a} and \reef{diff34}.

We summarize our results in table~\ref{sum-table} for the complexity of states with excitations in a single mode $\qq$. The table includes all the states discussed in section~\ref{excited1} and the present section. This covers all states with an even number of excitations in a single mode $\qq$ where the spin axis can be different from the momentum orientation. This could be generalized further by allowing differently oriented spin axes for the different particle and antiparticle excitations. Our methods readily apply to this scenario, but it will be hard to find analytical expressions as one needs to find closed expressions for the eigenvalues of $\Delta$. Instead, it will be easy to evaluate the eigenvalues numerically for any specific choice of spin orientations.

\subsubsection{Excitations in many modes}
At this stage, we are essentially prepared to consider general excited states of the form
\begin{align}\label{excite77}
|\tilde\psi\rangle=\prod_{i,j} \,a^{\br_i\dagger}_{\qq_i} \ b^{\br_{j}\dagger}_{-\qq_{j}}\,|0\rangle\,,
\end{align}
where the only constraint is that the total number of excitations must be even to ensure that these states lie on the connected component of the space of fermionic Gaussian states, \ie 
\beq
i_\mt{max}+j_\mt{max}=2n\,.
\label{top}
\eeq
Implicitly, we also assume for simplicity that in all of the different momentum sectors, that the spins are oriented along the momentum axis of the respective modes. 

As in the previous examples of excited states, the above states will still be tensor product over all modes $\pp$ where only the sectors with $\pp=\qq_i$ differ from Dirac ground state.\footnote{We use $\qq_i$ here and in the following to refer to  the momentum labels of both the particle or antiparticle excitations. As in our previous discussions, we refer to a momentum sector $\qq_i$ as states spanned by the four creation operators, $a^{\br\dagger}_{\qq_i}$ or $b^{\br\dagger}_{-\qq_i}$ with $\br=1,2$.} This means it will be sufficient to compute the covariance matrix $\Omega(m,\qq_{i})$ for each excited sector $\qq_{i}$ to find the change in complexity for this class of excited states. In particular, we would first examine the state to determine the number of excitations in each of these sectors. In many sectors, there would be an even number of excitations and for these sectors, we can use the results in table~\ref{sum-table} to evaluate their contribution to the complexity. Hence only the sectors with an odd number of excitations need further consideration. 

An important observation is that in each sector with an odd number of excitations, the state will lie in the disconnected component from the reference state $|\Omega(M,0)\rangle_{\qq_i}$ for this momentum $\qq_i$. This implies that any geodesic transforming the full reference state $|\bar 0\rangle$ to the excited target state \reef{excite77} will actually not preserve the tensor product structure over modes. In fact, this is the reason that it only works if we have an even number of excitations, because the generator of the geodesic will always mix two excitations intermediately along the path, and not until the end point is reached, will the tensor product over momentum modes be restored. Clearly, the geodesic accomplishing this transformation is not unique and arbitrary pairings are among the even number of excited modes are allowed. However, all these geodesics will have the same length, so it does not matter for the purpose of computing complexity. Further, our approach of evaluating the length of the geodesic(s) using the relative covariance matrix can still be applied, but clearly it does not depend on these details since it only refers to the endpoints of the geodesic where the tensor product structure holds.

Let us begin with the sectors with a single excitation.
We can compute the relative covariance matrix $\Delta(\qq_i)=\tilde\Omega(m,\qq_i)\,\omega(M,0)$ for each of these excited modes $\qq_i$ and find its eigenvalues. The calculations are a little tedious, but the result is rather simple and given by
\begin{align}\label{ice9}
\mathrm{spec}(\Delta(\qq_i))=\left(1,1,-1,-1,\frac{m+\ii |\qq_i|}{E_{\qq_i}},\frac{m+\ii |\qq_i|}{E_{\qq_i}},\frac{m-\ii |\qq_i|}{E_{\qq_i}},\frac{m-\ii |\qq_i|}{E_{\qq_i}}\right)\,.
\end{align}
Hence rather than finding two identical quadruples, we find a single quadruple with the familiar form $(\ex^{2\ii\vt},\ex^{2\ii\vt},\ex^{-2\ii\vt},\ex^{-2\ii\vt})$ where $\vt=\tan^{-1}(|\qq_i|/m)$ and two pairs $(1,1)$ and $(-1,-1)$. Note that this result is independent of the type of excitation (particle or antiparticle) and of the orientation of the spin ($\br_i=1,2$). Based on our discussion in section~\ref{sec:general}, the eigenvalue pair $(-1,-1)$ implies that the reference and target states in the $\qq_i$ sector lie on disconnected components. However, if we combine two sectors $(\qq_i,\qq_j)$ with an odd number of excitations, these additional pairs combine into two quadruples given by $(-1,-1,-1,-1)$ with $\vt=\pi$ and $(1,1,1,1)$ with $\vt=0$. For example, combining two sectors with a single excitation, the complexity contributions are described by four angles
\begin{align}
2\vt_1=\pi\,,\quad
2\vt_2=0\,,\quad
2\vt_3=\tan^{-1}\!\left(\frac{|\qq_i|}{m}\right)\,,\quad
2\vt_4=\tan^{-1}\!\left(\frac{|\qq_j|}{m}\right)\,.
\end{align}

Our discussion above implied that an unmatched eigenvalue pair $(-1,-1)$ also appears for the sectors with three excitations. In such a situation, there must be a particle-antiparticle pair with the same spin label, \ie a pair of creation operators as appear in the (A) states in eq.~\reef{excite8}. This pair can be dealt with separately as in section \ref{excited1} and a quadruple of eigenvalues appears specified by the angle $2\vt=\pi-\tan^{-1}\!\left(\frac{|\qq_i|}{m}\right)$, as in eq.~\reef{march2}. Similarly, as in the previous case, the unpaired creation operator then yields an eigenvalue quadruple of the form $(1,1,-1,-1)$. Hence we see the unmatched pair $(-1,-1)$ appearing here, and as discussed above, if the reference and target state are connected within the set of Gaussian states, this pair can be matched with another such pair from one of the other sectors with an odd number excitations.

While we have outlined the steps needed of the evaluation of the complexity of a general state \reef{excite77}, writing out the final complexity would require a rather elaborate expression since there are many different possibilities for the different sectors. Instead let us focus on the special case where in each excited sector, there is only a single excitation. Then following the analysis in section \ref{excited1}, it is straightforward to evaluate the difference in the complexity of the excited state and that of the vacuum. With the $\kappa=1,2$ measures, we find
\begin{align}
\Delta\mathcal{C}_{\kappa=2}(|\bar{0}\rangle\to |\tilde\psi\rangle)&=n\,\pi^2-\sum_i \left[ \tan^{-1}\!\left(\frac{|\qq_j|}{m}\right)\right]^2
\label{hotpot}\\
\Delta\mathcal{C}_{\kappa=1}(|\bar{0}\rangle\to |\tilde\psi\rangle)&=n\,\pi-\sum_i  \tan^{-1}\!\left(\frac{|\qq_j|}{m}\right)\,,
\nonumber
\end{align}
where $n$ is defined in eq.~\reef{top}, \ie $n$ is half of the total number of excitations. We repeat the usual caveat that implicitly, for $\Delta\mathcal{C}_{\kappa=1}$, we are aligning the basis in each $\qq_i$ sector with the generators which produce the desired unitary transformation. In comparing to our previous results, we see that these results match $\Delta\mathcal{C}_{\kappa=2}$ and $\Delta\mathcal{C}_{\kappa=1}$ in eqs.~\reef{diff33a} and \reef{diff34}, respectively, for the states in eq.~\reef{excite85} (or the corresponding differences in for the more general states in eq.~\reef{excite88a}).

\section{Discussion and outlook} \label{discuss}

As was reviewed in section \ref{preamble}, Nielsen's perspective  \cite{Nielsen:2005mn1,Nielsen:2006mn2,Nielsen:2007mn3} allows one to bring the power of differential geometry to bear on the problem of constructing optimal quantum circuits, and also provides an objective manner in which to measure the complexity as the `length' of extremal paths in this geometry. The present paper extends the study of \cite{Jefferson:2017sdb}, which applied Nielsen's geometric approach to investigate the ground state complexity of a free scalar field theory. In particular, we examined the complexity of Gaussian states in a free fermionic quantum field theory. Let us reiterate that some aspects of our work overlaps with the recent studies in \cite{simonR,Khan:2018rzm}.  As discussed in section \ref{sec:prelude}, we can think of the Gaussian states for $N$ fermionic degrees of freedom, as those annihilated by a family of $N$ destruction operators $a_i$,
and hence the transformations carrying us from one state to another can be identified with the Bogoliubov transformations amongst the annihilation and creation operators. This perspective readily reveals a group structure for the family of transformations of interest, namely $\mathrm{O}(2N)$ for free fermions. This group structure can be connected to Nielsen's construction of the unitary circuits which prepare these states by observing that the operators $\op_I$ in eq.~\reef{controlY} form a representation of the corresponding Lie algebra $\mathrm{so}(2N)$ and the circuits are then trajectories in the corresponding group manifold. 

In contrast, with \eg \cite{Jefferson:2017sdb,simonR,Khan:2018rzm}, our approach was to emphasize this group structure and in doing so the precise representation appearing in the construction of the circuits becomes less important. Rather we focused on the action of the group transformation on the covariance matrix \reef{covmat}, which can be used to completely characterize the Gaussian states for either fermionic or bosonic degrees of freedom. With this representation, we were able to equip the group with a natural positive definite right-invariant metric, which allowed us to find all geodesics and their path lengths analytically by exploiting the underlying $\mathrm{U}(N)$ symmetry of this metric, as explained in appendix~\ref{app:minimal}. We found this more abstract group theoretic perspective yields an extremely powerful approach to apply Nielsen's construction to this problem. In particular, we were able to prove that our unitary circuits in fact correspond to minimal geodesics in the corresponding geometry on the space of states.\footnote{We might point out that such a proof was not provided in \cite{Khan:2018rzm}, which studied the same problem. However, we emphasize that where our work overlapped with the latter paper, \eg the complexity of the ground state for a free Dirac fermion in four dimensions in section \ref{ground1}, our results agreed.} Further, we evaluated the complexity of the ground state for a variety of different disentangled reference states, and we also evaluated the complexity of various families of excited states.

In evaluating the complexity, one must choose a cost function \reef{costco} and in our analysis, we focused on three choices, the $F_2$ measure and the $\kappa=1$ and $\kappa=2$ measures. Of course, the $F_2$ measure can be recognized as the proper distance in the Riemannian geometry, which we defined for the $SO(2N)$ group manifold. With a physicist's perspective, we can regard the $\kappa=2$ measure as a `standard' test particle action on the same geometry and of course, it yields the same optimal trajectories as the $F_2$ measure. Given the interpretation of the $Y^I$ functions in eq.~\reef{controlY} as indicating when particular gates appear in the circuit, the $\kappa=1$ measure comes the closest amongst the examples in eq.~\reef{costco} to the original definition of simply counting the number of gates in the circuit. Unfortunately, in the practical situation where the relevant trajectories are constructed by many orthogonal generators, this measure defines a `Manhattan metric' on the relevant submanifold and does not provide the most useful measure to distinguish different circuits, \eg \cite{Nielsen:2005mn1,Jefferson:2017sdb}. Furthermore, as discussed in \cite{Jefferson:2017sdb}, the precise value of the complexity depends on the precise choice of the generators ${\cal O}_I$ appearing in the construction. An advantage of the previous two measures is that they do not suffer from this basis dependence,\footnote{Strictly speaking, this statement applies for any orthonormal basis of generators, however, a completely basis independent framework is produced by extending these measures to $\tilde{F}_{2}(U,\vec Y)=\sqrt{\sum_{I,J} g_{IJ}\, Y^I\, Y^J}$ and $\tilde{F}_{\kappa=2}(U,\vec Y)=\sum_{I,J} g_{IJ}\, Y^I\, Y^J$ where $g_{IJ}$ is a frame metric on the space of generators --- see the discussion of penalty factors below.}  and of course, our analysis of the geodesics, \eg in appendices \ref{app:geodesics} and \ref{app:minimal}, made reference to the $F_2$ measure (but as noted above, the $\kappa=2$ measure yields identical geodesics). Implicitly, our results for the $\kappa=1$ measure assume that the basis of generators is aligned with the generators producing the desired transformation. For higher values of $\kappa$, \ie $\kappa>2$, one finds a similar basis dependence \cite{Jefferson:2017sdb}, which is unfortunate because the $\kappa$ cost functions were introduced because of the close parallels between the resulting complexity for free fields and the results found for holographic complexity \cite{Jefferson:2017sdb,Chapman:2017rqy}.  However, at least for $\kappa=1$, this situation can be remedied by making use of the Schatten norm (\eg see \cite{bhatia2013matrix,watrous2018theory}).  This norm actually provides a family of measures based on computing the singular value decomposition of the desired transformation and in the present case, it reduces to $(\sum_I |2\vti|^p)^{1/p}$, where $p$ is a positive integer and the $\vti$ are precisely the eigenvalues of the generator producing the desired transformation. With $p=2$, this reduces to the standard Frobenius-Hilbert-Schmidt norm, and we recover the $F_2$ measure which we were studying in the main text. More importantly with $p=1$, we will recover our results for the $\kappa=1$ measure, however, they are now basis independent when framed in terms of the Schatten norm. We return to discuss this issue in more detail in \cite{lfh-unpublished,comparison-paper}.

As discussed in section \ref{twofermi}, we chose a natural metric \reef{gonzo} defined by the group structure of the present problem, namely one proportional to the Killing form on the $O(2N)$ group manifold. This choice simplified the calculation of the geodesics and their lengths, as explained in appendix~\ref{app:minimal}. Alternatively, one may wish to use a different right-invariant metric which involves ``penalty factors'', \eg as in the $F_{1p}$ measure in \reef{costco} or in an elaboration of the $F_2$ cost function of the form $\tilde{F}_{2}(U,\vec Y)=\sqrt{\sum_{I,J} g_{IJ}\, Y^I\, Y^J}$.  With such a penalized metric, one favours certain directions in the circuit space over others, \ie a higher cost is given to certain classes of gates (\ie Lie algebra generators). For example, in \cite{Jefferson:2017sdb}, it was suggested that such an approach could be used to restore a notion of locality for circuit complexity in quantum field theories by increasing the cost of gates that correlate degrees of freedom separated by larger distances.\footnote{Let us note however that ref.~\cite{nonlocal} recently argued that the microscopic model underlying holographic complexity must be nonlocal.} However, in the initial exploration of the latter for the scalar field theory, the shortest paths were dramatically changed and in general, evaluating the geodesics relied on numerical computations. We expect a similar situation will arise for the present fermionic systems, especially if the penalty factors  break the $\mathrm{U}(N)$ symmetry of the present metric. We leave the study of such penalized metrics for a future project. 

An alternative approach was introduced in \cite{Chapman:2017rqy} to compute circuit complexity by assigning a geometry to the space of states rather than the space of unitaries. In particular, this approach is based on the fact that the set of normalized vectors in a Hilbert space\footnote{More precisely, we also divide by the complex phase to get a representation of the ``space of rays'' in the Hilbert space.} form a Riemannian manifold whose metric is inherited from the positive definite inner product of the Hilbert space, \ie the Fubini-Study metric \cite{fubini1904sulle,study1905kurzeste}. As it turns out, the minimal geodesic between two states $|\psi_{\mathrm{R}}\rangle$ and $|\psi_{\mathrm{T}}\rangle$ is just given by the arc of a circle that connects the two states in the two-dimensional plane spanned by them. The geodesic length is therefore trivially given by the angle $\vartheta$ between the two state vectors which can be computed as
	$\cos\vartheta=\langle \psi_{\mathrm{R}}|\psi_{\mathrm{T}}\rangle$.
However, as stressed in \cite{Susskind:2014jwa,Brown:2016wib,Brown:2017jil}, this geometry alone is inappropriate to define a notion of circuit complexity for states in quantum field theory since the maximum separation of any two states is only $\pi$. Instead, it was proposed in \cite{Chapman:2017rqy} to restrict the manifold of states to a subset, in particular to the set of Gaussian states in the context of free field theories. In this case, the geodesics are forced to lie on a submanifold with a more intricate geometry. In the case of a free scalar field, the complexity determined with this alternative approach was actually found to agree with that determined with the Nielsen approach in \cite{Jefferson:2017sdb}.  Similarly, for a free fermionic field, one can show that the length of the minimal geodesics found in the present paper agrees with the lengths found with the Fubini-Study metric restricted to the submanifold of Gaussian states. We believe that this is a general feature which relates minimal Lie group geodesics with respect to a ``natural metric'' with the Fubini-Study geodesics on the quotient manifold where we divide the Lie group $\mathrm{Sp}(2N,\mathbb{R})$ or $\mathrm{O}(2N)$ by their subgroup $\mathrm{U}(N)$ \cite{lfh-unpublished,comparison-paper}. 

A primary motivation to develop techniques to evaluate complexity in simple quantum field theories is to better understand holographic complexity. Of course, there is no \textit{a priori} reason to expect the results for free field theories to agree with those in holography, which necessarily describes strongly coupled quantum field theories with a large number of degrees of freedom. Nevertheless, it was found that if the cost function is chosen appropriately, the scalar field complexity exhibits some remarkable similarities with holographic complexity \cite{Jefferson:2017sdb,Chapman:2017rqy}. In particular, the leading divergences in both the CV and CA proposals are extensive, \ie they are proportional to the volume of the time slice on which the boundary state is evaluated, as shown in \cite{Carmi:2016wjl}. Just as for the scalar field in \cite{Jefferson:2017sdb}, we found here that complexity of a fermionic field yields an analogous leading divergence with the $\kappa$ cost functions, in particular, with $\kappa=2$ and $\kappa=1$ as shown in eqs.~\reef{eq:comp-kappa=1-exact} and \reef{eq:comp-asymp}, respectively. In contrast, the $F_2$ cost function gives a result proportional to $V^{1/2}$ which does not match the holographic results.\footnote{An alternative approach \cite{Brown:2017jil} would be to assign the cost $(V \Lambda^{d-1})^{1/2}$ to each gate in the $F_2$ measure. However, this choice would be problematic, \eg in comparing complexities for different UV cutoffs.} Hence, our fermionic results reinforce the previous insights provided by the complexity calculations for a free scalar field with regards to the form of the cost functions that implicitly underly holographic complexity.

Another interesting feature of the complexity results for the free scalar \cite{Jefferson:2017sdb} was that the leading contribution with the $\kappa$ measures contained an extra logarithmic factor proportional to $\log^\kappa(\Lambda/\omega_0)$, where $\omega_0$ was the frequency specifying the unentangled reference state. Surprisingly, a similar logarithmic factor was found in the leading divergence in the holographic complexity \cite{Carmi:2016wjl} for the complexity=action proposal. In the latter case, the logarithmic factor came from joint terms \cite{Lehner:2016vdi} in the gravitational action, and the argument of the logarithm was ambiguous because of the freedom in the normalization of the null normals on the boundary of the WDW patch.  Whereas this ambiguity had originally been seen as problematic for the CA conjecture, the scalar field results indicated that it is a perfectly natural feature in the complexity of QFT states. 

However, we find that no such logarithmic factors appear in the leading divergences of the complexities evaluated here for a fermionic field, \eg see eqs.~\reef{eq:comp-kappa=1-exact} and \reef{eq:comp-asymp}. This motivated our study of alternative reference states in section \ref{sec:alternative}. However, we found that for any reference state which was translationally invariant and spatially unentangled, the leading singularity in the ground state complexity takes the same form. That is, with the $\kappa=1$ measure, the leading divergence is precisely the same for all such reference states, \ie it is independent of $\hat{q}=q/M$, as shown in eq.~\reef{frank2}. On the other hand, eq.~\reef{frank1} shows that while the numerical coefficient varies with the reference state, \ie with $\hat q$, the leading singularity is still proportional to $V\Lambda^3$ in all cases. Of course, this is not in contradiction with holographic complexity. Typically, holography involves a supersymmetric boundary theories and so there will be both bosonic and fermionic degrees of freedom. Further, as explained in \cite{Jefferson:2017sdb}, if one were to choose the reference scale to be proportional to the cutoff, \ie $\omega_0=\ex^{-\sigma} \Lambda$, then the logarithmic factor would simply appear as some numercal factor, \ie $\sigma^\kappa$, multiplying the usual $V\Lambda^{d-1}$ divergence.\footnote{With this observation, we can also see that no obvious tension between the different structure of the UV divergences found with complexity=action and complexity=volume proposals for holographic complexity \cite{Carmi:2016wjl}.} However, with regards to supersymmetry, it may be interesting to investigate if the relative strength of the logarithmic factor in the leading divergence found with complexity=action changes as the amount of supersymmetry in the boundary theory changes. 

As noted above, while there was no reference frequency that appeared in the ground state complexity of a fermionic field, we were able parametrize a whole family of reference states (all of which were spatially unentangled and translationally invariant). These states were characterized by a momentum $\qq$ and a mass scale $M$, although we found that the complexity only depended on the dimensionless vector $\hat\qq=\qq/M$. We can make a parallel with these reference states in the free scalar field theory as follows: In notation analogous to eq.~\reef{boil2}, the ground state of scalar is given by
\begin{align}
	|0\rangle=\otimes_\pp\, |m,\pp\rangle_\pp\,.
\end{align}
Here, the state $|M,\qq\rangle_\pp$ indicates the ground state of the Hamiltonian of a single degree of freedom (in the mode $\pp$)
\begin{align}
	H_\pp(M,\qq)=\frac{1}{2}\left(\pi_\pp^2+(M^2+|\qq|^2)\phi_\pp^2\right)\,.
\end{align}
Now we can choose our reference state, the state where we put every mode $\pp$ into the ground state of $H_\pp(M,\qq)$ with the same $M$ and $\qq$. The resulting reference state,
\begin{align}\label{rope}
	|\psi_{\mathrm{R}}\rangle=\otimes_\pp |M,\qq\rangle_\pp\,,
\end{align}
is both spatially unentangled and translationally invariant, as desired.
Now in the scalar theory, it just so happens that the only relevant quantity is the frequency $\omega_0=\omega(M,\qq)=\sqrt{M^2+|\qq|^2}$ and eq.~\reef{rope} is precisely the family of reference states considered in \cite{Jefferson:2017sdb}. Here, we are constructing them in a way that parallels our construction of the fermionic reference states in section \ref{sec:alternative}.

However, let us note that it is straightfoward to extend this family of reference states for the scalar field as follows: We can choose any fixed Gaussian state $|G\rangle_\pp$ to construct a reference state analogous to that in eq.~\reef{rope}. Hence for a scalar field, with a single bosonic degree of freedom at each spatial point (or in each momentum mode), we can form a two-dimensional family of reference states corresponding to $\mathcal{M}_{b,1}=\mathrm{Sp}(2,\mathbb{R})/\mathrm{U}(1)$ by performing Bogoliubov transformations to a fixed $|G_0\rangle_\pp$ (\eg with $G_0=\mathbb{1}$), as discussed in section \ref{boso}. In terms of the language used in the previous paragraph, we can say that the most general $|G\rangle_\pp$ can be labeled by a reference frequency $\omega_0$ and an angle $\theta_0$. The corresponding state $|\omega_0,\theta_0\rangle_\pp$ is the ground state of the following Hamiltonian,
\begin{align}
	H_\pp(\omega_0,\theta_0)=\frac{1}{2}\big[(\cos(\theta_0)\pi_\pp+\sin(\theta_0)\phi_\pp)^2+\omega_0^2(\cos(\theta_0)\phi_\pp
	-\sin(\theta_0)\pi_\pp)^2\big]\,.
\end{align}
This general family of reference states extends the coefficient matrix $A_{ab}$ of \cite{Jefferson:2017sdb} to have complex values. Further, with this extension, it would not suffice to construct the unitary circuit with only entangling gates. That is, one would have to extend the analysis of geodesics on the group $\mathrm{GL}(N,\mathbb{R})$ in \cite{Jefferson:2017sdb} to geodesics on the full $\mathrm{Sp}(2N,\mathbb{R})$ group of Bogoliubov transformations discussed in section \ref{boso}.\footnote{Of course, here $N$ denotes the total number of degrees of freedom in the (regulated) scalar field theory.} It would be interesting to study the effect of this generalization on the complexity of the ground state. Further, it is amusing to note that in considering a theory of $n$ scalar fields, the general family of reference states would be $n(n+1)$-dimensional, \ie the full family corresponds to $\mathcal{M}_{b,n}=\mathrm{Sp}(2n,\mathbb{R})/\mathrm{U}(n)$.

Of course, the above generalization can also be applied to a fermionic field. In particular, given a system with $n$ fermionic degrees of freedom at each spatial point or in each momentum mode,  the corresponding space of Gaussian states is $n(n-1)$-dimensional, \ie recall $\mathcal{M}_{f,n}=\mathrm{O}(2n)/\mathrm{U}(n)$ as discussed in section \ref{twofermi}. Again, we can choose any of the corresponding Gaussian states to define $\Omega_0$ in constructing the reference state $|\Omega_{\mathrm{R}}\rangle=\otimes_\pp |\Omega_0\rangle_\pp$, as described in section \ref{sec:alternative}. For example, with the four-dimensional Dirac field studied in the main text, we have $n=4$ degrees of freedom at each spatial point and in principle, we can construct a twelve-dimensional family of spatially unentangled and translationally invariant reference states. In section \ref{sec:alternative}, in fact, we only considered the three-dimensional subspace labeled by the vector $\hat\qq=\qq/M$.
It would be interesting to study the effect of choosing reference states throughout this full family on the ground state complexity of the Dirac field.

In sections \ref{excited1} and \ref{geee}, we were also able to study the complexity of a broad variety of excited states. This study was facilitated by the fact that in a free fermionic theory, acting with products of creation operators $a^{r\dagger}_\qq$ transforms the vacuum to another Gaussian state, which allowed us to apply the general framework developed in section~\ref{sec:general}. While the complexity depended on the details of which modes were excited (\eg see table \ref{sum-table}), a general feature was that when using the $\kappa$ cost functions, the difference between the complexities of the excited state and the vacuum state was finite,\footnote{Assuming that we are only exciting a finite number of momentum modes.} \eg see eqs.~\reef{diff22a} or \reef{hotpot}. This result is perhaps not unexpected but we note that it is in keeping with our expectations for holographic conjecture. That is, low energy excitations in the bulk will not effect the structure of the UV divergences, which is determined by contributions coming from the asymptotic regions of the bulk spacetime. One explicit example of this behaviour is provided with the complexity of formation in \cite{Chapman:2016hwi,Carmi:2017jqz}. However, our new results for excited states in the free fermionic field theory provide some additional motivation to study the complexity of excited states in a holographic framework more carefully. In any event, the finite difference in the complexities of the excited states and the vacuum reinforces the suggestion that the cost functions which are implicit in the microscopic rules governing holographic complexity are similar to the $\kappa$ cost functions used in our free field studies. 

We should note that while we considered a broad family of excited states in our complexity calculations, these only represent discrete points in the full $N(N-1)$-dimensional space of Gaussian states.\footnote{Hence in terms of the UV cutoff, the dimension of this space is of the order $(V\Lambda^3)^2$.} While our excited states form a physically interesting case, where one considers states with an arbitrary (but even) number of particles (and antiparticles) that have well-defined momentum, Gaussian states are far more complicated in general, \eg $(1+\alpha\,a^{\br\dagger}_\qq\,b^{\br\dagger}_{-\qq})|0\rangle$ is a Gaussian state without a definite particle number.  However, we want to emphasize that our framework also applies to these more complicated, coherent excitations. For instance, we can consider a set of coherent creation operators
\begin{align}
	A^\dagger_i=\int \frac{d^3p}{(2\pi)^2}\sum_s\left( f_i(\pp,s)\,a^{s\dagger}_{\pp}+g_i(\pp,s)\,b^{s\dagger}_{\pp}\right)\,,
\end{align}
with smearing functions $f_i$ and $g_i$, such that $\{A_i,A^\dagger_j\}=\delta_{ij}$. In this case, each excited state
\begin{align}
	|\tilde \psi\rangle=\prod_i A^\dagger_i|0\rangle\,,
\end{align}
will still be Gaussian (since as before, they are annihilated by the $A^\dagger_i$ themselves). As long as the number of such excitations is even, $|\tilde\psi\rangle$ will live on the same connected component as the Dirac vacuum (and appropriate choices of the reference state), such that the complexity can be computed. In general, $|\tilde\psi\rangle$ will not be a tensor product over momentum modes implying that it is not translationally invariant and it will have not just spatial, but also momentum entanglement. Computing the complexity of such states will thus require more work in the continuum than the energy eigenstates considered in this paper, because one needs to find a generalized normal mode basis, such that both $|\tilde\psi\rangle$ and the reference state are tensor products with respect to this basis. A simple solution for such examples in practice will be to put the field theory onto a finite lattice and compute the circuit complexity from the eigenvalues of $\Delta$, which will be a large but finite matrix.

The techniques introduced in this paper are quite general. One could easily extend the present discussion to Dirac fields in higher (or lower) spacetime dimensions. It may also be interesting to study circuit complexity for states in a theory of chiral fermions. A more challenging extension of the present work would be to evaluate the complexity of Gaussian states with odd fermion number. As noted before, the corresponding geodesics could not remain within the space of Gaussian states because they must reach the disconnected component of $\mathcal{M}_{f,\ssc N}=\mathrm{O}(2N)/\mathrm{U}(N)$. Hence, it may be just as simple to consider the complexity of more general states, \ie non-Gaussian states. Another possibility would be to analytically continue our formula for length of the minimal geodesics by just defining the circuit complexity to be computed in the same way from the relative covariance matrix $\Delta$. This would be an analytical continuation where we allow paths in the complexification of $\mathrm{O}(2N)$ which is a connected Lie group. However, it is not clear how such a path could be related to a sequence of intermediate quantum states between reference and target state.

Another interesting direction would be connect the present model for the complexity for fermionic states (as well as in \cite{simonR,Khan:2018rzm}) to previous discussions of simulating fermionic systems on a universal quantum computer, \eg \cite{abramX,bravyY,Jordan}. In particular, ref.~\cite{Jordan} bounded the complexity of quantum algorithms computing scattering amplitudes in interacting fermionic field theories. In particular, they studied the two-dimensional Gross-Neveu model \cite{Gross:1974jv} describing $N$ species of fermions with quartic interactions. The first step in this process is to prepare the vacuum of the free theory (using adiabatic state preparation) and the upper bound on the number of gates required for this step is proportional to $V^3\Lambda^6/m^3$. Of course, the complexity of preparing the vacuum of a free fermion is a central question in the present paper but our result for a two-dimensional theory would be of order $V\Lambda$, \eg see explicit calculations for $d=2$ in \cite{Khan:2018rzm}. The latter is dramatically smaller than the bound found by \cite{Jordan} and hence it would be interesting to investigate if the present construction based on Nielsen's geometric approach \cite{Nielsen:2005mn1,Nielsen:2006mn2,Nielsen:2007mn3} can be adapted to provide practical quantum algorithms.

\acknowledgments
We thank Jose Barbon, Eugenio Bianchi, Shira Chapman, Jens Eisert, Michal Heller, Ro Jefferson, Hugo Marrochio, and Fernando Pastawski  for useful discussions. LH thanks Mark Penney for a helpful email exchange regarding the topology of type DIII symmetric spaces. Research at Perimeter Institute is supported by the Government of Canada through Industry Canada and by the Province of Ontario through the Ministry of Research \& Innovation. LH thanks the Visiting Graduate Fellow program for hospitality at the Perimeter Institute during this project. LH was also supported by a Frymoyer fellowship, a Mebus fellowship and by the National Science Foundation under Grant No.~PHY-1404204 awarded to Eugenio Bianchi. RCM also thanks the KITP for hospitality during this research. Research at the KITP was supported in part by the National Science Foundation under Grant No. NSF PHY11-25915. RCM is supported by funding from the Natural Sciences and Engineering Research Council of Canada and from the Simons Foundation through the ``It from Qubit'' collaboration.

\appendix

\section{Geodesics on Lie groups}\label{app:geodesics}
We show under which conditions, the one-parameter subgroup $\ex^{sA}$ is a geodesics on a Lie group $\mathcal{G}$. In particular, this allows us to prove that \emph{any} such subgroup is geodesic if we equip $\mathrm{SO}(2N)$
with the unpenalized right-invariant metric that we use in the body of this paper. These results are standard material from Lie group geometry and the study of symmetric spaces \cite{mimura1991topology}, but we will review them here for completeness and to match our conventions.

\subsection{Geodesics for right-invariant metric}
We consider the setting of a general Lie group $\mathcal{G}$ with Lie algebra $\mathfrak{g}$ and positive definite metric $\langle\cdot,\cdot\rangle_{\mathbb{1}}: \mathfrak{g}\times\mathfrak{g}\to\mathbb{R}$. We can extend this metric to all tangent spaces of the Lie group by requiring right-invariance, which means at the point $M\in G$ (we think of $M$ as a matrix in the fundamental representation of the group), we have the inner product
\begin{align}
\langle X,Y\rangle_{M}=\langle X M^{-1},YM^{-1}\rangle_{\mathbb{1}}\,,
\end{align}
where $X,Y\in T_MG$ and $XM^{-1},YM^{-1}\in T_\mathbb{1}G=\mathfrak{g}$. In this context, we can ask the question for which $A\in\mathfrak{g}$ is the trajectory $\ex^{s A}$ a geodesic.

Given a Lie group $\mathcal{G}$ with right-invariant metric $\langle\cdot,\cdot\rangle_{M}$ and a Lie algebra element $A\in\mathfrak{g}$, the path
	\begin{align}
	\gamma: \mathbb{R}\to \mathcal{G}: s\mapsto \ex^{s A}
	\end{align}
in the one-parameter subgroup generated by $A$ is a geodesic if and only if the generator $A$ is a critical point of the norm function $\lVert A\rVert=\sqrt{\langle A,A\rangle_{\mathbb{1}}}$ under the adjoint action of the group. This can be derived in the following three steps:
	\begin{enumerate}
		\item Left-invariant vector fields are Killing vector fields\\
		A Lie group with right-invariant metric has automatically its left-invariant vector fields as Killing vector fields. A left-invariant vector field $X$ on $G$ is completely determined by its value $X_\mathbb{1}=B\in T_\mathbb{1}G=\mathfrak{g}$ at the identity, while at any other point $M\in G$, the vector field will have the value $X_M=MB\in T_MG$. Left-invariant vector fields are the infinitesimal generators of right translations in the group, so that a group with right-invariant metric is invariant under those. As generators of a diffeomorphism that leaves the metric invariant, left-invariant vector fields are Killing vector fields.
		\item Left-invariant vector fields give rise to conserved charges\\
		For every Lie algebra element $B\in \mathfrak{g}$, the left-invariant vector field $X_M=MB$ gives rise to conserved quantities along any geodesic $\gamma(s)$. Namely, we have the conserved charge
		\begin{align}
		Q_B(s)=\langle \dot{\gamma}(s),X_{\gamma(s)}\rangle_{\gamma(s)}=\langle \dot{\gamma}(s),\gamma(s)B\rangle_{\gamma(s)}
		\end{align}
		with $\frac{d}{ds}Q_B(s)=0$. Due to the fact that we have as many linearly independent Killing vector fields $X_B$ as we have generators $B\in\mathfrak{g}$, the combination of conserved quantities and initial point $\gamma(0)$ characterize the geodesic uniquely. Put differently, if we know $\gamma(0)$ and $\dot{\gamma}(0)$, we can use the $\dim\mathfrak{g}$ linearly independent charges $Q_{B_i}$ to rewrite the geodesics equation as a first order-equation with a unique solution. In particular, if we find trajectory $\gamma(s)$ that preserves all charges $Q_B(s)$, we can be certain that we found a geodesic.
		\item Critical points on the adjoint orbits generate geodesics\\
		Based on our previous considerations, we can check what the conditions on $A\in\mathfrak{g}$ are, so that $\gamma(s)=\ex^{sA}$ is a geodesic. We can compute for $\gamma(s)$, the charges
		\begin{align}
		Q_B(s)=\langle \dot{\gamma}(s),X_{\gamma(s)}\rangle_{\gamma(s)}=\langle \ex^{sA}A,\ex^{sA}B\rangle_{\ex^{sA}}=\langle A,\ex^{sA}B\ex^{-sA}\rangle_{\mathbb{1}}\,,
		\end{align}
		where we used right-invariance of the metric to compute the inner product at the identity. From this equation we can prove that $\langle A,[A,B]\rangle_{\mathbb{1}}=0$ for all $B\in\mathfrak{g}$ is a necessary and sufficient condition for all charges being conserved:
		\begin{itemize}
			\item Necessary. We just evaluate $\frac{d}{ds}Q_B(s)|_{s=0}=\langle A,[A,B]\rangle_{\mathbb{1}}$. All charges will be conserved if this equation vanishes for all $B\in\mathfrak{g}$.
			\item Sufficient. Baker-Campbell-Hausdorff implies $\frac{d}{ds}Q_B(s)=\langle A,[A,\sum^{\infty}_{n=0}\frac{s^n}{n!}[A,B]_{(n)}]\rangle_{\mathbb{1}}$ where $[A,B]_{(n)}=[A,[A,B]_{(n-1)}]$ with $[A,B]_{(0)}=B$. Clearly, the charge is conserved if $\langle A,[A,C]\rangle_{\mathbb{1}}=0$ holds for all $C\in\mathfrak{g}$.
		\end{itemize}
		The condition $\langle A,[A,B]\rangle_{\mathbb{1}}=0$ for all $B\in \mathfrak{g}$ has an elegant geometric interpretation. The adjoint action of $B$ on $A$ can be written as $\mathrm{Adj}_{\ex^{sB}}A=\ex^{sB}A-A\ex^{sB}$. How does the norm of $A$ change under the adjoint action? We need to compute
		\begin{align}
		\frac{d}{ds}\lVert \mathrm{Adj}_{\ex^{sB}}A\rVert^2|_{s=0}=\frac{d}{ds}\langle\mathrm{Adj}_{\ex^{sB}}A,\mathrm{Adj}_{\ex^{sB}}A\rangle_{\mathbb{1}}|_{s=0}=-\langle A,[A,B]\rangle_{\mathbb{1}}\,.
		\end{align}
		This reproduces above condition, but provides a geometrical interpretation of it. If we start to move the Lie algebra element around under the adjoint action of an arbitrary $B$, its norm should not change to linear order. This means the point $A$ is a critical point of the norm function on its adjoint orbit.
	\end{enumerate}

\subsection{Geodesics for bi-invariant metric}
A metric is bi-invariant, if we can compute the inner product $\langle X,Y\rangle_{M}$ by either left- or right-translation to the identity. This implies
\begin{align}
\langle X,Y\rangle_{M}=\langle M^{-1}X,M^{-1}Y\rangle_{\mathbb{1}}=\langle XM^{-1},YM^{-1}\rangle_{\mathbb{1}}\,.
\end{align}
Given a bi-invariant inner product $\langle A,B\rangle_{\mathbb{1}}$ at the identity, we can first right-translate it to $M$ and then left-translate it back, which must agree with the original product:
\begin{align}
\langle A,B\rangle_{\mathbb{1}}=\langle MAM^{-1},MBM^{-1}\rangle_{\mathbb{1}}\,.
\end{align}
From this, we see that the requirement of bi-invariance is equivalent to the condition that the metric $\langle\cdot,\cdot\rangle_{\mathbb{1}}$ is invariant under the adjoint action $\mathrm{Adj}_MA=MAM^{-1}$ of any group element $M$. If we have a Lie group $\mathcal{G}$ with a right-invariant metric that satisfies this condition, the metric is bi-invariant. Moreover, invariance under the adjoint action of any group element $M$ includes $M=\ex^{sB}$ and thus implies
\begin{align}
\frac{d}{ds}\lVert \mathrm{Adj}_{\ex^{sB}}A\rVert^2|_{s=0}=-\langle A,[A,B]\rangle_{\mathbb{1}}=0
\end{align}
for all $B\in\mathfrak{g}$. This means every path $\gamma(s)=\ex^{s A}$ is a geodesic.

We recall that the natural metric that we chose on $\mathrm{SO}(2N)$ was given by
\begin{align}
\langle A,B\rangle_{\mathbb{1}}=\mathrm{Tr}(AGB^\intercal g)=A^a{}_bG^{bc}(B^\intercal)_c{}^dg_{da}\,,
\end{align}
where $G$ and $g$ refer to the metric governing the anticommutation relations $\{\xi^a,\xi^b\}=G^{ab}$. This metric is positive definite, but it is also invariant under the adjoint action, which we can check by computing
\begin{align}
\langle MAM^{-1},MBM^{-1}\rangle_{\mathbb{1}}&=\mathrm{Tr}\big(MAM^{-1}G(MBM^{-1})^\intercal g\big)\\
&=\mathrm{Tr}\big(A\underbrace{M^{-1}G(M^{-1})^\intercal}_{=G} B^\intercal \underbrace{M^\intercal g M}_{=g}\big)=\langle A,B\rangle_{\mathbb{1}}\,,
\end{align}
where we used cyclicity of the trace to move $M$ from the front to the end. With this in hand, we know that extending $\langle\cdot,\cdot\rangle_{\mathbb{1}}$ to a right-invariant metric gives actually rise to a bi-invariant metric. In particular, all geodesics departing from the identity are given by one-parameter subgroups $\gamma(s)=\ex^{sA}$.

\section{Minimal geodesics in $\mathrm{SO}(2N)$}\label{app:minimal}
\begin{figure}
	\begin{center}
		\includegraphics{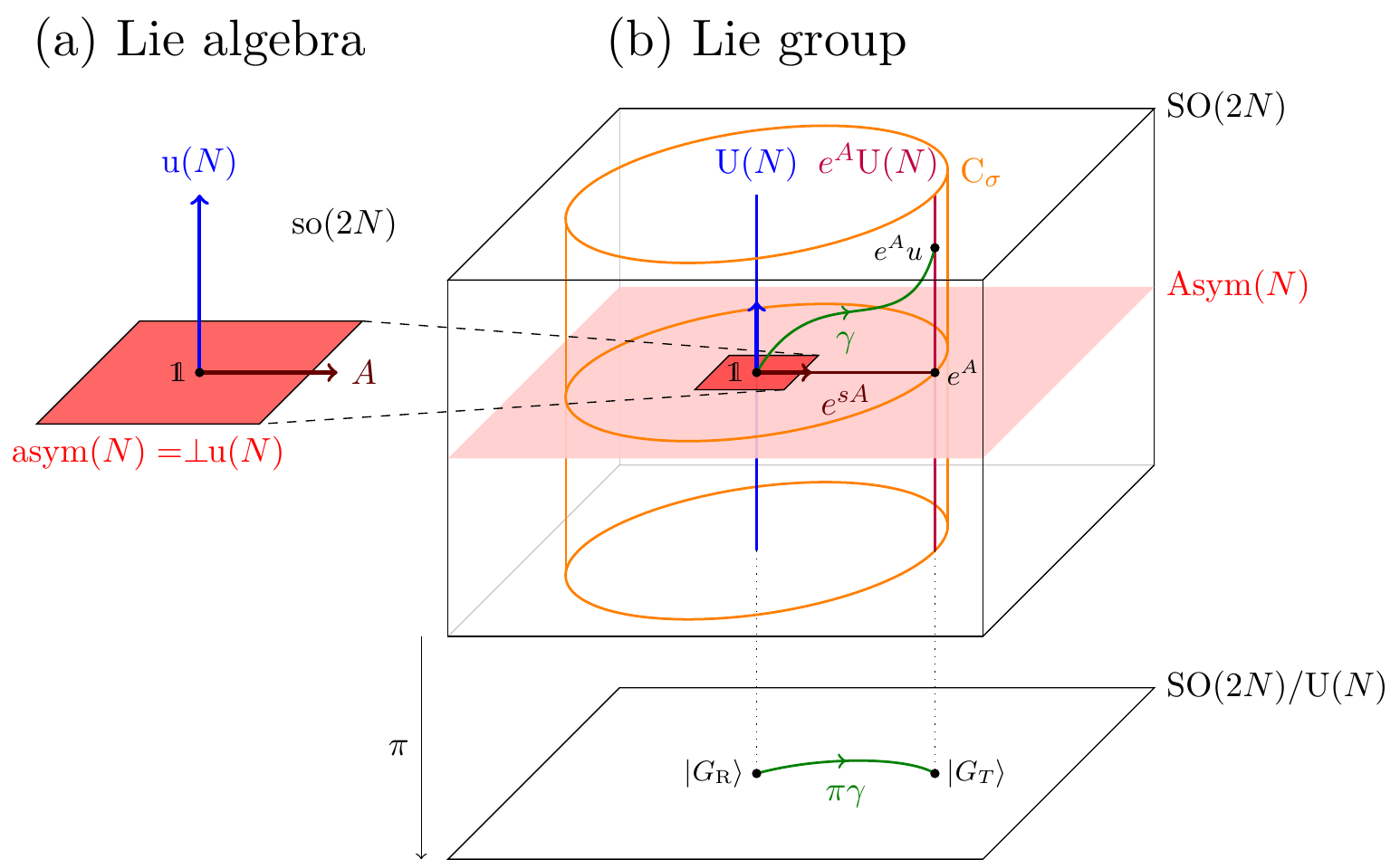}
	\end{center}
	\caption{This sketch illustrates the geometry of the Lie algebra $\mathrm{so}(2N)$ and the Lie group $\mathrm{SO}(2N)$. (a) The Lie algebra can be decomposed as $\mathrm{so}(2N)=\mathrm{u}(N)\oplus \mathrm{asym}(N)$, such that $\mathrm{asym}(N)$ is the orthogonal complement $\perp\!\!\mathrm{u}(N)$ of $\mathrm{u}(N)$. In particular, we can choose a vector $A\in\mathrm{asym}(N)$ to find the path $\ex^{sA}$ that connects $\mathbb{1}$ with $\ex^{A}\in\mathrm{Asym}(N)\subset\mathrm{SO}(2N)$. (b) The Lie group can be represented as fiber bundle over its quotient given by the symmetric space $\mathrm{SO}(2N)/\mathrm{U}(N)$. This base manifold can be interpreted as the space of Gaussian quantum states. The fiber over the reference state $|G_\mathrm{R}\rangle$ is given by the subgroup $\mathrm{U}(N)\subset\mathrm{SO}(2N)$, while the fiber $\ex^A\mathrm{U}(N)$ over any target state $|G_T\rangle$ is not a subgroup. We consider a path $\gamma$ in the group that connects $\mathbb{1}$ to some other group element $M=\ex^Au$. Such a point lies on the cylinder $\mathrm{C}_\sigma$ for $\sigma=\lVert A\rVert$.  Every curve $\gamma$ in the group can be projected down to a curve $\pi\gamma$ in the manifold of Gaussian states. The vertical submanifold $\mathrm{Asym}(N)=\exp\big(\mathrm{asym}(N)\big)$ is generated by exponentiating $\mathrm{asym}(N)$ and it plays an important role because it contains the minimal geodesics. Note that the $\mathrm{Asym}(N)$ has a complicated topology and intersects the fibers several times, but we only sketched a single layer corresponding to the region of $\mathrm{Asym}(N)$ near the identity. In particular, the straight line $\ex^{sA}$ connecting $\mathbb{1}$ with $\ex^A$ will turn out to be the minimal geodesic between $\mathbb{1}$ and the fiber $\ex^A\mathrm{U}(N)$. \emph{Note that we do not show the vector field $R$ consisting of radially outwards pointing unit vectors on the cylindric surfaces $\mathrm{C}_\sigma$, such that the curves $\ex^{sA}u$ are its integral curves. }
\label{geometry}	}
\end{figure}

We will give a general proof on which geodesics are the minimal ones connecting the identity with an equivalent class of unitaries that prepare the same state. In particular, we show that any such geodesic is nothing else than a collection of fermionic two-mode squeezing operations in fermionic normal modes. These results are the fermionic analogues to the derivation for bosons presented in the appendix of \cite{prep}. The geometry discussed in the following for the Lie algebra $\mathrm{so}(2N)$ and for the Lie group $\mathrm{SO}(2N)$ is illustrated in figure \ref{geometry}.

\subsection{Lie group geometry}
In the Nielsen approach to complexity, we equip the Lie group $\mathrm{SO}(2N)$ with right-invariant and positive metric. Such a metric is completely characterized by its value at the identity where we identity the tangent space $T_\mathbb{1}\mathrm{SO}(2N)$ with its Lie algebra $\mathrm{so}(2N)$. We represent a generator $A\in \mathrm{so}(2N)$ as matrices, namely linear maps $A^a{}_b$.

A natural choice for the invariant metric on $\mathrm{so}(2N)$ is given by
\begin{align}
\langle A,B\rangle_{\mathbb{1}}=A^a{}_bG^{bc}(B^\intercal)_c{}^d(g)_{da}=\mathrm{Tr}\left(AGB^\intercal g\right)=-\mathrm{Tr}(AB)\,,\label{eq:app_metric_identity}
\end{align}
where we used $GB^\intercal g=-B$ to find the RHS, which makes explicit that our natural inner product is just minus the Killing form on $\mathrm{SO}(2N)$. In particular, extending this metric to a right-invariant metric over the whole group will give rise to a bi-invariant metric. Given two tangent vectors $X,Y\in T_M\mathrm{Sp}(2N,\mathbb{R})$ represented as matrices at point $M\in\mathrm{SO}(2N)$, we can compute their inner product by multiplying with $M^{-1}$ from the right, leading to
\begin{align}
\langle X,Y\rangle_M=\langle XM^{-1},YM^{-1}\rangle_{\mathbb{1}}=-\mathrm{Tr}(XM^{-1}YM^{-1})\,.
\end{align}

\subsection{Fiber bundle structure}
The choice of the reference state $|\Omega_\mathrm{R}\rangle$ equips the Lie group $\mathrm{SO}(2N)$ with a fiber bundle structure. There exist group elements $M$ that leave the reference state invariant, such that $\Omega_\mathrm{R}=M\Omega_\mathrm{R}M^\intercal$. Such group elements are both orthogonal (with respect to $G$) and symplectic (with respect to $\Omega_\mathrm{R}$), so that they form the subgroup
\begin{align}
\mathrm{U}(N)=\mathrm{SO}(2N)\cap\mathrm{Sp}(2N,\mathbb{R})\,.
\end{align}
The different choices of subgroups are in one-to-one correspondence to the different choices of metrics $G_\mathrm{R}$.\par
We define the equivalence relation $M\sim\tilde{M}$ if and only if $M\Omega_\mathrm{R}M^\intercal=\tilde{M}\Omega_\mathrm{R}\tilde{M}^\intercal$. This means acting with $M$ and $\tilde{M}$ on $\Omega_\mathrm{R}$ will give the same target state. In particular, the subgroup $\mathrm{U}(N)$ is equal to the equivalence class $[\mathbb{1}]$ of the identity. Moreover, for every pair $M\sim\tilde{M}$, there exists a $u\in \mathrm{U}(N)$, such that $Mu=\tilde{M}$. Therefore, $\mathrm{SO}(2N)$ becomes a fiber bundle where the fibers correspond to the different equivalence classes diffeomorphic to $\mathrm{U}(N)$ and the base manifold is given by the quotient
\begin{align}
\mathcal{M}=\mathrm{SO}(2N)/\!\sim\,=\mathrm{SO}(2N)/\mathrm{U}(N)\,.
\end{align}
For general $N$, this space has some non-trivial topology and is generally referred to as symmetric space of type DIII. We will refer to it as $\mathcal{M}$, the space of pure Gaussian states, and identify a point $[M]\in\mathcal{M}$ with the Gaussian state $|M\Omega_\mathrm{R}M^\intercal\rangle$ up to an overall complex phase.

\subsection{Cartan decomposition}
Identifying the Lie algebra $\mathrm{so}(2N)$ with the tangent space at the identity, we have a natural ``vertical'' subalgebra $\mathrm{u}(N)\subset\mathrm{so}(2N)$ that is tangential to the fiber $[\mathbb{1}]=\mathrm{U}(N)$. A priori, there is no natural ``horizontal'' complement to write the Lie algebra as a direct sum of a vertical and a horizontal part. However, by equipping the Lie algebra with the inner product $\langle\cdot,\cdot\rangle_{\mathbb{1}}$, we can choose the orthogonal complement
\begin{align}
\mathrm{asym}(N):=\left\{A\in\mathrm{so}(2N)\big|\langle A,B\rangle_{\mathbb{1}}=0\,\forall\, B\in\mathrm{u}(N)\right\}
\end{align}
In contrast to $\mathrm{u}(N)$, $\mathrm{asym}(N)$ is not a subalgebra. Its name stems from the fact that the decomposition
\begin{align}
\mathrm{so}(2N)=\mathrm{asym}(N)\oplus \mathrm{u}(N)
\end{align}
is equivalent to splitting the set of generators into symmetric and antisymmetric matrices with respect to the symplectic form $\Omega_\mathrm{R}$ of the reference state:
\begin{itemize}
	\item \textbf{Vertical subspace $\mathrm{u}(N)$}\\
	A generator $B$ in the subspace $\mathrm{u}(N)$ must generate transformations that preserve the reference state $|\Omega_\mathrm{R}\rangle$. It must therefore be both orthogonal (\eg $BG=GB^\intercal$) and symplectic with respect $\Omega_{\mathrm{R}}$
	\begin{align}
	B\Omega_\mathrm{R}=\Omega_\mathrm{R}^\intercal B^\intercal\,,
	\end{align}
	which is equivalent to $B\Omega_{\mathrm{R}}$ being a symmetric matrix in a basis where $\Omega_\mathrm{R}$ takes the standard form from eq. (\ref{asymb}).
	\item \textbf{Horizontal subspace $\mathrm{asym}(N)=\perp\!\!\mathrm{u}(N)$}\\
	A generator $A$ that is orthogonal to all elements $B\in\mathrm{u}(N)$ satisfies
	\begin{align}
	0=\langle A,B\rangle_{\mathbb{1}}=\mathrm{Tr}(AGB^\intercal g)\,.
	\end{align}
	Using $GB^\intercal g=-A$, we can rewrite this expression as $-\mathrm{Tr}(AB)$. We will go a step further by inserting $\mathbb{1}=\Omega_{\mathrm{R}}\omega_{\mathrm{R}}$, leading to
	\begin{align}
	0=\langle A,B\rangle_{\mathbb{1}}=-\mathrm{Tr}(A\Omega_{\mathrm{R}}\omega_{\mathrm{R}}B)=-(A\Omega_{\mathrm{R}})^{ab}(\omega_{\mathrm{R}}B)_{ba}\,.
	\end{align} 
	In this expression, $(\omega_{\mathrm{R}}B)_{ba}$ is symmetric as a consequence of $B\Omega_{\mathrm{R}}=\Omega_{\mathrm{R}}^\intercal B^\intercal$. The fact that the inner product vanishes is therefore equivalent to $(A\Omega_{\mathrm{R}})$ being antisymmetric with respect $\Omega_{\mathrm{R}}$, namely
	\begin{align}
	A\Omega_\mathrm{R}=-\Omega_\mathrm{R}^\intercal A^\intercal\,.
	\end{align}
	We can refer to $\mathrm{asym}(N)$ as orthogonal complement $\perp\!\!\mathrm{u}(N)$ of $\mathrm{u}(N)$.
\end{itemize}
Exponentiating $\mathrm{asym}(N)$ defines the $N(N-1)$-dimensional submanifold
\begin{align}
\mathrm{Asym}(N)=\exp\left(\mathrm{asym}(N)\right)=\left\{\ex^A\big|A\in\mathrm{asym}(N)\right\}
\end{align}
consisting of all special-orthogonal group elements that are antisymmetric with respect to $\Omega_\mathrm{R}$.\par
The Cartan decomposition of a orthogonal group element $M$ is given by
\begin{align}
M=Tu\quad\text{with}\quad T=\sqrt{M\Omega_\mathrm{R}M^\intercal \omega_\mathrm{R}}\in\mathrm{Asym}(N)\quad\text{and}\quad u=T^{-1}M\in\mathrm{U}(N)\,.
\end{align}
It is unique and provides \emph{locally} (around the identity) a diffeomorphism between the special orthogonal group and the Cartesian product $\mathrm{Asym}(N)\times\mathrm{U}(N)$. In particular, it provides a \emph{local} trivialization of the fiber bundle $\mathrm{SO}(2N)$ where the base manifold is identified with the surface $\mathrm{Asym}(N)$ from which we can move up and down along the fiber by multiplying with group elements $u\in\mathrm{U}(N)$. Due to the fact that $\mathrm{Asym}(N)$ is \emph{locally} diffeomorphic to $\mathrm{asym}(N)$, we can use the pair $(A,u)$ as generalized coordinates for group elements $M(A,u)$ in a neighborhood around the identity:
\begin{align}
M(A,u)=\ex^Au\quad\text{with}\quad A\in \mathrm{asym}(N)\quad\text{and}\quad u\in\mathrm{U}(N)\,.
\end{align}

\subsection{Cylindrical foliation}
We can foliate the symplectic group by generalized cylinders defined as
\begin{align}
\mathrm{C}_\s=\left\{\ex^{A}u\big|A\in\mathrm{asym}(N),\lVert A\rVert=\s,u\in\mathrm{U}(N)\right\}
\end{align}
with the topology $S^{N(N-1)-1}\times\mathrm{U}(N)$. Moreover, we will define the radial vector field $R$ at point $M(A,u)\in\mathrm{SO}(2N)$ given by
\begin{align}
R_{M(A,u)}=\frac{\ex^AAu}{\lVert A\rVert}\,.
\end{align}
We will prove that this vector fields points radially outwards and is everywhere orthogonal to the cylindrical surfaces $\mathrm{C}_\s$.
Therefore, we need to show that $R$ is indeed orthogonal to the surfaces $\mathrm{C}_\s$. We will prove this individually for different directions. Note that the normalization $1/\lVert A\rVert$ is irrelevant here.
\begin{itemize}
	\item \textbf{Orthogonality to the $\mathrm{U}(N)$ fiber:}\\
	We show that $R$ is orthogonal to any vector pointing along the $\mathrm{U}(N)$ fiber. Let $X\in\mathrm{u}(N)$, so that $\ex^{A}uX$ points in the direction of the $\mathrm{U}(N)$ fiber at point $M(A,u)$. We can compute the inner product
	\begin{align}
	\langle R_{M(A,u)},\ex^{A}uX\rangle=\frac{1}{\Vert A\rVert}\langle \ex^{A}Au,\ex^{A}uX\rangle_{\ex^{A}u}
	\end{align}
	We define $Y=uXu^{-1}$ which lies in $\mathrm{u}(N)$ because $\mathrm{u}(N)$ is a subgroup. This implies $uX=Yu$. We can therefore compute
	\begin{align}
	\langle \ex^{A}uX,\ex^{A}Au\rangle_{\ex^{A}u}=\langle \ex^{A}Yu,\ex^{A}Au\rangle_{\ex^{A}u}=\langle \ex^{A}Y,\ex^{A}A\rangle_{\ex^{A}}=\langle \ex^{A}Y\ex^{-A},A\rangle_{\mathbb{1}}\,.
	\end{align}
	At this point, we can use the explicit form of the metric at the identity given by
	\begin{align}
	\langle \ex^{A}Y\ex^{-A},A\rangle_{\mathbb{1}}=\mathrm{Tr}\left(\ex^{A}Y\ex^{-A}GA^\intercal g\right)=-\mathrm{Tr}\left(\ex^{A}Y\ex^{-A}A\right)=\mathrm{Tr}\left(YA\right)=0\,,
	\end{align}
	where we used $GA^\intercal g=-A$ for all $A\in\mathrm{so}(2N)$. The vanishing trace
	\item \textbf{Orthogonality to a generator $A\in \mathrm{asym}(N)$ preserving $\mathrm{C}_\s$:}\\
	This second computation is slightly more involved. Let us look at a point $M=\ex^{A}u$ and ask what are the directions in $T_{M}\mathrm{SO}(2N)$ that are tangential to the surface $\mathrm{C}_\s$, but also to the surface $\exp\big(\mathrm{asym}(N)\big)u$. We can describe such elements by choosing a second generator $B\in\mathrm{asym}(N)$ that is orthogonal to $A$ with $\lVert A\rVert=\lVert B\rVert$. The circle
	\begin{align}
	\gamma(t)=\ex^{(\cos{(t)}\,A+\sin(t)\,B)}u
	\end{align}
	lies in $\mathrm{Asym}(N)$ and on $\mathrm{C}_\s$ with $\s=\lVert A\rVert=\lVert B\rVert$. This gives rise to the tangent vector
	\begin{align}
	\dot{\gamma}(0)=\frac{d}{dt}\ex^{(A+t\,B)}|_{t=0}\,.
	\end{align}
	We can compute the inner product with $R_{M(A,u)}$ using $\langle A,B\rangle_{\mathbb{1}}=-\mathrm{Tr}(AB)$:
	\begin{align}
	\langle R_{M(A,u)},\dot{\gamma}(0)\rangle_{\ex^Au}=\frac{1}{\lVert A\rVert}\langle A,\dot{\gamma}(0)u^{-1}\ex^{-A}\rangle_{\mathbb{1}}=-\frac{d}{dt}\mathrm{Tr}(A\,\ex^{(A+t\,B)}\ex^{-A})\,.
	\end{align}
	At this point, we can write out the full exponential as
	\begin{align}
	\sum^\infty_{n,m=0}\frac{d}{dt}\frac{\mathrm{Tr}[A\left(A+tB)^n(-A)^m\right]_{t=0}}{n!m!}&=\mathrm{Tr}\left[AB\!\!\!\!\sum^\infty_{n=1,m=0}\!\!\!\!\frac{(A)^{n-1}(-A)^m}{(n-1)!m!}\right]\\
	&=\mathrm{Tr}(AB)=0\,,
	\end{align}
	where we used the fact that trace is cyclic and that $B$ was chosen orthogonal to $A$. Note that the sum just gives the identity.
\end{itemize}
This proves that we have indeed a vector field $R$ that is everywhere orthogonal to the cylindrical surfaces $\mathrm{C}_\s$. Furthermore, we can quickly confirm that this vector field has indeed constant length equal to $1$, by computing
\begin{align}
\langle R_{M(A,u)},R_{M(A,u)}\rangle_{M(A,u)}=\frac{\langle \ex^AAu,\ex^AAu\rangle_{\ex^Au}}{\lVert A\rVert^2}=\frac{\langle A,A\rangle_{\mathbb{1}}}{\lVert A\rVert^2}=1\,.
\end{align}
Given a trajectory $\gamma: [0,1]\to \mathrm{SO}(2N): t\mapsto \gamma(t)$, we can compute how the coordinate $\s(\gamma)$ changes. Due to the fact that the vector field $R$ is orthogonal to the surface $\mathrm{C}_\s$ of constant $\s$ and correctly normalized, we have
\begin{align}
d\s=\langle R_{\gamma(t)},\dot{\gamma}(t)\rangle_{\gamma(t)}\,.
\end{align}

\subsection{Inequality for the geodesic length}
We will now use the cylindrical structure to bound the geodesic length from below. Given an arbitrary point $M(A,u)=\ex^{A}u$ on the cylinder $\mathrm{C}_\sigma$, let us assume that we have already found the shortest path connecting the identity $\mathbb{1}$ with $M(A,u)$. This path may be given by $\gamma(s)$ with $\gamma(0)=\mathbb{1}$ and $\gamma(1)=M(A,u)$. We can compute the change $d\s$ as the inner product
\begin{align}
d\s(s)=ds\,\langle \dot{\gamma}(s),R_{\gamma(s)}\rangle_{\gamma(s)}\,.
\end{align}
Clearly, if we integrate this inner product we find how far we move in the $\s$-direction. This follows directly from the fact that moving in the direction of $R$ increases $\s$ with a constant rate, while moving along any orthogonal direction does not change $\s$. Therefore, we have
\begin{align}
\s=\int^1_{0}\!\!\!d\s(s)=\int^1_{0}\!\!\!ds\,\langle\dot{\gamma}(s),R_{\gamma(s)}\rangle_{\gamma(s)}\,.
\end{align}
We can compare this with the actual length of the geodesic given by
\begin{align}
\lVert \gamma\rVert:=\int^1_{0}\!\!\!ds\,\lVert\dot{\gamma}(s)\rVert\,.
\end{align}
At this point, we should note that $\langle\dot{\gamma}(s),R_{\gamma(s)}\rangle_{\gamma(s)}\leq \lVert\dot{\gamma}(s)\rVert$ for all $s$. This follows from the fact that we are projecting onto the unit vector $R$, so this projection is at most the length of $\dot{\gamma}(s)$. We can combine these two equations to find the important inequality
\begin{align}
\s\leq \lVert\gamma\rVert\,,
\end{align}
stating compactly that any path connecting $\mathbb{1}$ with $M\in\mathrm{C}_\s$ must have a length of $\s$ or more.

At this point, we have not proven that for every $M\in\mathrm{C}_\s\subset\mathrm{SO}(2N)$ there exists a path with length $\s$ connecting $\mathbb{1}$ with $M$ and there certainly are points $M$ where we cannot find such a shortest path. However, we are interested in the minimal geodesic that connects the identity $\mathbb{1}$ with an arbitrary point in the the fiber $[M]$. This means if we find a single path that does this with length $\s$, we have proven that this is indeed \emph{the} optimal path and there is no shorter one.

\subsection{Shortest path to a fiber $\ex^{A}\mathrm{U}(N)$}
We will now show explicitly that for every fiber $\ex^{A}\mathrm{U}(N)$ with $\sigma=\lVert A\rVert$, there exists a path of length $\sigma$ that connects the identity with the point $\ex^A$ on this fiber. This path is given by
\begin{align}
\gamma(s)=\ex^{sA}
\end{align}
and reaches the representative $\ex^{A}$ at $s=1$. This path has length $\lVert\gamma\rVert=\lVert A\rVert=\s$. At this point, we have proven that for our chosen inner product $\langle A,B\rangle_{\mathbb{1}}=-\mathrm{Tr}(AB)$, the shortest path is indeed always given by $\ex^{sA}$ with $A\in\mathrm{asym}(N)$.

We can now ask how $A$ is related to the target state $|\Omega_\mathrm{T}\rangle$. We must have
\begin{align}
\Omega_{\mathrm{T}}=\ex^{A}\Omega_\mathrm{R}\ex^{A^\intercal}\,.
\end{align}
Now requiring that $A\in\mathrm{asym}(N)$ implies that $M\Omega_{\mathrm{R}}=\ex^A\Omega_{\mathrm{R}}$ is antisymmetric. In an invariant language, we have equivalently
\begin{align}
\omega_\mathrm{R}M=M^\intercal \omega_\mathrm{R}\,.
\end{align}
With this in hand, we can claim that the linear map $M=\sqrt{\Omega_{\mathrm{T}}\omega_\mathrm{R}}$ will do the job. Importantly, $M$ satisfies $M\Omega_\mathrm{R}=\Omega_\mathrm{R}M^\intercal$. We can check explicitly
\begin{align}
\sqrt{\Omega_{\mathrm{T}}g_\mathrm{R}}\Omega_\mathrm{R}\sqrt{\Omega_{\mathrm{T}}\omega_\mathrm{R}}^\intercal=\sqrt{\Omega_{\mathrm{T}}\omega_\mathrm{R}}\sqrt{\Omega_{\mathrm{T}}\omega_\mathrm{R}}\Omega_\mathrm{R}=\Omega_{\mathrm{T}}\omega_\mathrm{R}\Omega_\mathrm{R}=\Omega_{\mathrm{T}}\,.
\end{align}
The algebra element that generates $M$ is given by $A=\log{M}=(\log{\Omega_{\mathrm{T}}\omega_{\mathrm{R}}})/2$. We have $\s=\lVert A\rVert=\lVert\log{\Omega_{\mathrm{T}}\omega_\mathrm{R}}\rVert/2$. Let us note at this point that all expressions, such as $\log \Omega_{\mathrm{T}}\omega_\mathrm{R}$ and $\sqrt{\Omega_{\mathrm{T}}\omega_\mathrm{R}}$ are well defined, because $\Omega_{\mathrm{T}}\omega_\mathrm{R}$ is an orthogonal matrix. This fact implies that $\Omega_{\mathrm{T}}\omega_\mathrm{R}$ is (a) diagonalizable and (b) has conjugate complex eigenvalues $\pm \ex^{\ii \varphi}$.\footnote{Note that the definition of square root involves a small subtlety: If we describe the conjugate eigenvalues $(\ex^{\ii r_i},\ex^{-\ii r_i})$ using $r_i\in [0,\pi]$, the square root $\sqrt{\Omega_{\mathrm{T}}\omega_\mathrm{R}}$ has the same eigenvectors, but eigenvalues $(\ex^{\ii r_i/2},\ex^{-\ii r_i/2})$ which defines $\sqrt{\Omega_{\mathrm{T}}\omega_\mathrm{R}}$ uniquely for $r_i\neq\pi$. For $r_i=\pi$, the square root $\sqrt{\Omega_{\mathrm{T}}\omega_\mathrm{R}}$ has eigenvalues $(\ex^{\ii r_i/2},\ex^{-\ii r_i/2})$, but the assignment to the eigenvectors could be interchanged. However, either choice will lead to a valid definition for the square root $\sqrt{\Omega_{\mathrm{T}}\omega_\mathrm{R}}$ for the purpose here.} The linear map $\Omega_{\mathrm{T}}\omega_\mathrm{R}$ encodes the invariant information about the relation between the reference state $|\Omega_\mathrm{R}\rangle$ and the target state $|\Omega_{\mathrm{T}}\rangle$, which we can refer to as
\begin{align}
\Delta^a{}_b=(\Omega_{\mathrm{T}})^{ac}(\omega_\mathrm{R})_{cb}\,.
\end{align}
The eigenvalues of $\Delta$ come in conjugate pairs $(\ex^{\ii r_i},\ex^{-\ii r_i})$. We can compute the geodesic distance, which is equal to the norm $\lVert A\rVert$, directly from $\Delta$:
\begin{align}
\lVert A\rVert=\frac{\sqrt{\mathrm{Tr}(\ii\log\Delta)^2}}{2}\,.
\end{align}

\newpage
\bibliographystyle{JHEP}
\bibliography{bib_references}

\providecommand{\href}[2]{#2}\begingroup\raggedright\begin{thebibliography}{10}

\bibitem{Susskind:2014rva}
L.~Susskind, {\it {Computational Complexity and Black Hole Horizons}},  {\em
  Fortsch. Phys.} {\bf 64} (2016) 24--43,
  [\href{http://arxiv.org/abs/1403.5695}{{\tt arXiv:1403.5695}}].

\bibitem{Susskind:2014jwa}
L.~Susskind and Y.~Zhao, {\it {Switchbacks and the Bridge to Nowhere}},
  \href{http://arxiv.org/abs/1408.2823}{{\tt arXiv:1408.2823}}.

\bibitem{Susskind:2014moa}
L.~Susskind, {\it {Entanglement is not enough}},  {\em Fortsch. Phys.} {\bf 64}
  (2016) 49--71, [\href{http://arxiv.org/abs/1411.0690}{{\tt
  arXiv:1411.0690}}].

\bibitem{Stanford:2014jda}
D.~Stanford and L.~Susskind, {\it {Complexity and Shock Wave Geometries}},
  {\em Phys. Rev.} {\bf D90} (2014), no.~12 126007,
  [\href{http://arxiv.org/abs/1406.2678}{{\tt arXiv:1406.2678}}].

\bibitem{Brown:2015bva}
A.~R. Brown, D.~A. Roberts, L.~Susskind, B.~Swingle, and Y.~Zhao, {\it
  {Holographic Complexity Equals Bulk Action?}},  {\em Phys. Rev. Lett.} {\bf
  116} (2016), no.~19 191301, [\href{http://arxiv.org/abs/1509.07876}{{\tt
  arXiv:1509.07876}}].

\bibitem{Brown:2015lvg}
A.~R. Brown, D.~A. Roberts, L.~Susskind, B.~Swingle, and Y.~Zhao, {\it
  {Complexity, action, and black holes}},  {\em Phys. Rev.} {\bf D93} (2016),
  no.~8 086006, [\href{http://arxiv.org/abs/1512.04993}{{\tt
  arXiv:1512.04993}}].

\bibitem{Alishahiha:2015rta}
M.~Alishahiha, {\it {Holographic Complexity}},  {\em Phys. Rev.} {\bf D92}
  (2015), no.~12 126009, [\href{http://arxiv.org/abs/1509.06614}{{\tt
  arXiv:1509.06614}}].

\bibitem{Cai:2016xho}
R.-G. Cai, S.-M. Ruan, S.-J. Wang, R.-Q. Yang, and R.-H. Peng, {\it {Action
  growth for AdS black holes}},  {\em JHEP} {\bf 09} (2016) 161,
  [\href{http://arxiv.org/abs/1606.08307}{{\tt arXiv:1606.08307}}].

\bibitem{Lehner:2016vdi}
L.~Lehner, R.~C. Myers, E.~Poisson, and R.~D. Sorkin, {\it {Gravitational
  action with null boundaries}},  {\em Phys. Rev.} {\bf D94} (2016), no.~8
  084046, [\href{http://arxiv.org/abs/1609.00207}{{\tt arXiv:1609.00207}}].

\bibitem{Yang:2016awy}
R.-Q. Yang, {\it {Strong energy condition and complexity growth bound in
  holography}},  {\em Phys. Rev.} {\bf D95} (2017), no.~8 086017,
  [\href{http://arxiv.org/abs/1610.05090}{{\tt arXiv:1610.05090}}].

\bibitem{Chapman:2016hwi}
S.~Chapman, H.~Marrochio, and R.~C. Myers, {\it {Complexity of Formation in
  Holography}},  {\em JHEP} {\bf 01} (2017) 062,
  [\href{http://arxiv.org/abs/1610.08063}{{\tt arXiv:1610.08063}}].

\bibitem{Carmi:2016wjl}
D.~Carmi, R.~C. Myers, and P.~Rath, {\it {Comments on Holographic Complexity}},
   {\em JHEP} {\bf 03} (2017) 118, [\href{http://arxiv.org/abs/1612.00433}{{\tt
  arXiv:1612.00433}}].

\bibitem{Carmi:2017jqz}
D.~Carmi, S.~Chapman, H.~Marrochio, R.~C. Myers, and S.~Sugishita, {\it {On the
  Time Dependence of Holographic Complexity}},  {\em JHEP} {\bf 11} (2017) 188,
  [\href{http://arxiv.org/abs/1709.10184}{{\tt arXiv:1709.10184}}].

\bibitem{Reynolds:2016rvl}
A.~Reynolds and S.~F. Ross, {\it {Divergences in Holographic Complexity}},
  {\em Class. Quant. Grav.} {\bf 34} (2017), no.~10 105004,
  [\href{http://arxiv.org/abs/1612.05439}{{\tt arXiv:1612.05439}}].

\bibitem{Zhao:2017iul}
Y.~Zhao, {\it {Complexity, boost symmetry, and firewalls}},
  \href{http://arxiv.org/abs/1702.03957}{{\tt arXiv:1702.03957}}.

\bibitem{Reynolds:2017lwq}
A.~Reynolds and S.~F. Ross, {\it {Complexity in de Sitter Space}},  {\em Class.
  Quant. Grav.} {\bf 34} (2017), no.~17 175013,
  [\href{http://arxiv.org/abs/1706.03788}{{\tt arXiv:1706.03788}}].

\bibitem{Couch:2017yil}
J.~Couch, S.~Eccles, W.~Fischler, and M.-L. Xiao, {\it {Holographic complexity
  and non-commutative gauge theory}},
  \href{http://arxiv.org/abs/1710.07833}{{\tt arXiv:1710.07833}}.

\bibitem{Swingle:2017zcd}
B.~Swingle and Y.~Wang, {\it {Holographic Complexity of
  Einstein-Maxwell-Dilaton Gravity}},
  \href{http://arxiv.org/abs/1712.09826}{{\tt arXiv:1712.09826}}.

\bibitem{Fu:2018kcp}
Z.~Fu, A.~Maloney, D.~Marolf, H.~Maxfield, and Z.~Wang, {\it {Holographic
  complexity is nonlocal}},  \href{http://arxiv.org/abs/1801.01137}{{\tt
  arXiv:1801.01137}}.

\bibitem{CHM}
H.~Casini, M.~Huerta, and R.~C. Myers, {\it {Towards a derivation of
  holographic entanglement entropy}},  {\em JHEP} {\bf 05} (2011) 036,
  [\href{http://arxiv.org/abs/1102.0440}{{\tt arXiv:1102.0440}}].

\bibitem{Lewkowycz:2013nqa}
A.~Lewkowycz and J.~Maldacena, {\it {Generalized gravitational entropy}},  {\em
  JHEP} {\bf 08} (2013) 090, [\href{http://arxiv.org/abs/1304.4926}{{\tt
  arXiv:1304.4926}}].

\bibitem{Dong:2016hjy}
X.~Dong, A.~Lewkowycz, and M.~Rangamani, {\it {Deriving covariant holographic
  entanglement}},  {\em JHEP} {\bf 11} (2016) 028,
  [\href{http://arxiv.org/abs/1607.07506}{{\tt arXiv:1607.07506}}].

\bibitem{Jordan:2011ne}
S.~P. Jordan, K.~S.~M. Lee, and J.~Preskill, {\it {Quantum Algorithms for
  Quantum Field Theories}},  {\em Science} {\bf 336} (2012) 1130--1133,
  [\href{http://arxiv.org/abs/1111.3633}{{\tt arXiv:1111.3633}}].

\bibitem{Jordan:2011ci}
S.~P. Jordan, K.~S.~M. Lee, and J.~Preskill, {\it {Quantum Computation of
  Scattering in Scalar Quantum Field Theories}},
  \href{http://arxiv.org/abs/1112.4833}{{\tt arXiv:1112.4833}}.

\bibitem{Jordan:2014tma}
S.~P. Jordan, K.~S.~M. Lee, and J.~Preskill, {\it {Quantum Algorithms for
  Fermionic Quantum Field Theories}},
  \href{http://arxiv.org/abs/1404.7115}{{\tt arXiv:1404.7115}}.

\bibitem{Jordan:2017lea}
S.~P. Jordan, H.~Krovi, K.~S.~M. Lee, and J.~Preskill, {\it {BQP-completeness
  of Scattering in Scalar Quantum Field Theory}},
  \href{http://arxiv.org/abs/1703.00454}{{\tt arXiv:1703.00454}}.

\bibitem{osborne2012hamiltonian}
T.~J. Osborne, {\it Hamiltonian complexity},  {\em Reports on Progress in
  Physics} {\bf 75} (2012), no.~2 022001.

\bibitem{gharibian2015quantum}
S.~Gharibian, Y.~Huang, Z.~Landau, S.~W. Shin, et~al., {\it Quantum hamiltonian
  complexity},  {\em Foundations and Trends{\textregistered} in Theoretical
  Computer Science} {\bf 10} (2015), no.~3 159--282.

\bibitem{Orus:2013kga}
R.~Orus, {\it {A Practical Introduction to Tensor Networks: Matrix Product
  States and Projected Entangled Pair States}},  {\em Annals Phys.} {\bf 349}
  (2014) 117--158, [\href{http://arxiv.org/abs/1306.2164}{{\tt
  arXiv:1306.2164}}].

\bibitem{vidal2009entanglement}
G.~Vidal, {\it {Entanglement Renormalization: an introduction}},  in {\em
  {Understanding Quantum Phase Transitions}} (L.~D. Carr, ed.).
\newblock CRC press, 2010.
\newblock \href{http://arxiv.org/abs/0912.1651}{{\tt arXiv:0912.1651}}.

\bibitem{Hashimoto:2017fga}
K.~Hashimoto, N.~Iizuka, and S.~Sugishita, {\it {Time Evolution of Complexity
  in Abelian Gauge Theories - And Playing Quantum Othello Game -}},
  \href{http://arxiv.org/abs/1707.03840}{{\tt arXiv:1707.03840}}.

\bibitem{Jefferson:2017sdb}
R.~A. Jefferson and R.~C. Myers, {\it {Circuit complexity in quantum field
  theory}},  {\em JHEP} {\bf 10} (2017) 107,
  [\href{http://arxiv.org/abs/1707.08570}{{\tt arXiv:1707.08570}}].

\bibitem{Chapman:2017rqy}
S.~Chapman, M.~P. Heller, H.~Marrochio, and F.~Pastawski, {\it {Towards
  Complexity for Quantum Field Theory States}},
  \href{http://arxiv.org/abs/1707.08582}{{\tt arXiv:1707.08582}}.

\bibitem{Yang:2017nfn}
R.-Q. Yang, {\it {A Complexity for Quantum Field Theory States and Application
  in Thermofield Double States}},  \href{http://arxiv.org/abs/1709.00921}{{\tt
  arXiv:1709.00921}}.

\bibitem{Kim:2017qrq}
R.-Q. Yang, C.~Niu, C.-Y. Zhang, and K.-Y. Kim, {\it {Comparison of holographic
  and field theoretic complexities for time dependent thermofield double
  states}},  {\em JHEP} {\bf 02} (2018) 082,
  [\href{http://arxiv.org/abs/1710.00600}{{\tt arXiv:1710.00600}}].

\bibitem{simonR}
A.~P. Reynolds and S.~F. Ross, {\it {Complexity of the AdS Soliton}},
  \href{http://arxiv.org/abs/1712.03732}{{\tt arXiv:1712.03732}}.

\bibitem{Khan:2018rzm}
R.~Khan, C.~Krishnan, and S.~Sharma, {\it {Circuit Complexity in Fermionic
  Field Theory}},  \href{http://arxiv.org/abs/1801.07620}{{\tt
  arXiv:1801.07620}}.

\bibitem{prep}
S.~Chapman, J.~Eisert, L.~Hackl, M.~P. Heller, R.~Jefferson, H.~Marrochio,
  R.~C. Myers, and F.~Pastawski, {\it {Circuit Complexity for Thermofield
  Double States}},  {\em to appear} (2018).

\bibitem{Nielsen:2005mn1}
M.~A. Nielsen, {\it {A geometric approach to quantum circuit lower bounds}},
  \href{http://arxiv.org/abs/quant-ph/0502070}{{\tt quant-ph/0502070}}.

\bibitem{Nielsen:2006mn2}
M.~A. Nielsen, M.~R. Dowling, M.~Gu, and A.~M. Doherty, {\it {Quantum
  Computation as Geometry}},  {\em Science} {\bf 311} (2006) 1133--1135,
  [\href{http://arxiv.org/abs/quant-ph/0603161}{{\tt quant-ph/0603161}}].

\bibitem{Nielsen:2007mn3}
M.~A. Nielsen and M.~R. Dowling, {\it {The geometry of quantum computation}},
  \href{http://arxiv.org/abs/quant-ph/0701004}{{\tt quant-ph/0701004}}.

\bibitem{Brown:2016wib}
A.~R. Brown, L.~Susskind, and Y.~Zhao, {\it {Quantum Complexity and Negative
  Curvature}},  {\em Phys. Rev.} {\bf D95} (2017), no.~4 045010,
  [\href{http://arxiv.org/abs/1608.02612}{{\tt arXiv:1608.02612}}].

\bibitem{Brown:2017jil}
A.~R. Brown and L.~Susskind, {\it {The Second Law of Quantum Complexity}},
  \href{http://arxiv.org/abs/1701.01107}{{\tt arXiv:1701.01107}}.

\bibitem{alvarez1999comment}
J.~Alvarez and C.~G{\'o}mez, {\it A comment on fisher information and quantum
  algorithms},  {\em arXiv preprint quant-ph/9910115} (1999).

\bibitem{CCintroduction}
S.~Arora and B.~Barak, {\em Computational Complexity: A Modern Approach}.
\newblock Cambridge University Press, 2009.

\bibitem{CCintroduction2}
C.~Moore and S.~Mertens, {\em The Nature of Computation}.
\newblock Oxford University Press, 2011.

\bibitem{Aaronson:2016vto}
S.~Aaronson, {\it {The Complexity of Quantum States and Transformations: From
  Quantum Money to Black Holes}},  2016.
\newblock \href{http://arxiv.org/abs/1607.05256}{{\tt arXiv:1607.05256}}.

\bibitem{watrous}
J.~Watrous, {\it Quantum computational complexity},  in {\em Encyclopedia of
  complexity and systems science}, pp.~7174--7201.
\newblock Springer, 2009.

\bibitem{mimura1991topology}
M.~Mimura and H.~Toda, {\em Topology of Lie groups, I and II}, vol.~91.
\newblock American Mathematical Soc., 1991.

\bibitem{BH-BFGaussians}
E.~Bianchi and L.~Hackl, {\it {Bosonic and fermionic Gaussian states from
  K\"ahler structures}}, . To appear (2018).

\bibitem{Bianchi:2015fra}
E.~Bianchi, L.~Hackl, and N.~Yokomizo, {\it {Entanglement entropy of squeezed
  vacua on a lattice}},  {\em Phys. Rev.} {\bf D92} (2015), no.~8 085045,
  [\href{http://arxiv.org/abs/1507.01567}{{\tt arXiv:1507.01567}}].

\bibitem{bump}
D.~Bump, {\em Lie groups}.
\newblock Springer, 2004.

\bibitem{dutta1995real}
B.~Dutta, N.~Mukunda, R.~Simon, et~al., {\it The real symplectic groups in
  quantum mechanics and optics},  {\em Pramana} {\bf 45} (1995), no.~6
  471--497.

\bibitem{kirillov2008introduction}
A.~Kirillov, {\em An introduction to Lie groups and Lie algebras}, vol.~113.
\newblock Cambridge University Press, 2008.

\bibitem{peskin1995introduction}
M.~E. Peskin and D.~V. Schroeder, {\em An introduction to quantum field
  theory}.
\newblock Addison-Wesley Publishing Company, 1995.

\bibitem{bhatia2013matrix}
R.~Bhatia, {\em Matrix analysis}, vol.~169.
\newblock Springer Science \& Business Media, 2013.

\bibitem{watrous2018theory}
J.~Watrous, {\em Theory of Quantum Information}.
\newblock Cambridge University Press, 2018.
\newblock (See section 1.1).

\bibitem{lfh-unpublished}
L.~Hackl, {\it Notes on circuit complexity of bosonic and fermionic gaussian
  states},  {\em unpublished}.

\bibitem{comparison-paper}
S.~Chapman, L.~Hackl, M.~P. Heller, H.~Marrochio, and R.~C. Myers, {\it
  Geometry of circuit complexity: Nielsen versus fubini-study},  {\em in
  preparation}.

\bibitem{nonlocal}
Z.~Fu, A.~Maloney, D.~Marolf, H.~Maxfield, and Z.~Wang, {\it {Holographic
  complexity is nonlocal}},  {\em JHEP} {\bf 02} (2018) 072,
  [\href{http://arxiv.org/abs/1801.01137}{{\tt arXiv:1801.01137}}].

\bibitem{fubini1904sulle}
G.~Fubini, {\em Sulle metriche definite da una forma hermitiana: nota}.
\newblock Office graf. C. Ferrari, 1904.

\bibitem{study1905kurzeste}
E.~Study, {\it K{\"u}rzeste wege im komplexen gebiet},  {\em Mathematische
  Annalen} {\bf 60} (1905) 321--378.

\bibitem{abramX}
D.~S. Abrams and S.~Lloyd, {\it Simulation of many-body fermi systems on a
  universal quantum computer},  {\em Physical Review Letters} {\bf 79} (1997),
  no.~13 2586.

\bibitem{bravyY}
S.~B. Bravyi and A.~Y. Kitaev, {\it Fermionic quantum computation},  {\em
  Annals of Physics} {\bf 298} (2002), no.~1 210--226.

\bibitem{Jordan}
S.~P. Jordan, K.~S.~M. Lee, and J.~Preskill, {\it {Quantum Algorithms for
  Quantum Field Theories}},  {\em Science} {\bf 336} (2012) 1130--1133,
  [\href{http://arxiv.org/abs/1111.3633}{{\tt arXiv:1111.3633}}].

\bibitem{Gross:1974jv}
D.~J. Gross and A.~Neveu, {\it {Dynamical Symmetry Breaking in Asymptotically
  Free Field Theories}},  {\em Phys. Rev.} {\bf D10} (1974) 3235.

\end{thebibliography}\endgroup

\end{document}